%% file: Main_Document_HHG.tex
\documentclass[a4paper,11pt]{article}

\usepackage[pages=all, color=black, position={current page.south}, placement=bottom, scale=1, opacity=1, vshift=5mm]{background}

\SetBgContents{} 

\usepackage[margin=0.85in]{geometry} 

\usepackage{amsmath}
\usepackage{amsthm}
\usepackage{amssymb}

\usepackage[utf8]{inputenc}
\usepackage{hyperref}
\hypersetup{
	unicode,
	pdfauthor={Author One, Author Two, Author Three},
	pdftitle={Technical design report of a complete and compact broadband high-harmonics femtosecond beamline},
	pdfsubject={Technical design report of a complete and compact broadband high-harmonics femtosecond beamline},
	pdfkeywords={technical report, HHG},
	pdfproducer={LaTeX},
	pdfcreator={pdflatex}
}
\usepackage{graphicx, color}
\usepackage[sort&compress,numbers,square]{natbib}
\usepackage{braket}
\usepackage{lineno}
\usepackage[table]{xcolor}
\usepackage{caption}
\usepackage{multirow}
\usepackage{subcaption}
\usepackage{amsfonts}
\usepackage{mathrsfs} 
\usepackage{dutchcal} 

\title{\boldmath Technical design report of a complete and compact broadband high-harmonics femtosecond beamline based on a modular hollow waveguide for photons generation centered on the upper region of the extreme ultraviolet spectral range - A thorough analysis}
\author{Y. Brelet{*} \and A. Marquette \and N. Beyer \and G. Versini \and J. Faerber \and M. Vomir \and V. Halte \and M. Barthelemy}

\date{
	Institut de Physique et Chimie des Materiaux, CNRS, IPCMS, UMR 7504, 67000 Strasbourg, France, \\ 
	{*}now at Université de Caen Normandie, ENSICAEN, CNRS/IN2P3, LPC Caen UMR6534, F-14000 Caen, France\\\texttt{brelet@lpccaen.in2p3.fr}\\[2ex]%
}

\begin{document}
\maketitle
\flushbottom
\maketitle

%
\begin{abstract}
We have successfully developed and implemented an entire and compact table-top high-order harmonics generation (HHG) setup from monochromatic and intense femtosecond ($10^{-15}$ s) laser pulses launched in a target composed of a high-purity monoatomic noble gas specie, which can be Argon or Helium, distinctively. Its frequency arrangement is distributed both in the full eXtreme UltraViolet (XUV, $22-124$ eV) spectral region and in the bottom part of the Soft-X Ray range (SXR, $124-132$ eV), at once. Specifically, the core of this coherent secondary light source is based solely on a homemade, modular, affordable, though sturdy, design. We take advantage of this opportunity to present our design guidance of the XUV generation from a hollow capillary waveguide apparatus, and our simple recipe regarding the alignment process of the latter, which is easily carried out thanks to our adjustable design. Then, a comprehensive description of our entire XUV beamline is described, and participate in adding essential contents to the existing literature. Concurrently, we conducted theoretical studies, in order to anticipate or explain our experimental results. Overall, we found very good consistency between the experimental and cost-effective time-consuming numerical results. Finally, our setup provides very good vacuum performance under high gas load pressures, to a few atmospheres. All of these attributes fulfill the requirements regarding ultrafast time-resolved pump-probe configuration in table-top element-sensitive spectroscopy of complex and integrated optoelectronic devices made of magnetic materials. This setup will contribute to address pivotal aspects of electrons, spins and lattice dynamics arising in future data storage devices tailored for high-speed communications and high-capacity data storage. It finds a resonance in the growing fields of Nanophotonics and Spintronics, subsets of the thrilling and promising attosecond Science, the natural time scale of electron motion in atomic or molecular photoemission dynamics.
%
\end{abstract}


	\tableofcontents
\include{Main_Text}
\include{M_Supplemental_Material}





\bibliographystyle{unsrt}
\bibliography{biblio.bib}


\end{document}

%% file: Main_Text.tex
\section{Introduction}
\subsection{Motivation}
High-order harmonic generation (HHG) is an upconversion process which results from an extreme nonlinear coherent interaction of an intense fundamental laser field with atoms or molecules, giving rise to ultrashort extreme ultraviolet radiation. It was first observed by McPherson \textit{et al}, and dates back to $1987$ \cite{McPherson:87}. This pioneer discovery heralded a new era in nonlinear optics and spurred Scientists to search for uncharted underpinnings, and mechanically gave an impetus to tremendous insights and leaps forward in various disciplines ranging from chemistry, biology, biomedical to condensed matter. 
Among the many salient works that have been carrying out since this breakthrough, one can dig out that the HHs can impressively (even with laser of energy $< 1$ eV) extend to the biologically relevant Oxygen K-shell absorption edge, within the so-called water transparency window ($284-543$ eV) \cite{Schmidt:18}, \cite{Chevreuil:21}, \cite{Cardin_2018}, \cite{Stein_2016} or \cite{Smith2020}, in the soft X-ray (SXR) region, or else, that the dawn of attosecond Science \cite{Krausz2001}, \cite{Paul_2001} has paved the way to Spectroscopy methods with a cutting-edge temporal resolution \cite{doi:10.1126/sciadv.adp5805}. Recently, tabletop-scale or laboratory-scale apparatus has kept on developing worldwide \cite{Nishimura2021} and HHG sources are even turning into turnkey and accessible for user-oriented applications \cite{Hort:19}, \cite{Hilbert2020}, or being aei very compact experiment where nonlinear pulse self-compression to few-cycles regime is combined with phase-matched harmonic generation into a helium-filled hollow-core fibre \cite{Gebhardt2022}. 
While HH are usually, and historically, delivered from a target composed of a rare gas (sometimes a mixture of rare gasses), it is also apropos to quote but a few unconventional, but established, approaches to generate high-harmonics radiation, using solid-state materials, even though some issues (post-generation re-absorption of the XUV photons, for instance) have yet to be alleviated. Indeed, for instance, the authors in \cite{Shcherbakov2021} demonstrate the generation of HH driven by intense mid-infrared laser pulses in an ultra-thin resonant gallium phosphide (GaP) metasurface. Meanwhile, hybrid metasurface uses (resonant) plasmonic nanostructures, called nano-antennas with dielectric nanoparticles, which enable strong confinement and high enhancement of the local electric field (see \cite{JongKwan_2021} and references therein regarding different methods of HHG from condensed matter, including media such as a two-dimensional topological insulator and graphene). 
\\
Regardless of the employed device, they have all in common the fact that they constitute a unique scientific apparatus for ultrafast pump-probe layouts for dynamical interrogation of the matter, and round up under the single banner of Ultrafast Science. Coherent table-top X-ray sources from HHG are also a complementary tool to large-scale facilities such as Synchrotrons or FELs (Free-Electrons Laser), offering advantages such as the availability of the source, routine access to lower photon energies, shorter pulse duration, and easier pump-probe synchronization accompanied by a potentially attosecond resolution \cite{Cardin_2018}. Thus, these plateforms equip Scientists with an additional and decisive appliance to question material structures with unprecedented nanometric sharpness and catch the dynamics of electronic processes shaking at sub-femtosecond timescale, a long-standing elusive conundrum, until the advent of nowadays ultrafast lasers. 
Yet, all these considered setups are complex assemblies involving vacuum environment, which, in our case, coexists with gas flow, sophisticated optics or precision mechanics, and pose consequential technical challenges that one must strive to overcome. One of these tasks concerns the gas delivery system, for XUV generation, which must subtly supply a localized high pressure (up to a few atmospheres) in the light-matter interaction region, while not contaminating the adjoining vessels, which would degrade the vacuum level at their heart or worse, would harm sensitive and expensive sensors or other dedicated items. This is because the entire beamline is inescapably windowless in the axis of the optical propagation path, as XUV are absorbed in glass. How these difficulties are managed will be exposed throughout the paper.
\\
From a fundamental understanding, the phenomenon of HHG in gases can be explained in a semi-classical way considering the well-known and intuitive three-step model \cite{Corkum93}, the semi-analytical quantum formalism (strong-field approximation) in \cite{Lewenstein94} or in a full-quantum picture using the time dependent Schrödinger equation (TDSE) \cite{Gorlach2020}, \cite{Gonoskov2016}, to cite but a few. As mentioned earlier, in most schemes, HHG relies on the interaction of a strong ultrashort laser field with a gaseous target filled with a noble gas (or a mixture). Three main gas targets are commonly reported in the community: gas jet (from a nozzle and skimmer assembly), gas-filled cell or gas-filled waveguide, of which a few seminal papers are by \cite{PhysRevLett.83.2187}, \cite{doi:10.1126/science.280.5368.1412}. Regarding more specifically the latter, the authors in \cite{Cassou:14} have successfully demonstrated that HHG photon flux can be enhanced by using waveguide compared with free propagation like in a gas jet. Thus, noticeably, it provides a high gas density confined in a waveguide configuration, a plane wave-like structure and a higher interaction length than gas jet configuration. In fact, laser light is guided by glancing-incidence reflection from the walls of the capillary waveguide (mode confinement), thus allowing the laser and high-harmonics beams to co-propagate over an extended interaction length with well-controlled intensity and phase profile \cite{1564377}. In addition, it minimizes the XUV virtual source size compared with a high-density gas jet, enabling an interesting demagnification (ouptut divergence is a few mrad in a non-guided geometry, compared with $< 1 $mrad in a waveguide \cite{Chen:09}). 
Nevertheless, that being said, papers, such as \cite{KIM2022107803}, keep on publishing, showing that there is still a quest for mastering the HHG process driven by a kHz and multi-mJ class femtosecond laser. At last, one key challenge and hot-topic that comes to the fore is the route towards attosecond ($10^{-18}$ s) Science, which the femtosecond barrier shattered more than twenty years ago to create the first isolated attosecond pulse \cite{Krausz2001}. 
\\
Our setup is built from scratch, except that it is driven by a commercial $800$ nm femtosecond laser, at low-frequency repetition rate ($1$ kHz), focused in a gas target of Argon (Ar) or Helium (He). Efforts have been undertaken to compact the beamline as much as possible, resulting in a length of $2.5$ m, disregarding the laser case. We also used a modest driving laser energy of $3.5$ mJ in order to ionize the interaction medium. Significantly, our investigations exhibit that, in Helium, the generated XUV energy comb of odd harmonics spans over $\sim2.5$ octaves, with a low cut-off at $\sim 22.4$ eV ($q=15$th harmonic $\equiv \lambda_q\approx 53.4$ nm) and a high cut-off at $\sim 131.7$ eV ($q=85$th harmonic $\equiv \lambda_q\approx9.4$ nm), which is at the foot of the Soft X-rays range. We show that by simply rinsing thoroughly the circuits to rid of the prior gas, we are able to shift the device quickly and easily from Argon to Helium and vice versa. Interestingly, it covers several resonances (or element-specific absorption edges) of $3$\textit{d} Transition Metals (TM) and $4$\textit{f} Rare Earths (RE, Lanthanides group), at once. Another key feature, as part of the technical specifications, is that we can manipulate a couple atmospheres of gasses, which turns out to be valuable for the future version of our apparatus. We have insisted on the modularity aspect of the hollow core waveguide (HCW), which allows one to manage gas consumption and adjust for the HHG intensity together with ease of optical alignment, as well as a fast and easily change of HCW without breaking a large volume of vacuum, which is often cumbersome and time-consuming. It is worth noting that our development could also be adapted for attosecond spectroscopy, particularly by incorporating optical components and vibration-compensated opto-mechanical devices\footnote{In essence, the tens of fs duration, near-infrared, pulses are spectrally broadened by focusing them into a meter-long stretched hollow-core fiber pressurized with a rare gas, and then temporally compressed with a set of chirped (dispersive) mirrors (or wedged-plates), resulting in a few ($<10$) fs pulses.} for shorter pulses duration and by installing a CEP-Stable module (successive Carrier-Envelope Phase-locked pulses) in the Laser. 
From a fundamental point of view, such a setup helps to investigate the origin of the ultrafast demagnetization dynamics arising when a complex magnetic structure is struck by an intense laser field (see, for instance, \cite{Richter2024_PRB} and references therein). To achieve this aim, the hereinafter presented instrument is envisioned for a femtosecond time-resolved Transverse Magneto-Optical Kerr Effect (T-MOKE) spectroscopy bench to observe the asymmetry in the reflectance (called the magnetic contrast) for two opposite magnetization directions (of a permanent magnet of tens of mT, for instance) aligned along the incidence plane of the vertically-polarized laser field. This will be performed on bulk dielectric samples with thin films of transition metals and rare-earths in binary or ternary heterostructures or alloys, to investigate the light-spin-lattice interaction and their coupling to electron and phonon baths \cite{Ferte2023}. The study of ultrafast demagnetization \cite{ChanLaoVo}, \cite{Mathias4792} is important because it addresses societal concerns as ferromagnetic systems are considered candidates for ultrafast spintronics applications. 
\\
The purpose of the present technical report is, therefore, threefold. First, in awareness of the above considerations, we present a single-pass, cost-effective, and scalable secondary XUV source using an interchangeable and segmented dielectric hollow core waveguide geometry, which is easy to assemble, yet reliable. But not only: indeed, the whole Probe (in the sense of a strobe) optical path (actually, the hardest part to design) of a typically IR-Pump/XUV-Probe setup is mounted. We have concentrated our efforts to internally design and build by ourselves, so as to own every definition and detailed parts in 2D and 3D blueprints, to operate with ease preventive or curative maintenance. Second, we show, as a road map, how we run iteratively through an ascending procedure to derive analytically and numerically all the features leading to a relevant sizing of the XUV photon source. At the same time, we endeavor to give a general outline of the microscopic backgrounds behind high harmonic generation, so as to paint a picture of the physical mechanisms in order to be as much exhaustive as possible.  However, going into more detail remains beyond the scope of this paper. 
Ultimately, this work is an attempt to tackle the problematic under all the physical and engineering aspects intrinsically tied to such a scientific instrument, as much in optics, plasma and mechanics, as instrumentation and hydrodynamics, relying on different numerical tools and methods, for completeness regarding our guideline. Hence, it explains our decision to decipher and dissect the entire manufacturing process, filling this gap in the existing literature, so as to give a basis and hints that propose a technical foundation for raising such a facility, and will serve for future research in this field by other scholars.
Lastly, note that the totality of the theoretical or experimental values derived is compared with the literature, and that all simulations and numerical codes presented in this work were designed, implemented, and validated by the authors.
\\
The paper is organized in three major sections, each divided into subsections. First, we briefly introduce and propose a step-by-step review of basic theoretical work that investigates the influence of gas pressure and strong-field ionization, from micro- to macroscopic reasoning, together with the supplemental material, in order, at last, to determine the sizing of our \textit{ad hoc} tailored XUV source. The next section thoroughly describes the overall experimental setup and devices, and the techniques proposed for diagnostics are exposed with exactness. The last section deals with the results achieved in successfully obtaining high-harmonics generation. 
\\
Finally, this paper is also accompanied with supplemental material, proposing additional simulations, especially in engineering, experimental results, and pairs references that will provide a richer overview and offer to the reader a full description of our development as well as detailed contents, which will hopefully feed the Ultrafast Science Community with complementary data while using HCWs filled of a rare gas.
\subsection{Tasks and workflow synoptic}
\begin{figure}[ht!]
\centering\includegraphics[width=16.7cm]{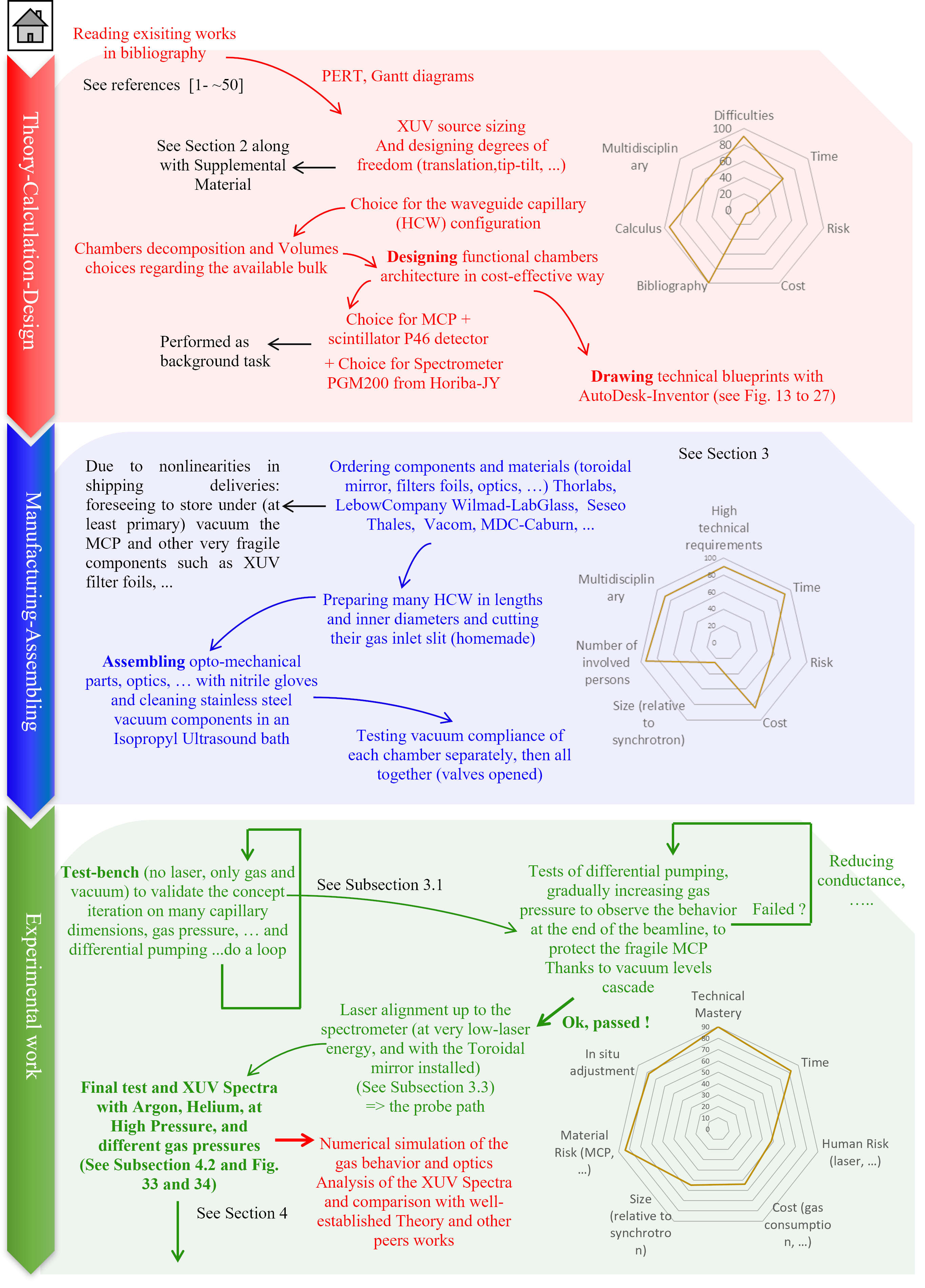}
\caption{Workflow chart showing the progression of the project. Web diagrams are proposed for each task. }
\label{fig:Flow_Chart}
\end{figure}

We show in Fig. \ref{fig:Flow_Chart} a comprehensive sequence of the important phases of the development, as a first glance and where the reader can fly over to his/her section of interest. 

%
%
%
\section{Basic theoretical framework for waveguide sizing}
\label{sec:Theoretical considerations}
\medskip
In this section, we review and use analytical models to guide the development and sizing of our high-order harmonics source, as well as to anticipate the experimental results. To support these analytical models, we performed numerical simulations, in Matlab (The MathWorks Inc.), containing both micro- and macroscopic features of HHG. This part comes with the section \ref{Sec:theory_basics_AtomsHighfield} of the Supplementary Material.
\subsection{Microscopic point of view: an atom in the strong field ionization regime}
\label{subsec:Theoretical considerations_SFA}
\begin{figure}[ht!]
\centering\includegraphics[width=16.1cm]{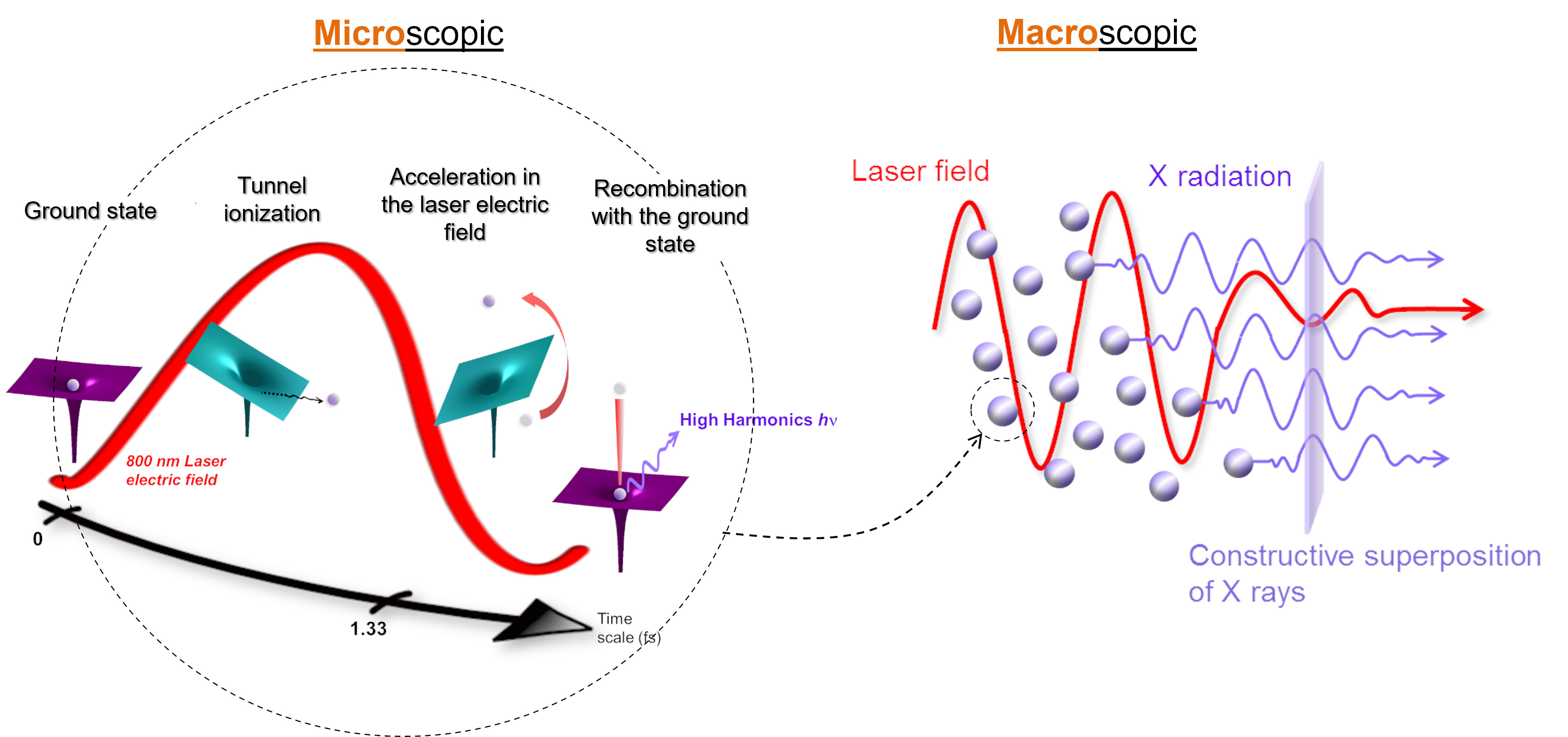}
\caption{Scrutinizing the HHG process. Left: illustration of the microscopic phenomenon of HHG for one emitter (one atom of noble gas). Right: coherent macroscopic building of HHG from the many atoms in presence, for an exploitable XUV signal.}
\label{fig:Micro_Macro}
\end{figure}

\begin{figure}[ht!]
\centering\includegraphics[width=13.4cm]{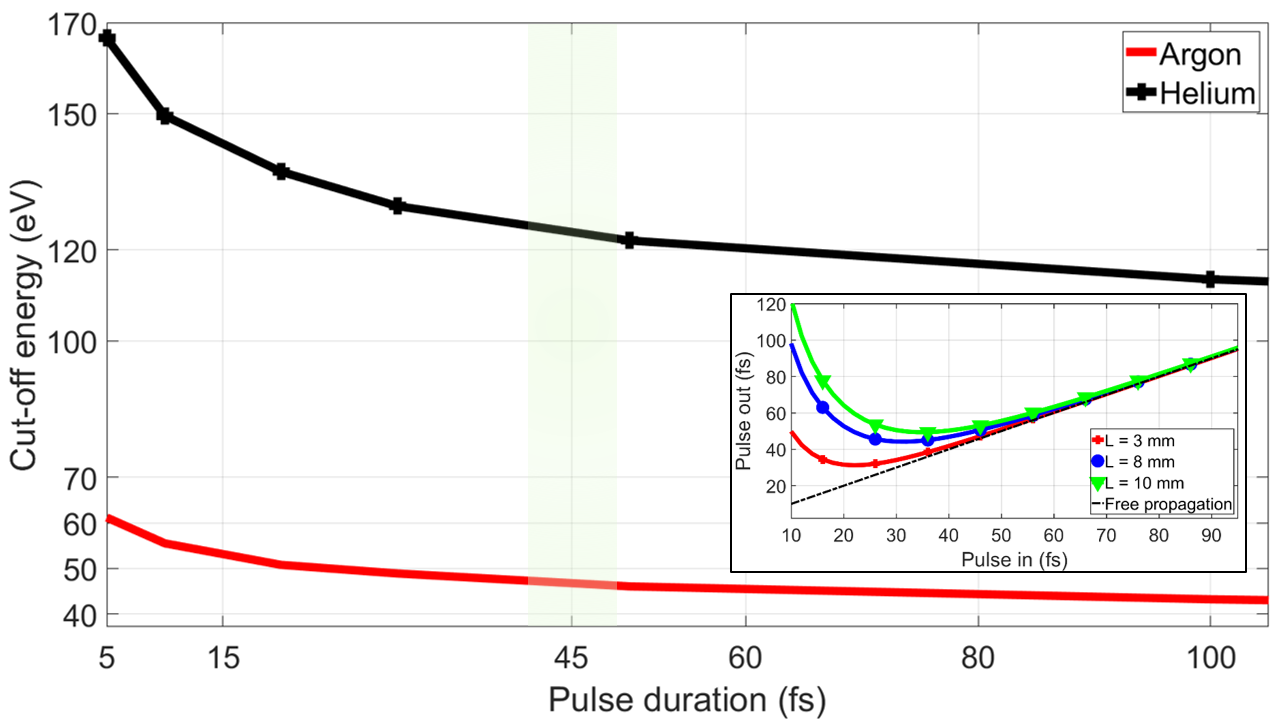}
\caption{Predicted high cut-off energy for argon and helium vs femtosecond laser pulse duration $t_p$, for $\lambda_0 = 800$ nm. Green shadowed region is the pulse duration range where our cut-off is expected, considering pulse propagation in the plasma, and pulse temporal broadening in the entrance window via self-phase modulation and group delay dispersion (GDD). To illustrate the latter, insert shows the theoretical temporal broadening of femtosecond pulses before (Pulse in) and after (Pulse out) propagation through a L$=8$ mm BK$7$ thickness, for a Gaussian unchirped pulse with $t_0$ pulse duration (Fourier transform limited) at FWHM, in the form $t_p = t_0\sqrt{1+(4\ln2 (\mathrm{GDD}_{BK7}/t_0^2))^2}$ with GDD$_{BK7}$ = $43.96$ fs$^2 / $mm.}
\label{fig:Cutoff_E}
\end{figure}

An atom exposed to a laser beam conveying a field strength comparable to the one of the Coulomb field (the attraction force between electrons and protons) experienced by the electron, undergoes various nonlinear processes. The most typical and likely processes where the electron is released from its parent atom are either through tunneling or through multiphoton ionization (MPI), even above-threshold ionization (ATI) (which is an extension of the MPI), the latter giving rise to a photoelectron spectrum composed of a series of peaks. With multi-cycle femtosecond driving laser pulses, pulse trains of attosecond high-photon energies in the XUV and X-ray energy regions can be generated, and are the cornerstone of high-order harmonics generation. 
\\ More details are given in the accompanying supplemental material. 
\\
High-order harmonics generation is well-described by the semi-classical three-step scenario, which in a single-atom assumption, reads \cite{Corkum93}: 
i) an electron at the bound-state leaves the atom by the onset of tunnel ionization, after that the laser radiation has disrupted the Coulomb atomic barrier potential, ii) it is accelerated, during its excursion in the continuum, by the laser electric field and gains kinetic energy, iii) when the time-varying laser electric field reverses, it eventually recombines (coherent elastic scattering) when driven back in the vicinity of its parent ion and releases its quiver (kinetic and potential) energy in the form of a XUV radiation depending on the excursion time (determined by the $e^-$ long/short trajectories \cite{Gaarde2008}) spent in the continuum (see Fig. \ref{fig:Micro_Macro}).
A meaningful aspect of the semi-classical picture for HHG is that it allows us to write the following universal and phenomenological cut-off law that successfully predicts the highest released photon energy for a single-atom or ion
\begin{equation}
E_\textrm{max} \approx I_p + 3.17U_p
\label{eqn:eqnCutoff}
\end{equation}
where, in SI units, $U_p = E_0^2 e^2/4\omega_0^2m_e\approx 9.33$ x $10^{-14} I_0$ (W/cm$^2$) $\lambda_0^2$ ($\mu$m$^2$) is the ponderomotive energy \cite{Salieres95}. In order to derive the maximum cut-off energy possible in our configuration, the highest intensity $I_0$ in the term $U_p$ is chosen such that the critical ionization is reached, and stems from the ADK  (Ammosov, Delone and Krainov) formulation Eq. (\ref{eqn:eqnTunnelRate}).
In Fig. \ref{fig:Cutoff_E}, the derived cut-off energy is plotted as a function of the pulse duration $t_p$. The expected cut-off for Argon is about $50$ eV, in agreement with observations in \cite{Hort:19} or \cite{doi:10.1063/1.4812266}, and about $130$ eV for Helium, consistent with \cite{Popmintchev10516}, given here a pulse duration of $45$ fs. 
The cut-off energy decreases monotonically with the pulse duration. Note that as our pulse duration is not continuously monitored, the precautionary principle leads us to consider a (arbitrary) slight variation, in Fig. \ref{fig:Cutoff_E}, of the pulse duration, which could be, for instance, either sensitive to day-to-day atmospheric humidity or ambient temperature fluctuations (albeit our experimental room is equipped and controlled), or after hours of laser working, which causes thermal (heat) events responsible for, even tiny, structural modification in the optical and opto-mechanical components inside the laser body. 
\begin{figure}[ht!]
\centering\includegraphics[width=13.8cm]{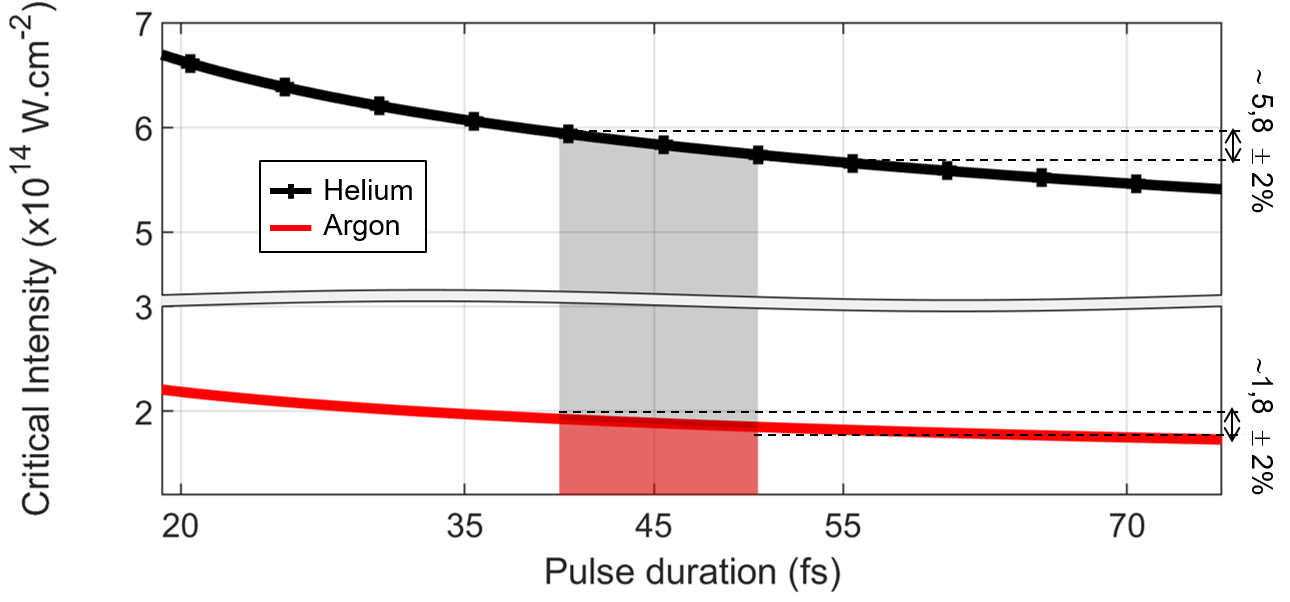}
\caption{The critical intensity (log-log scale) as a function of the laser pulse duration, for $\eta_{\textrm{cr}}=4\%$ and $\eta_{\textrm{cr}}=0.5\%$, for both Argon and Helium, respectively. Shaded areas (black $+$ for Helium, red for Argon) bounded in their upper part by the corresponding curve, with an intensity estimation $\pm 2 \%$ for both gas species, and in the lower part by the $x$-axis, show the region of interest in our experimental conditions, where we can apply a laser intensity. Vertical bounds indicate the range in which could possibly varied our pulse duration, giving a range of critical intensities we can reach. }
\label{fig:IntensityCritical_vs_PulseDuration}
\end{figure}
The critical ionization $I_c$ of the above ADK model is given in \cite{photonics10010024}
\begin{equation}
I^c_0(t_p,I_p,\lambda) = \bigg \lbrack \frac{g_0}{F_0} W_0 \bigg(-\frac{1}{g_0}\bigg(\frac{D_0}{t_p}\bigg)^{1/g_0}\bigg)\bigg\rbrack^{-2}, \: \: \; \mathrm{for} \: \: g_0 < 0,
\label{eqn:eqnCriticalIntensity}
\end{equation}
with
\begin{equation*}
D_0 = 2\sqrt{2 \mathrm{ln}(2)} \frac{\vert \mathrm{ln}(1-\eta_{cr}) \vert{}{}}{\kappa_0}
\end{equation*}
where $W_0$ is the real principal branch of the Lambert Function, $\eta_{cr}$ is the critical ionization derived from the ADK model, and the following atom-dependent parameters, in atomic unit:  $\kappa_0 = 6.12$ and $6.5714$, $F_0=0.8311$ and $1.6196$, $g_0=-0.1417$ and $-0.5122$ for Argon and Helium, respectively. The conversion relation between the atomic and SI units can be found in Appendix A of \cite{Maurer_2021}.
%
%
\subsection{Macroscopic phase matching for HHG}
\begin{figure}[ht!]
\centering\includegraphics[width=16.2cm]{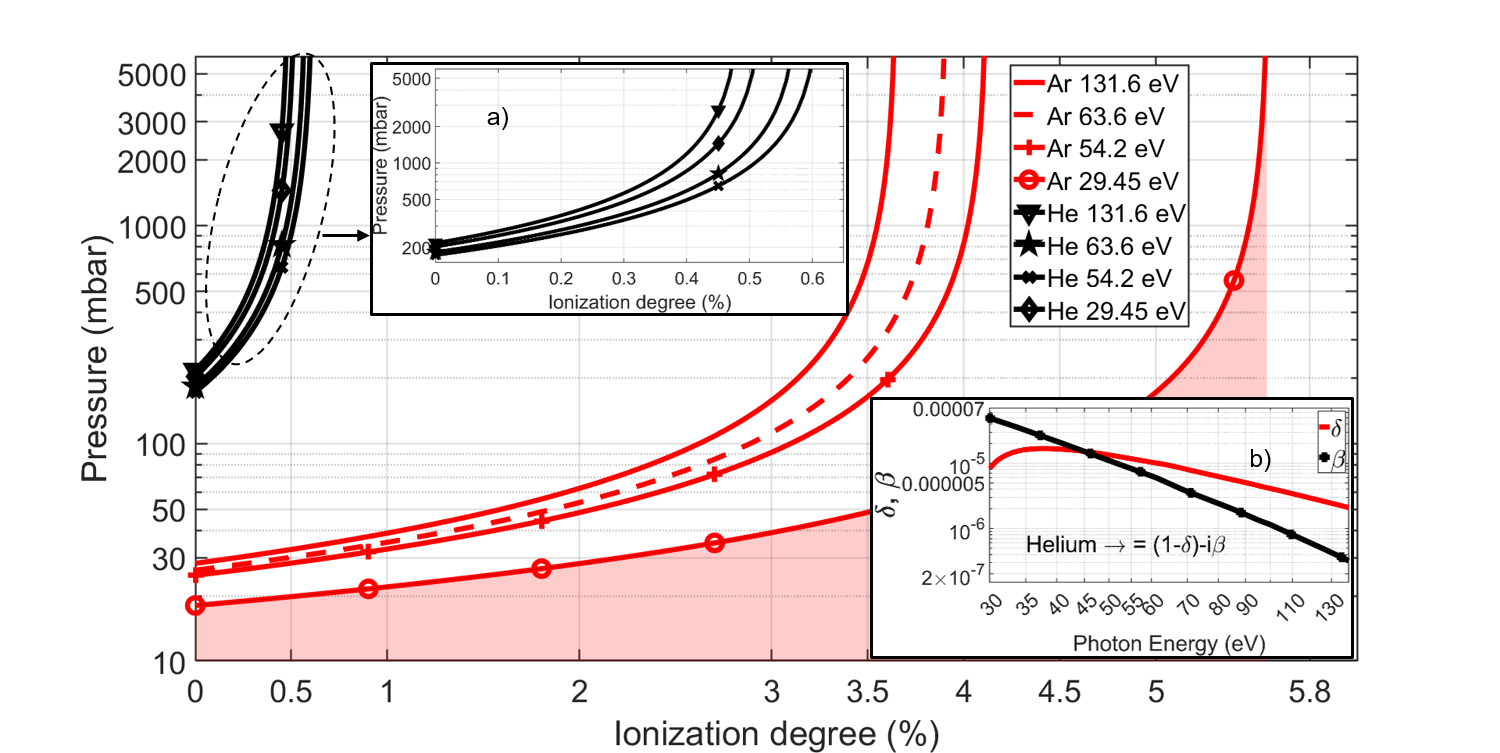}
\caption{Phase matching pressure $P_{\textrm{PhMa}}$, for a medium centered at the laser focus ($z=0$), as a function of ionization degree $\eta$ in Ar and in He for four different harmonics ($\#q = 85, 41, 35, 19$ or $9.42$, $19.5$, $22.9$, $42.1$ nm, or $131.6$, $63.6$, $54.2$, $29.45$ eV, respectively), reading curves from left to right. Data for $n_{\textrm{XUV}}$ are taken from \cite{CXRO}. The Argon under-curve for $\#q=19$ is shaded, showing the region where pressure must be applied for an, \textit{a minima}, on-axis phase-matching. Other curves are not shaded for visual convenience, but their interpretation remains the same. a) is a close-up of Helium part. b) is the index of refraction explaining why the sequence of Helium curves are not monotonic in energy. }
\label{fig:Ioniz_Degree}
\end{figure}

\begin{figure}[ht!]
\centering\includegraphics[width=12.8cm]{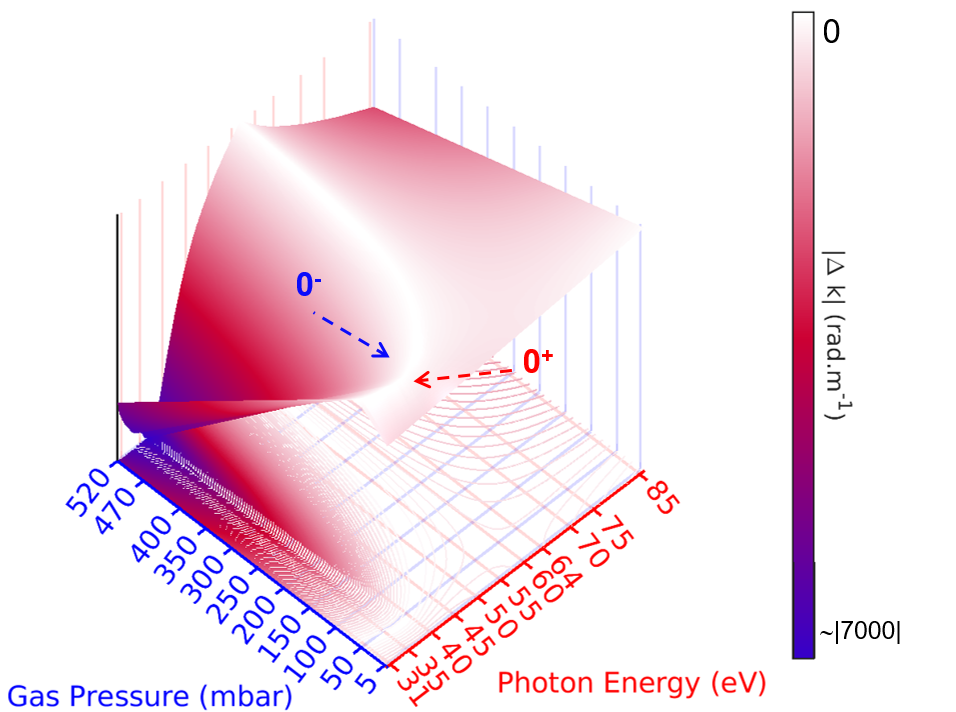}
\caption{Phase mismatching in Argon, for an ideal constant ionization fraction $\eta = 4.5\%$. $a=150$ µm, $L_{\textrm{med}} = 31$ mm, $\lambda_0 = 800$ nm. The white line indicates perfect phase matching, \textit{i.e.} $\Delta k = 0$, which tends asymptotically $\rightarrow \infty$. Experimentally, an acceptable close-phase matching occurs when $0^-<\Delta k<0^+$. Here, we set $I_0$ = $3.3$ x$10^{14}$ W/cm$^2$ and $z=-L_{\textrm{med}}/2+10$ mm, thus considering focus position is $10$ mm inside the capillary, at the laser entrance side (our experimental conditions), from the capillary center. Beam waist $w_0$ is estimated to $w_0 \approx 75$ µm. Here, we have chosen a constant $\alpha_q^{Ar} = 22.5$x$10^{-14}$ cm$^2$.W$^{-1}$ corresponding to the cut-off, the branch where both long and short trajectories gather in one single trajectory, in Fig. \ref{fig:QuantumTraj_Ar}.} 
\label{fig:fig1}
\end{figure}

\begin{figure}[ht!]
\centering\includegraphics[width=12.8cm]{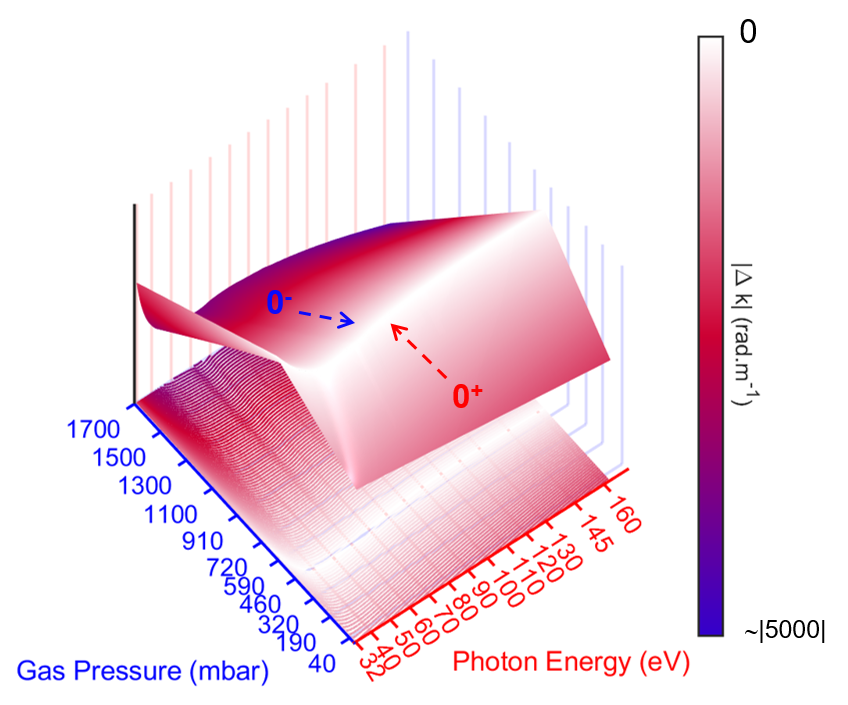}
\caption{Phase mismatching in Helium, for an ideal constant ionization fraction $\eta = 0.4\%$, $a=150$ µm, $L_{\textrm{med}} = 13$ mm, $\lambda_0 = 800$ nm. The white line indicates perfect phase matching, \textit{i.e.} $\Delta k = 0$, which tends asymptotically $\rightarrow \infty$. An acceptable close-phase matching occurs when $0^-<\Delta k<0^+$. Here, we set $I_0$ = $3.3$ x$10^{14}$ W/cm$^2$ and $z=-L_{\textrm{med}}/2+1$ mm, thus considering focus position is $1$ mm inside the capillary, at the laser entrance side (our experimental conditions), from the capillary center. Beam waist $w_0$ is estimated to $w_0 \approx 75$ µm. Here, we have chosen a constant $\alpha_q^{He} = 23.8$x$10^{-14}$ cm$^2$.W$^{-1}$  corresponding to the cut-off, the branch where both long and short trajectories gather in Fig. \ref{fig:QuantumTraj_He}. }
\label{fig:fig2}
\end{figure}

\begin{figure}[ht!]
\centering\includegraphics[width=12.8cm]{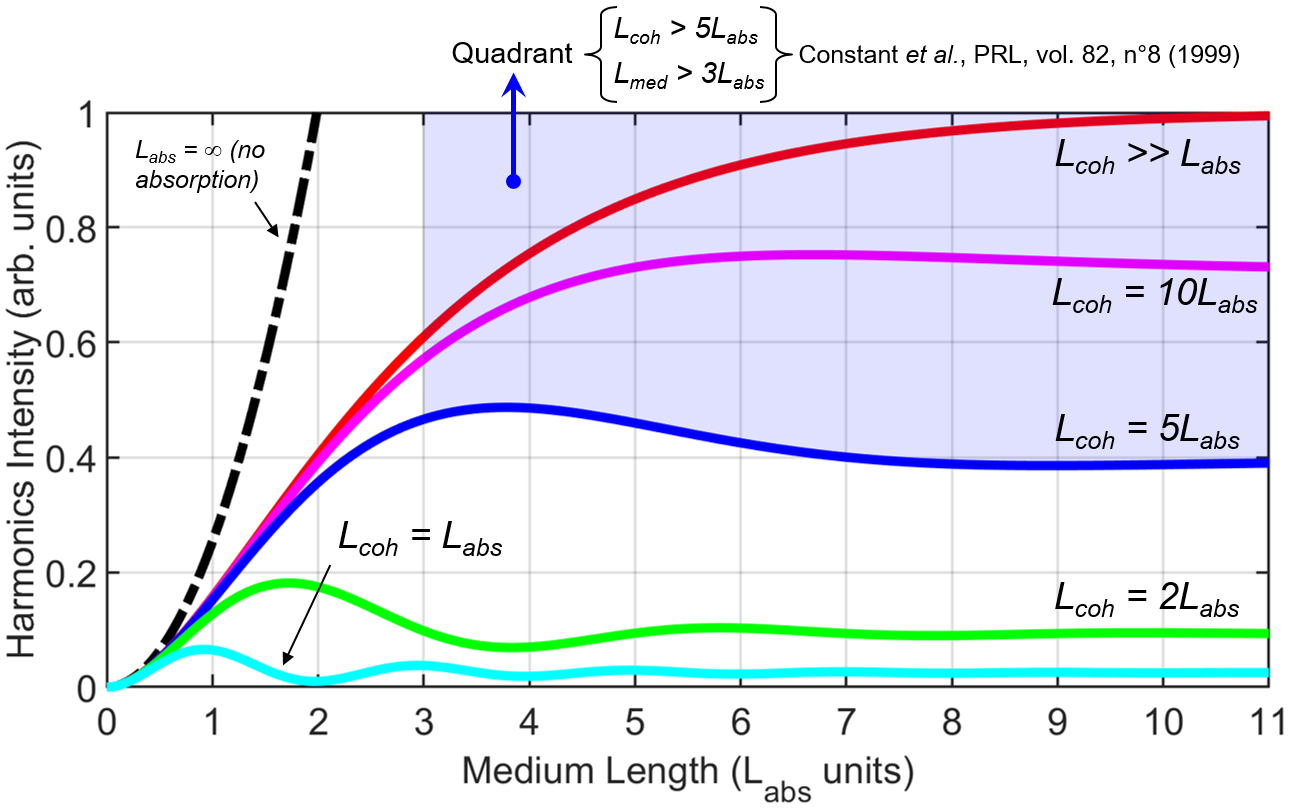}
\caption{Harmonic Intensity as a function of medium length $L_{\textrm{med}}$, for a given pressure $P$, and a given harmonics $q$, showing the influence of absorption in the building up of the phase matching process. Top-right (in blue watermark) quadrant indicates the optimal loci of the pair $\{ L_{\textrm{med}}; L_{\textrm{abs}} \}$ for phase-matching \cite{PhysRevLett.82.1668}. }
\label{fig:SHHG_vsLcoh}
\end{figure}

\begin{figure}[ht!]
\centering\includegraphics[width=13.6cm]{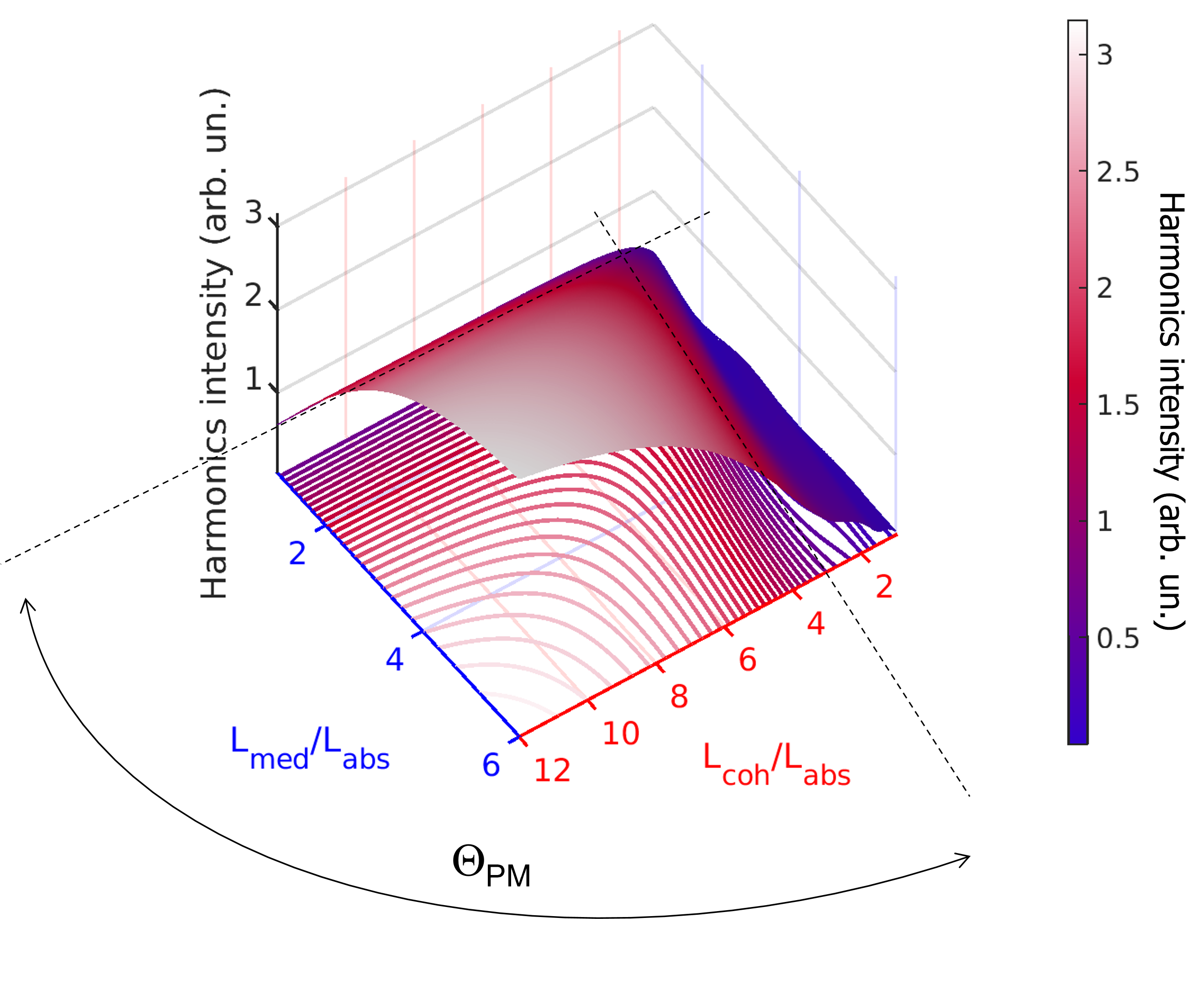}
\caption{Map of harmonic Intensities, in arb. un., as a function of $L_{\textrm{med}}/L_{\textrm{abs}}$ and $L_{\textrm{coh}}/L_{\textrm{abs}}$, for a given pressure $P$, and a given harmonics $q$, showing the influence of absorption in the building up of the phase matching process, \cite{PhysRevLett.82.1668}. Then, intensity saturates at optimized $L_{\textrm{med}}/L_{\textrm{abs}}$ and $L_{\textrm{coh}}/L_{\textrm{abs}}$ values, tallying to the white part. The angular sector $\Theta_{\textrm{PM}}$ delimiting the blue and the red parts corresponds to the blue quadrant in Fig. \ref{fig:SHHG_vsLcoh}.}
\label{HHG_Intensity_vs_Characteristic_Lengths}
\end{figure}

\begin{figure}[ht!]
\centering\includegraphics[width=14.8cm]{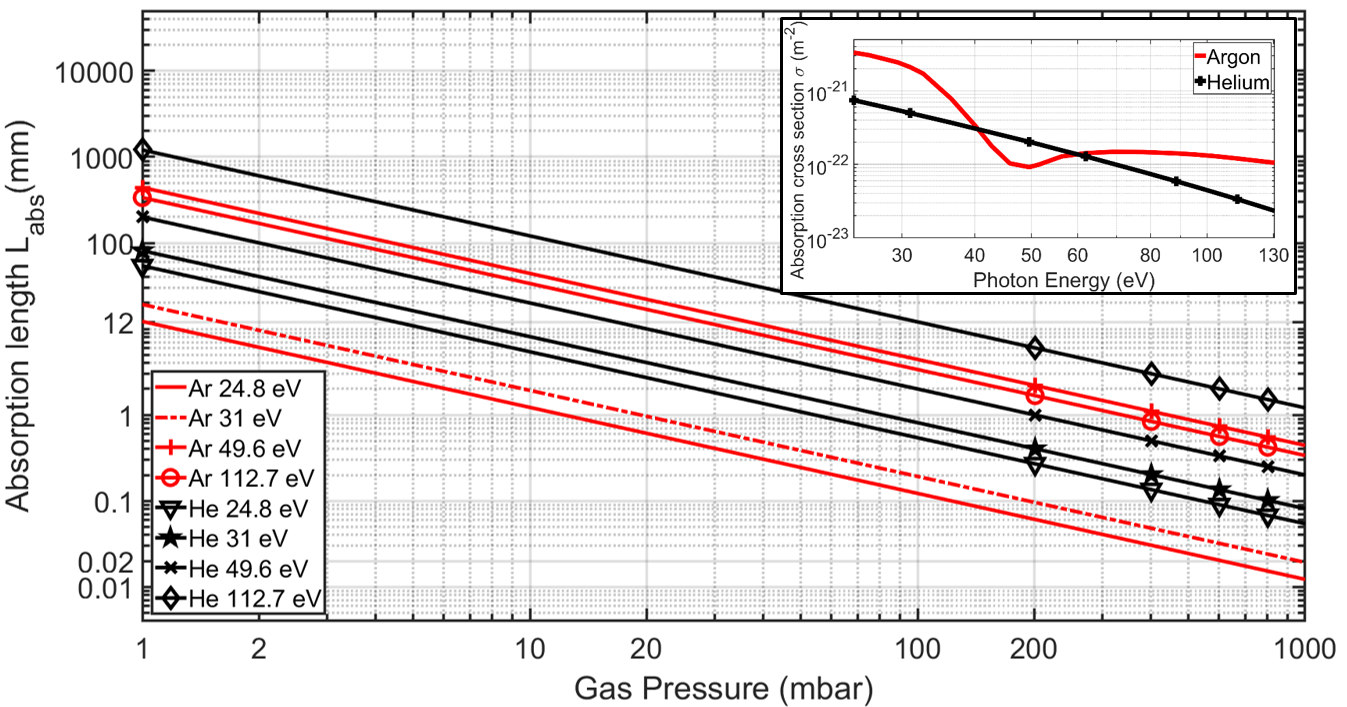}
\caption{Log-Log scale representation of the absorption length $L_{\textrm{abs}}$ with respect to the gas pressure (He and Ar) for a few photon energies $E_{\textrm{XUV}}$. $T=300^{\circ}$ K and system is assumed isothermal. Insert represents the absorption cross section as a function of photon energy. Data from \cite{MARR1976497}. The drop in the cross section of Ar, under the one of He, at $\sim50$ eV, is due to Cooper minimum \cite{Weissbilder:22}. The latter explains, together with Eq. (\ref{eqn:eqn2}) why curve for Ar $@49.6$ eV is above the one $@112.7$ eV, and does not follow the monotonic tendency observed for Helium. }
\label{fig:fig3}
\end{figure}

Besides the microscopic single-atom response, which gives, if taken alone, a weak HHG emission signal, macroscopic effects of the many atoms in play ensure the HH signal growth (which can be viewed as the coherent combination of numerous emitters along the propagation direction), leading to a bright, directed, thus exploitable, coherent beam (see Fig. \ref{fig:Micro_Macro}). Because the harmonic signal and fundamental laser must add in phase for a constructive coherent build-up \cite{1564377}, condition for phase matching is encountered when the driving laser pulse and the generated XUV bursts travel at the same phase velocity over an extended region in the nonlinear medium. A conventional prerequisite is to have a look at the 1D-macroscopic longitudinal phase mismatch $\Delta \mathbf{k}$, where $\mathbf{k}$ is the wave vector, in the on-axis\footnote{We limit our analysis to the contribution of the phase-mismatch on the laser propagation axis, $r=0$, where $\textbf{k}_{\omega q}$ is co-linear to $\textbf{$\hat{z}$}$.} of laser propagation, and in free-focus geometry (in not considering the Gouy phase term), which is the sum of the contributions from the pressure-dependent neutral atoms and free electrons dispersion, from the (geometric) mode dispersion, the quantum path (the so-called long and short trajectories of electrons in the continuum), and from the Gouy phase shift, and is formulated as a vectorial momentum conservation equation, defined by \cite{Popmintchev10516}
\begin{eqnarray}
\lefteqn{ \Delta \mathbf{k}  = \Delta \mathbf{k}_{\omega q} + \mathbf{k}_{\omega \textrm{IR}}= \Delta \mathbf{k}_{q,g} + \Delta \mathbf{k}_{q,n} +\Delta \mathbf{k}_{q,e} +\Delta \mathbf{k}_{q,G} +\Delta \mathbf{k}_{q,\Phi at} }
\nonumber\\
&  \underbrace{= q\frac{u_{11}^{2}\lambda_0}{4\pi a^2}}_{\textrm{geometric}\: \Delta k_{q,g} } - \underbrace{qP(1-\eta)\frac{2\pi}{\lambda_0}(\Delta\delta+n_2)}_{\textrm{atoms}\:\Delta k_{q,n}}+\underbrace{qP\eta N_a r_e\lambda_0}_{\textrm{free electrons}\:\Delta k_{q,e}} -\underbrace{q \frac{d}{d\textrm{z}} \tan^{-1}\frac{z}{z_R}}_{\textrm{Gouy phase shift} \:\Delta k_{q,G}} + \underbrace{\alpha_q \nabla I_0}_{\textrm{dipole phase}\:\Delta k_{q,\Phi at} } {} 
\label{eqn:eqn1}
\end{eqnarray}
where $q$ is the harmonic order, $z_R=\pi w_0^2 /\lambda_0$ (in our case $\rightarrow \sim 150$ mm) the Rayleigh length, $z$ the position of the capillary compare with the focus at $z=0$, in the propagation axis \textbf{$\hat{z}$}, $u_{11}=2.405$ is the mode factor and is the $1$st zero of the Bessel function $J_0$, $r_e$ is the electron radius, $a$ is the inner radius of the HCW, $P$ is the pressure in atmosphere, $N_a$ is the density of atoms per atmosphere, $\eta$ is the ionization level, $\lambda_0$ is the fundamental near-infrared (NIR) laser wavelength, $n_2 = \tilde{n}_2 I_0$ is the Kerr nonlinear index of refraction per atmosphere at $\lambda_0$, $\Delta \delta$ is the difference between the indices of refraction $n_{\textrm{NIR}}$ and $n_{\textrm{XUV}}$ of the gas per atmosphere at the fundamental and XUV wavelengths, respectively. Here, the nonlinear refractive index $n_2$ of the gas is neglected, as it is small (a few percent) compared to the $\Delta \delta$ contribution \cite{PhysRevLett.83.2187}. The refractive index $n_{\textrm{NIR}}$ is determined using the formula and coefficients derived in \cite{Dalg_King}. The refractive index $n_{\textrm{XUV}}$ is obtained via the formula $n_{\textrm{XUV}} = 1-\delta -i\beta = 1-r_e\lambda_0^2/(2\pi)\sum_j [N_j (f_1^j+if_2^j)]$ where $N_j$ is the number of atoms of type $j$ per unit volume \cite{XDB}. The parameters $\xi$ and $\beta$ stand for the refractive index decrement and the absorption index, respectively. Atomic scattering factors $f_1$ and $f_2$ are tabulated values found, for example, in the database \cite{CXRO}. The Gouy phase shift is rewritten as $qz_R/(z^2+z_R^2)$, and in our case is null for a guided geometry \cite{Weissbilder:22}.
%
\\From Eq. (\ref{eqn:eqn1}), phase matching is achieved if the ionization level $\eta$ is below the critical ionization $\eta_{\textrm{cr}} = [\lambda_0^2 N_a r_e/(2\pi\Delta\delta) (1-1/q^2)+1]^{-1}$ \cite{doi:10.1126/science.280.5368.1412, PhysRevLett.83.2187}. At $\lambda_0 = 800$ nm, values of $\eta_{\textrm{cr}}$ are on the order of about $4.8\%$ and $0.5\%$ for the $35$th harmonic ($\sim 22.8$ nm $@800$ nm) in Argon and $91$st harmonic ($\sim 8.9$ nm $@800$ nm) in Helium (see Fig. \ref{fig:Ioniz_Degree}), respectively \cite{Lytlethesis}. Under this condition, the low ionization limit, the laser is weakly disturbed by nonlinear propagation in the medium. This is confirmed by calculating the characteristic propagation lengths over which self-phase modulation (SPM) or self-steepening becomes dominant \cite{Wagner2004}.
The phase matching is fulfilled when the corresponding pressure is a function of $\eta$, at which phase matching is obtained, can be derived from Eq. (\ref{eqn:eqn1}), by solving $\Delta k = 0$, in the general form
\begin{equation}
P_{\textrm{PhMa}} \mathrm{(bar)} = \underbrace{\frac{\frac{u_{11}^{2}\lambda_0}{4\pi a^2} }{[(1-\eta)k_0 \Delta \delta +\eta N_a r_e\lambda_0] = \textrm{Den}}}_\text{\textrm{Typical term for a first approximation in a waveguide} }- \underbrace{ \frac{[\Omega_i(z,z_R,\alpha_q,I_0) ]}{\textrm{Den} }}_{\substack{\textrm{with Gouy shift, dipole phase and} \\ \textrm{position of the plasma source versus laser focus}}}
\label{eqn:eqn_PM_Pressure}
\end{equation}
where $k_0 = 2\pi / \lambda_0$. The first term in Eq. (\ref{eqn:eqn_PM_Pressure}) corresponds to the widely used formulation of the phase-matching pressure for a capillary waveguide. The second term, with the function $\Omega_i$, includes the Rayleigh length $z_R$ for a Gaussian beam in free-space and $z$ which is the position of the medium compared to the laser focus, where $z\neq 0$ if the generating medium is not centered at the laser focus spot. The dimensions $z$ and $z_R$ both appear in the Gouy phase and dipole phase terms in Eq. (\ref{eqn:eqn1}). 
But, as already said, the Gouy phase term $\Delta k_{q,G}$ is worth $0$ in the case of a guided geometry. 
For a laser focus position not located at the center of the interaction medium, the atomic dipole term $\Delta k_{q,\Phi at} =(\partial \phi_q/\partial I).(\partial I/\partial z)\simeq -\alpha_q \partial I/\partial z = -2z\alpha_qI_0/[z_R^2(1+(z/z_R)^2)^2]$ is not null.
From Eq. (\ref{eqn:eqn_PM_Pressure}), in Fig. \ref{fig:Ioniz_Degree} the phase matching pressures for a few Harmonics, in Ar and He are displayed. Fig. \ref{fig:Ioniz_Degree} shows that the needed phase matching pressure begins at some minimum pressure for zero ionization and then rises with increasing ionization fraction.
In this Fig. \ref{fig:Ioniz_Degree} the needed pressure for phase matching climbs asymptotically to infinity when reaching a critical ionization fraction $\eta_{cr}$, which is wavelength-dependent, and is worth about $0.5\%$ and $3.5\%$ to $5.6\%$ for Helium and Argon, respectively. The degree of critical ionization $\eta_{cr}$ corresponds to identical phase velocities of the fundamental driving laser and $q^{\textrm{th}}$ harmonics fields.  For higher ionization fractions, no phase matching is possible, in principle, because the refractive index of the gas becomes negative, and is not able to compensate for the waveguide dispersion. But we will briefly discuss the latter that enters in what it is agreed to call the overdriven regime, where it is still possible to generate harmonics. 
\\The conversion efficiency from NIR to XUV is optimized for $\Delta k \rightarrow 0$ and, as shown in Eq. (\ref{eqn:eqn1}), can then be achieved by tuning the gas pressure inside the HCW to balance the contribution of the neutral atoms and the beam defocusing by free-electron plasma, of opposite sign \cite{Bellini2001}. The pressure inside the capillary was then varied to control and tune the HHG phase-matching while keeping the driving laser parameters unchanged. 
\\The plot in Fig. \ref{fig:fig1} shows the phase mismatch $\Delta k$ in Argon, as a function of pressure versus photon energies, with a constant degree of ionization $\eta$ set arbitrarily to $ \sim 0.4\%$, thus inferior to $\eta_{cr}$. It is clear that perfect phase matching for energies below $45$ eV occurs for a pressure of about $20$ mbar, as already observed in \cite{PhysRevLett.83.2187} or \cite{Popmintchev:08}. However, there is a range of pressures $\in \rbrack 0;80\rbrack$ mbar, up to $50$ eV, where an approximate, but successful, quasi-phase matching ($\Delta k \rightarrow 0^{\pm}$) can occur. 
The plot in Fig. \ref{fig:fig2} shows the phase mismatch $\Delta k$ in Helium, as a function of pressure versus photon energies, with a constant degree of ionization $\eta$ set arbitrarily to $0.04\%$ inferior to $\eta_{cr}$. The phase matching pressure is on the order of a few hundreds of mbar, an order of magnitude consistent with \cite{Popmintchev:08}. Notably, (quasi-) phase matching in Helium, here, spans over a larger area of gas pressures compared to that in Argon. 
In fact, phase matching is not restricted to the 1D on-axis, but is rather endowed with a local $\Delta k$, taking into account the spatial gradient of the atomic (intensity-dependent) dipole phase $\Phi_{at}$, in the focal region, as studied for cylindrical spatial coordinates $(r,z)$ in \cite{PhysRevA.55.3204}. In other words, considering off-axis points ($r \neq 0$), is called noncollinear phase-matching. For the sake of completeness, it would appear implicitly in the additional quantity $\Delta \mathbf{k}_{q,\Phi at}$ in the sum of Eq. (\ref{eqn:eqn1}).
In Fig. \ref{fig:fig1} and Fig. \ref{fig:fig2}, the asymptotic behavior of the phase-matching pressure derived at critical ionization $\eta_{cr}$, from Fig. \ref{fig:Ioniz_Degree}, has a direct repercussion on the manner $\Delta k$ scales steeply with pressure as the XUV wavelength decreases. Also, it is seen that the phase matched harmonic order can be extended by increasing the pressure of the considered gas. These results demonstrate an ideal phase-matching configuration that results in an optimal HHG yield. For this to happen, the main experimental knobs to tune are: $1$- the intensity, $2$- the gas pressure and $3$- the focusing conditions. Therefore, for experimental convenience, we chose to fix $1$- and $3$- (but a few results in sweeping a bit $3$- are presented in the Supplementary Material).
\\
Now, in order to determine the geometrical characteristics of the HCW to be used for the experiments, the absorption length L$_{\textrm{abs}}$ as a function of the gas pressure $P$ is given by \cite{doi:10.1063/1.4812266}
\begin{equation}
L_{\textrm{abs}}=\frac{\kappa T}{P\sigma_{\textrm{ion}}},
\label{eqn:eqn2}
\end{equation}
where $\kappa$ is the Boltzmann constant, $T$ is the temperature and $\sigma_{\textrm{ion}}$ is the photoionization cross-section, given in tables \cite{MARR1976497}. 
In order to figure out the influence of the gas pressure and of the three characteristic lengths on phase matching in the tight focusing regime, the macroscopic growth of the harmonic signal can be expressed in terms of L$_\textrm{abs}$ derived in Eq. (\ref{eqn:eqn2}) and is given by 
\begin{equation}
\Psi_{\textrm{HHG}} \propto P^2A_q^2 \frac{4L_{\textrm{abs}}^2}{1+4\pi^2(L_{\textrm{abs}}^2/L_{\textrm{coh}}^2)}\bigg \lbrack 1 + \exp\bigg (-\frac{L_{\textrm{med}}}{L_{\textrm{abs}}}\bigg) -2\cos\bigg (\frac{\pi L_{\textrm{med}}}{L_{\textrm{coh}}}\bigg)\exp\bigg (-\frac{L_{\textrm{med}}}{2L_{\textrm{abs}}}\bigg) \bigg\rbrack
\label{eqn:FluxPhoVsLcoh}
\end{equation}
where $A_q$ (units: C.m) is the amplitude of the atomic dipole response at the harmonic $q$. Eq. (\ref{eqn:FluxPhoVsLcoh}) will find an interest in Section \ref{sec:Results}.
The harmonic yield is maximized, hence phase matching is optimized, when the medium length L$_\textrm{med}$ is at least three times the absorption length L$_\textrm{abs}$, L$_\textrm{med}$ $\geq$ $3$L$_\textrm{abs}$, and the coherence length L$_{\textrm{coh}}$ is at least five times the absorption length L$_\textrm{abs}$, L$_\textrm{coh}$ $\geq$ $5$L$_\textrm{abs}$, for a constant gas density \cite{PhysRevLett.82.1668}. In that ideal case, the coherence length L$_{\textrm{coh}}=\pi/\Delta k$ theoretically tends to infinity, see the black dashed line in Fig. \ref{fig:SHHG_vsLcoh}, traced with Eq. (\ref{eqn:FluxPhoVsLcoh}), thus the harmonic photon flux reaches its maximum. One can see that, under plausible experimental conditions, where one is rather potentially able to approach $\Delta k \rightarrow 0$ in the best case, L$_{\textrm{coh}}>> L_{\textrm{abs}}$. But, although the coherence length is infinite or very higher than, but finite, the absorption length, the HH emission saturates when the medium length L$_{\textrm{med}}$ is longer than a few L$_{\textrm{abs}}$, since HH emitted beyond are reabsorbed. Re-absorption limits the useful medium length to $\sim 5-10$ absorption lengths \cite{PhysRevLett.82.1668}. This can be roughly summarized as the following condition: L$_\textrm{coh}$ $>$ L$_\textrm{med}$ $>$ L$_\textrm{abs}$. This can be generalized as shown in Fig. \ref{HHG_Intensity_vs_Characteristic_Lengths}.
Following Eq. (\ref{eqn:eqn2}), in Fig. \ref{fig:fig3} is displayed the absorption length for different photon energies $E_{\textrm{XUV}}$ and as a function of the gas pressure $P$. For Argon in the range of about $P\approx80-100$ mbar and $E_{\textrm{XUV}}= 49.6$ eV (where $\Delta k \rightarrow 0$, with the help of Fig. \ref{fig:fig1}) it is found L$_\textrm{abs}\approx 6$ mm, so L$_\textrm{med}$ should be $\geq3$x$6$ mm.
For Helium in the range of about $P\approx400-500$ mbar and $E_{\textrm{XUV}}= 112.7$ eV (where $\Delta k \rightarrow 0$, with the help of Fig. \ref{fig:fig2}) it is found L$_\textrm{abs}\approx 3$ mm, so L$_\textrm{med}$ should be $\geq3$x$3$ mm. Thus, although these are approximations, one stays in capillary lengths of size on the order of the cm, which is consistent with lengths reported here and there in the literature, and with the gas transmission abacuses given in the CXRO database \cite{CXRO}. 
%
%
\section{Experimental setup and methods}
\label{sec:Exp_Setup}
\medskip
We now turn to the practical facet of the project. In this section, we give in detail a complete description of our beamline, including the laser system. The ensemble of CAD designs and technical drawings that will be presented in the following has been devised with Inventor\textregistered Autodesk\textregistered.
\subsection{Feasibility study with the pressures being at stake - Decision-making}
\label{sec:Test_Bench}
\begin{figure}[ht!]
\centering
\begin{subfigure}{.5\textwidth}
  \centering
  \includegraphics[width=1.0\linewidth]{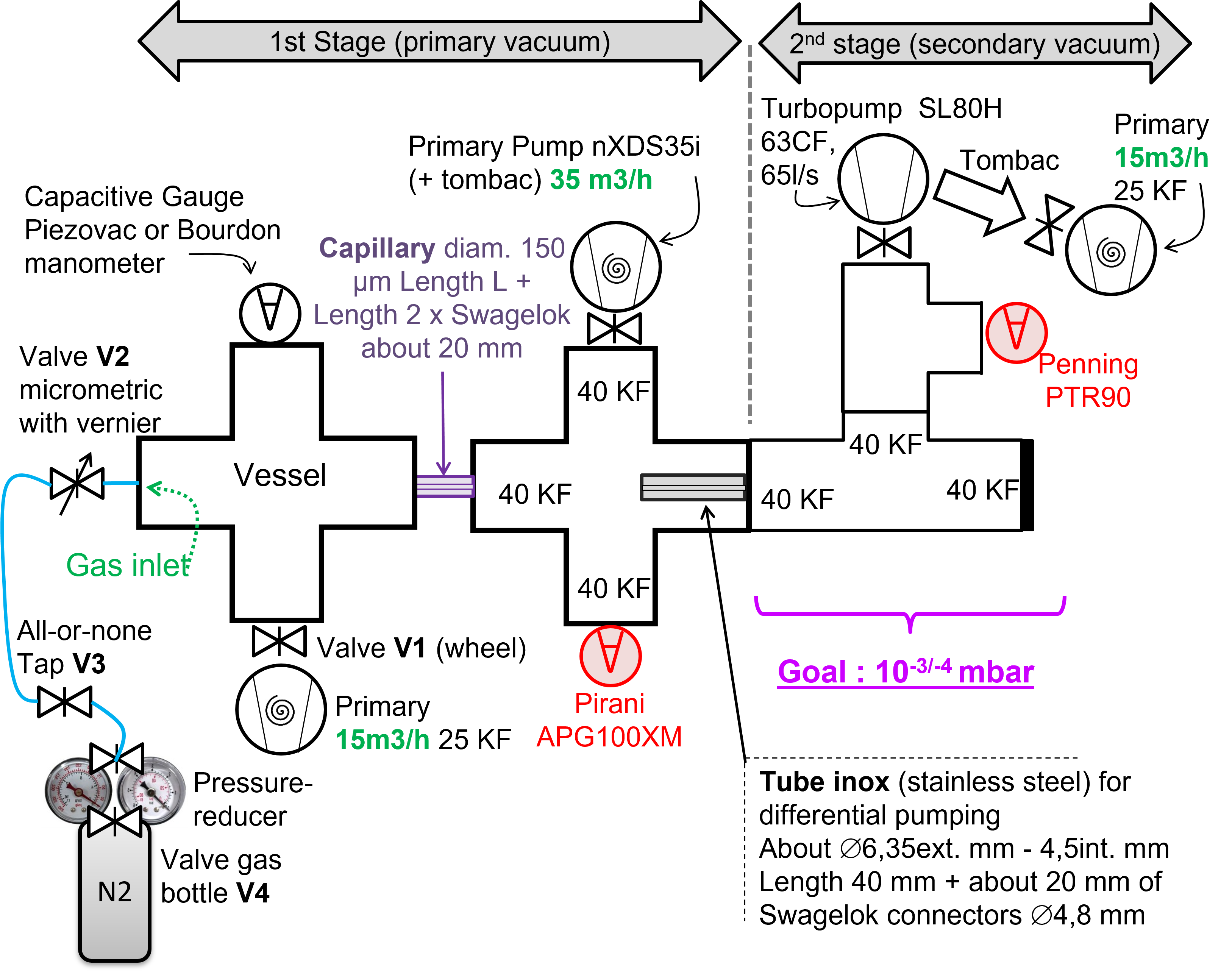}
  \label{fig:sub1}
\end{subfigure}%
\begin{subfigure}{.5\textwidth}
  \centering
  \includegraphics[width=1.0\linewidth]{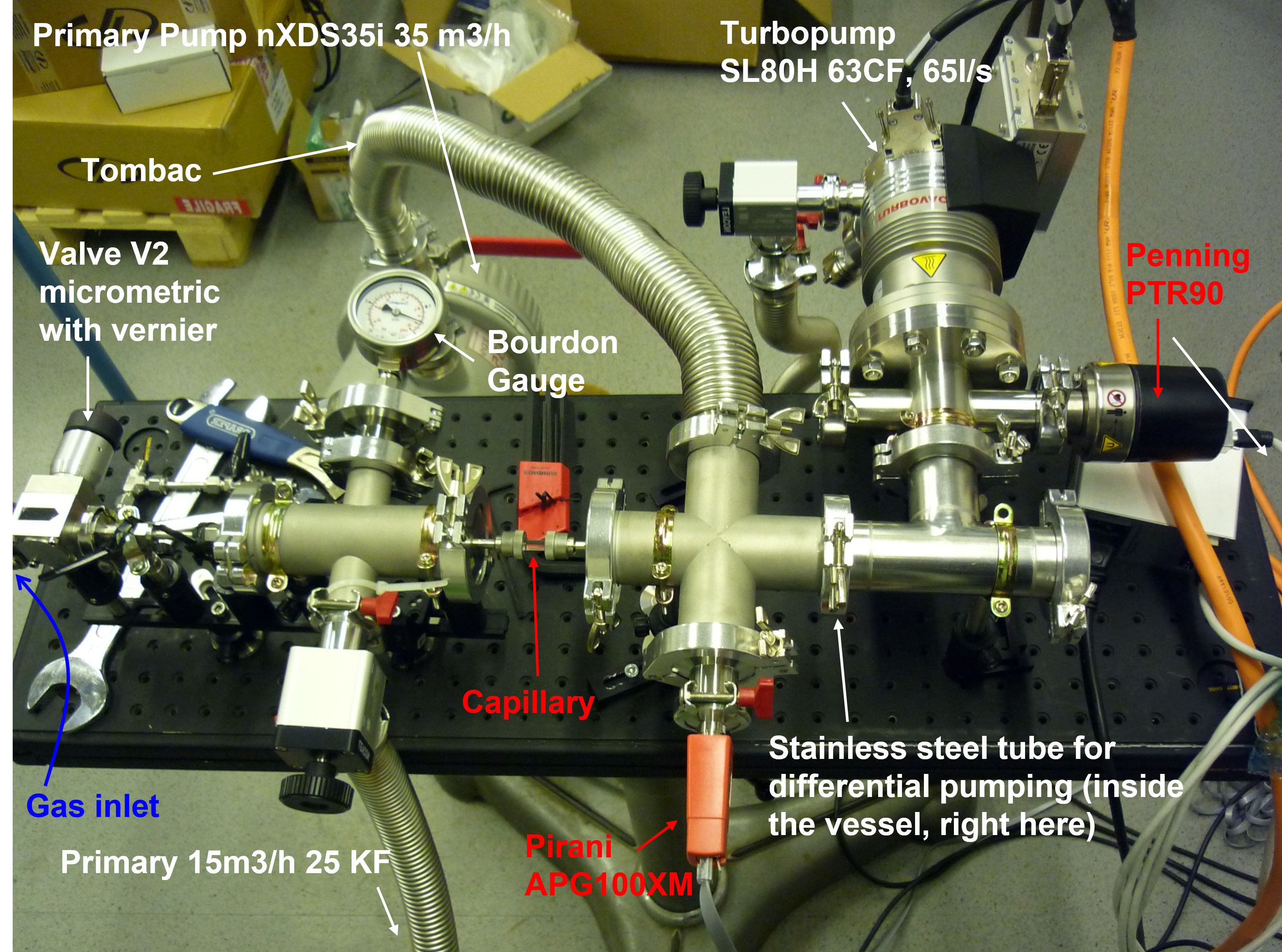}
  \label{fig:sub2}
\end{subfigure}
\caption{Left: plan of the experimental test bench. The capillary is $35$ mm long and $150$ µm diameter. The stainless steel tube for differential pumping is $40$ mm long, and diameter is variable. The first and second stages of vacuum levels are distinguished. Right: Photography of the gas delivery-vacuum bench, showing among others the measuring points of vacuum pressure: Bourdon, Pirani and Penning gauges. }
\label{fig:testBench}
\end{figure}

\begin{figure}[ht!]
\centering\includegraphics[width=14.5cm]{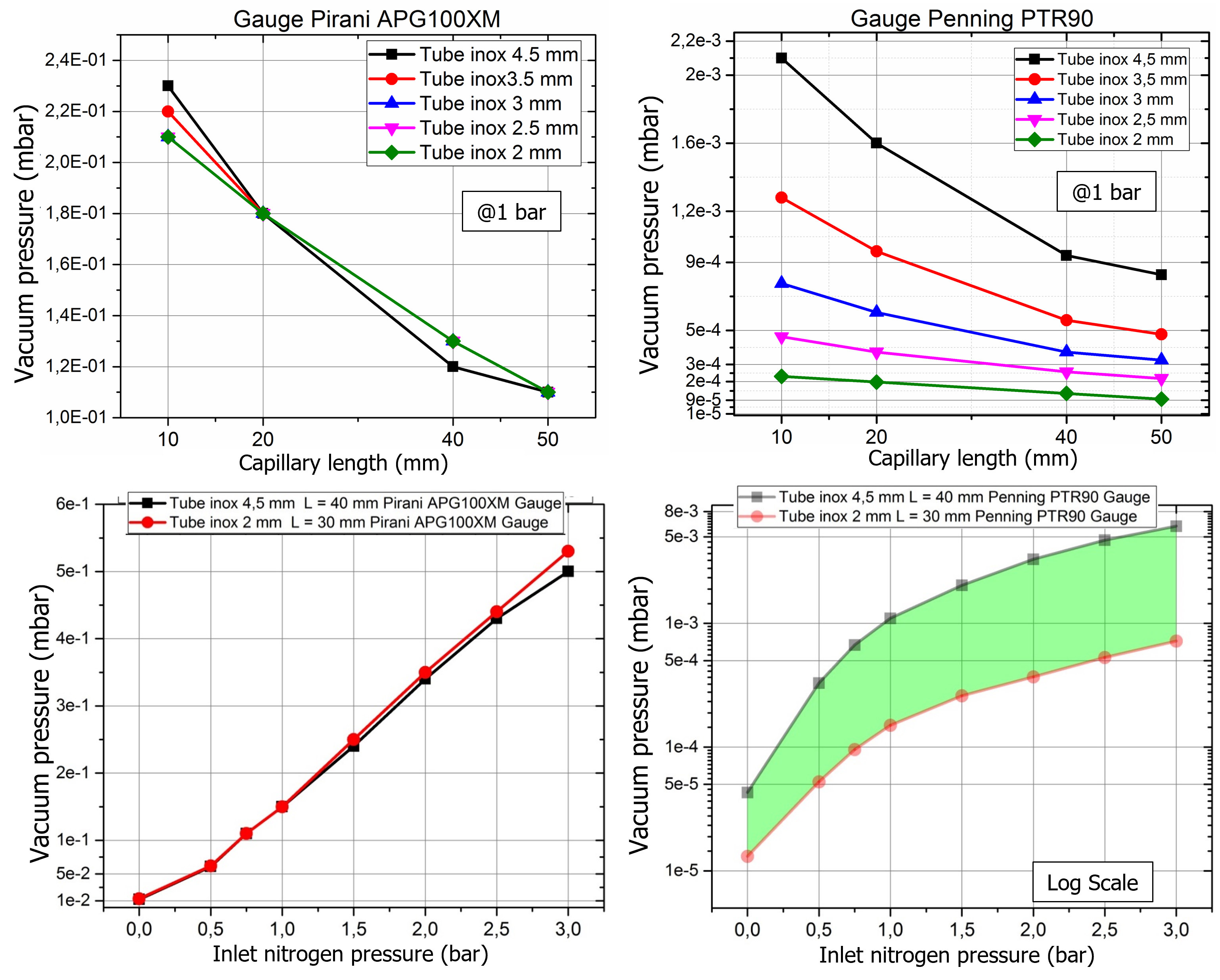}
\caption{Curves coming with Fig. \ref{fig:testBench}. Top-left: Vacuum measured at the first stage, from the Pirani Gauge APGX, for $1$ bar of ambient atmospheric air injected in the system, and for different capillary lengths. The term "inox" stands for stainless steel. Top-right: same, from the Penning gauge, in the second stage. Bottom-left: Vacuum measured at the first stage, from the Pirani Gauge APGX, given a single capillary of diameter $150$ µm and length $35$ mm, for two different stainless steel tubes of differential pumping, and in varying the inlet pressure of dry nitrogen. Bottom-right: log-scale of bottom-left figure, emphasizing the range of pressures, especially at higher inlet pressures, while considering two different tube diameters.  }
\label{fig:CurvesTestBench_Vacuum}
\end{figure}

We have first of all made a test-bench regarding the gas delivery unit and vacuum regulation, to acquaint ourselves with, and to validate, the gas flow behavior in the capillary. For this aim, a synoptic of the experiment is proposed in Fig. \ref{fig:testBench}. The whole bench is about $1$ m long, and is also shown in Fig. \ref{fig:testBench}. It is separated into two stages, the first corresponding to primary vacuum, the other comes after a stainless steel tube, accounting for differential pumping, and where the volume is pumped through a turbomolecular pump of $65$ l/s of molecular flow capacity and assisted with a scroll primary pump with pumping capability of $15$ m$^{3}$/h. 
In Fig. \ref{fig:CurvesTestBench_Vacuum}, we show pressure measurements in the first stage, given by the Pirani gauge, and in the second stage, read by the Penning gauge. Reading both of the top two graphs together, for a stable inlet pressure and a fixed capillary diameter, but a variable capillary length, one can see that the diameter of the differential pumping tube has here no significative influence on the (primary) vacuum level in the first stage, whereas in the second stage, it is valuable, since it can improve (degrade) the (secondary) vacuum level of one decade, by reducing (extending) the diameter of only $2$ mm.  
The lower graphics of Fig. \ref{fig:CurvesTestBench_Vacuum} are also results for the first and second stages, but with varying inlet pressure of nitrogen, while keeping constant the capillary length and diameter. It is first clear here that an input pressure of $3$ bars is close to the conceivable maximum, with regard to the nominal operating mode of the turbomolecular pump (about $10^{-3}$ mbar, and given by the manufacturer) to ensure, when using a tube of differential pumping of $4.5$ mm diameter. An empirical scaling law, in the molecular flow regime, stating that conductance is proportional to $d^3/l$, where $d$ is the inner diameter of the pipe and $l$ its length, tells us that lowering the hole diameter of the tube allows us to shorten the tube length while maintaining a satisfactory secondary vacuum (about $10^{-3/4}$ mbar) in the second stage. Note that the condition for the gas inlet (into the vessel, here) is not exactly the one that will be discussed in the following sections (into the capillary). 
\\
Henceforth, regarding this proof of functionality, this gave us a nudge, and we were confident to continue designing the XUV source, so as to integrate it into a larger setup. Particularly, the prior measurements suggest us to find a compromise between the smallest tube diameter and length for an adequate differential pumping. That, in such a way that it does trim enough the section of the laser beam (since it propagates through the aperture of the tube after having interact into the capillary filled of gas) to begin attenuating it before fully blocking it, as we will see below, but relatively large enough to compulsorily let the slightly diverging XUV radiation (substantially less diverging than the laser beam) travel through it in sustaining none loss.
\subsection{Description of the entire beamline and vacuum chambers architecture}
\begin{figure}[h!]
\centering\includegraphics[width=15.7cm]{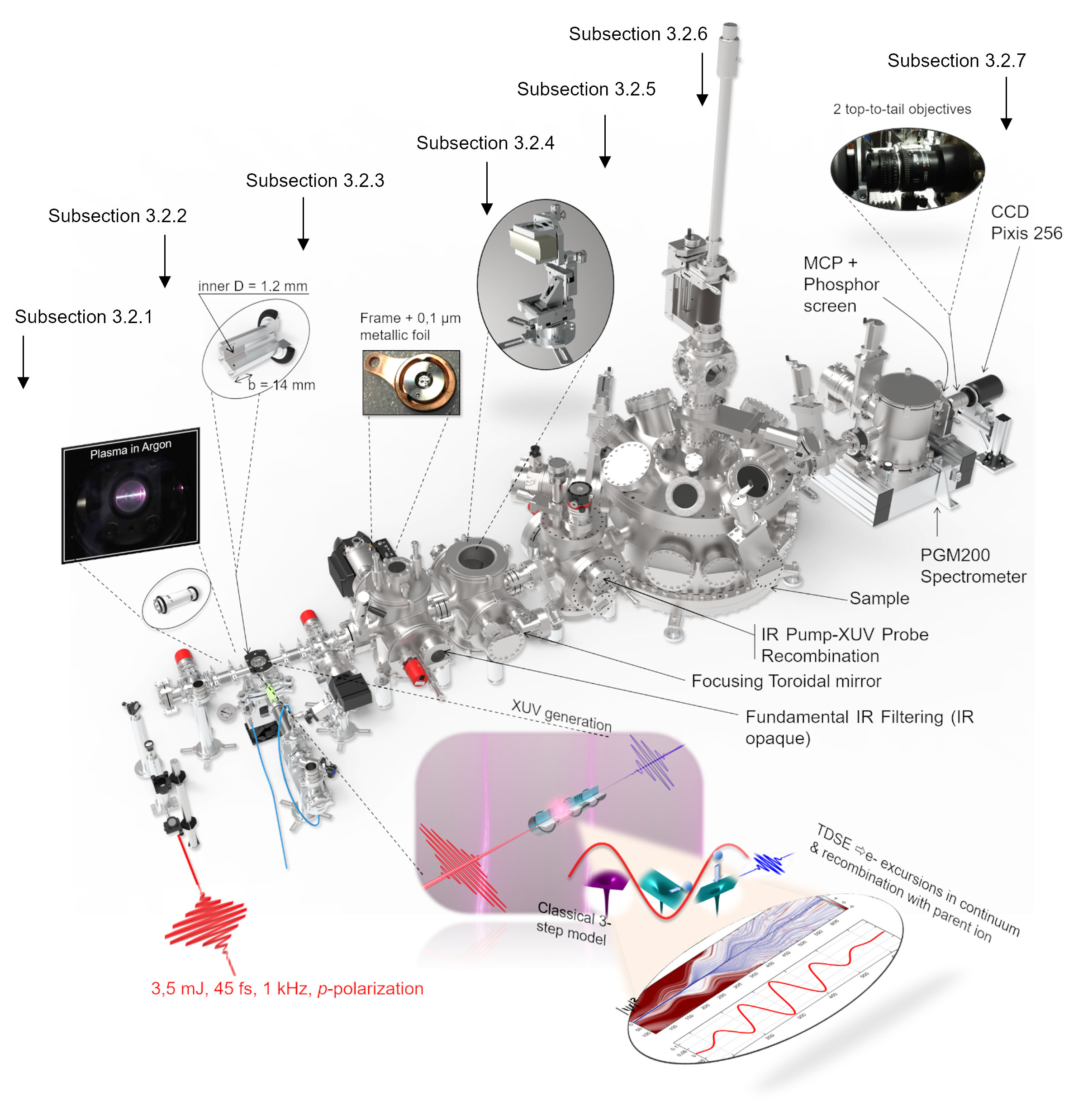}
\caption{Isometric view of the whole experimental beamline. A photography of the fluorescence of the plasma in Argon in the HCW is presented. A detailed view of the HCW sealed with o'rings is also shown. A Tree-structure sketches an iterative deep-inside from the HCW (artist view), the semi-classical three-step model, to a map of the $e-$ long/short spatial trajectories in the continuum, simulated with the TDSE by making use of the Crank-Nicolson Scheme (see Supplemental Material for a short review). Here, the beamline is in straight line, that-is-to-say, in "transmission" mode, in view of the sample under study. For Kerr measurement, the beamline is said to be set in "reflection", so the spectrometer at the end is turned at $90$ degrees with respect to the laser propagation axis, see also Fig. \ref{Kerr_Configuration}. }
\label{fig:fig4}
\end{figure}

In what follows, a detailed explanation of our XUV beamline in practice is provided. A general overview of the experimental setup is depicted in Fig. \ref{fig:fig4}. This current section is built in the sense of the laser propagation: it starts with the laser pulses arrival, on the left, and ends with the detection station, on the right. Note that it is constructed in straight line, but it is possible to turn the detection device at $90$ degrees, in order to perform the so-called time resolved Kerr magnetic measurements, as will be discussed in section \ref{sec:Interaction_Sample}. 
\subsubsection{Femtosecond laser beam characteristics }
The standalone femtosecond laser is based on a commercial system from Coherent. A detailed description can be retrieved in section \ref{sec:Laser_SuppMat} of the supplementary material. 
The system delivers $800$ nm, $9.5$ mJ, $45$ fs (about $17$ optical cycles, which gives a native frequency of $\sim 0.0222 $ PHz or $22.2$ THz), characterized with a Femtometer$^{\textrm{TM}}$ interferometric autocorrelator, linearly $p-$polarized pulses and operates at $1$ kHz repetition rate. A variable diameter iris and a set of two beamsplitters (in reflection, to avoid for managing additional aberrations or dispersion in transmission) attenuate the initial energy down to $\sim 3.5$ mJ which are used to seed the HH process. The transmitted rays through the beamsplitters end their course in a beam dump. It should be noted that while our laser fs is horizontally polarized, before entering the beamline, polarization is turned into vertical polarization (its quadrature) by a $90$ degrees lift of two mirrors (periscope), a fulfilling condition for studies of the magnetic asymmetry (ultrafast demagnetization) arising in T-MOKE spectroscopy \cite{PhysRevB.109.184440}.
We set the peak intensity voluntary (so, it is no more a degree of freedom), which is estimated to be $\sim 4.4$ x $10^{14}$ W/cm$^2$ (at the focus), taking into account our focal length $f=400$ mm. After the laser beam has passed through the entrance window, the propagating losses (its complementary: the HCW transmission) in the capillary waveguide were estimated by measuring the laser power (Thermopile detector $818P-040-25$ from Newport) before and after the waveguide, without gas, and we found about $25\%$ (so, its complementary, $75\%$ transmission). Then, the intensity available in the HCW to drive the HH process is $\sim 3.3$ x $10^{14}$ W/cm$^2$, with a spatial pattern enhanced thanks to a slightly closed diaphragm. We will keep and work with this experimental value for the following.
\begin{figure}[ht!]
\centering\includegraphics[width=15.1cm]{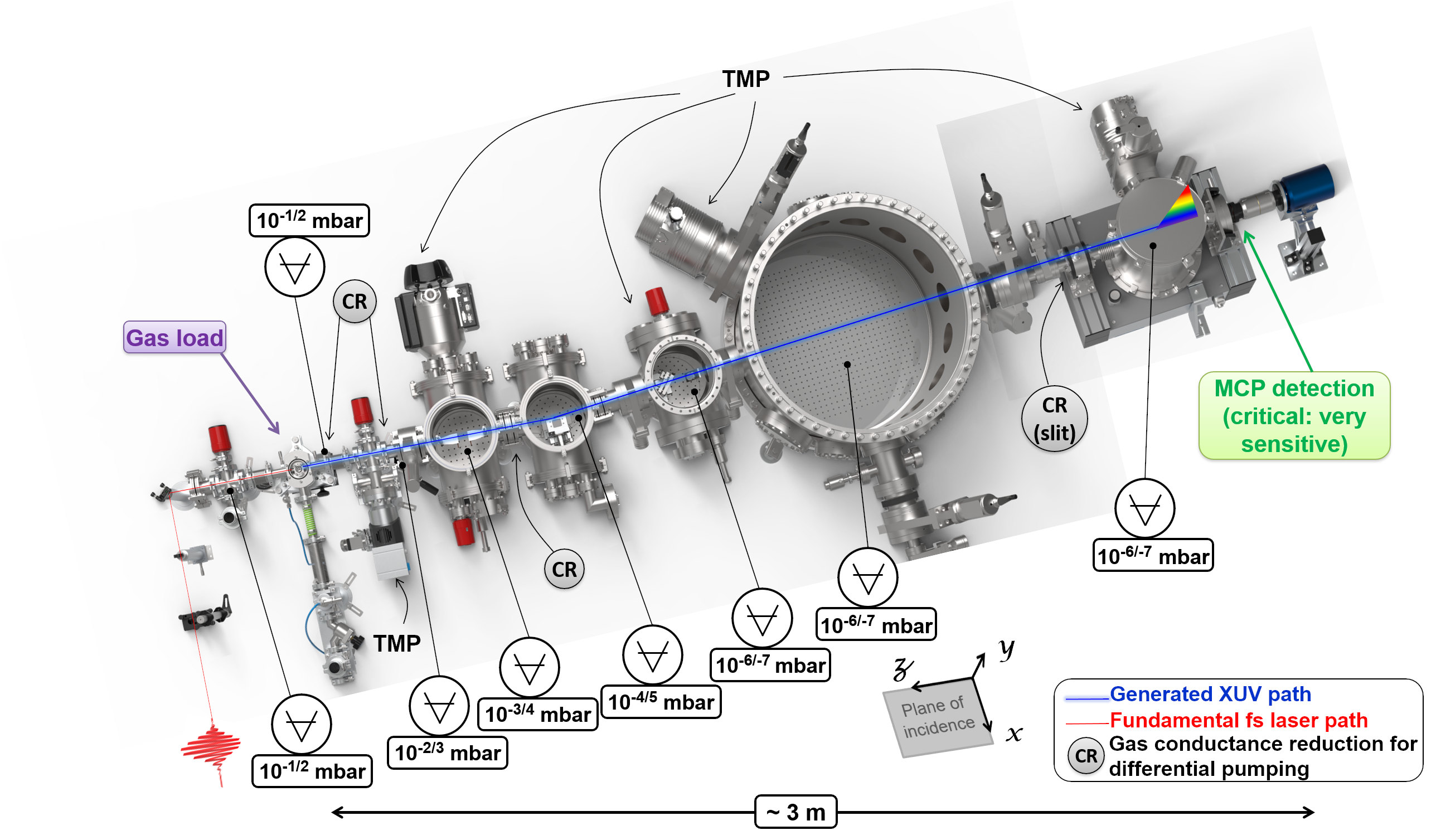}
\caption{Top-view of the whole experimental beamline shown in Fig. \ref{fig:fig4}. Hats of chambers are removed. Nominal vacuum range pressures are indicated in each chamber, when gas load is applied into the capillary, here for about $300$ mbar of He. The complete set of relevant components (TMPs, differential pumping, slit), and their position in the beamline, and contributing to a highly-efficient gas pumping, is shown in detail. Plane of laser incidence is $\{x,z\}$.}
\label{fig:fig4_b}
\end{figure}
\begin{figure}[ht!]
\centering\includegraphics[width=15.3cm]{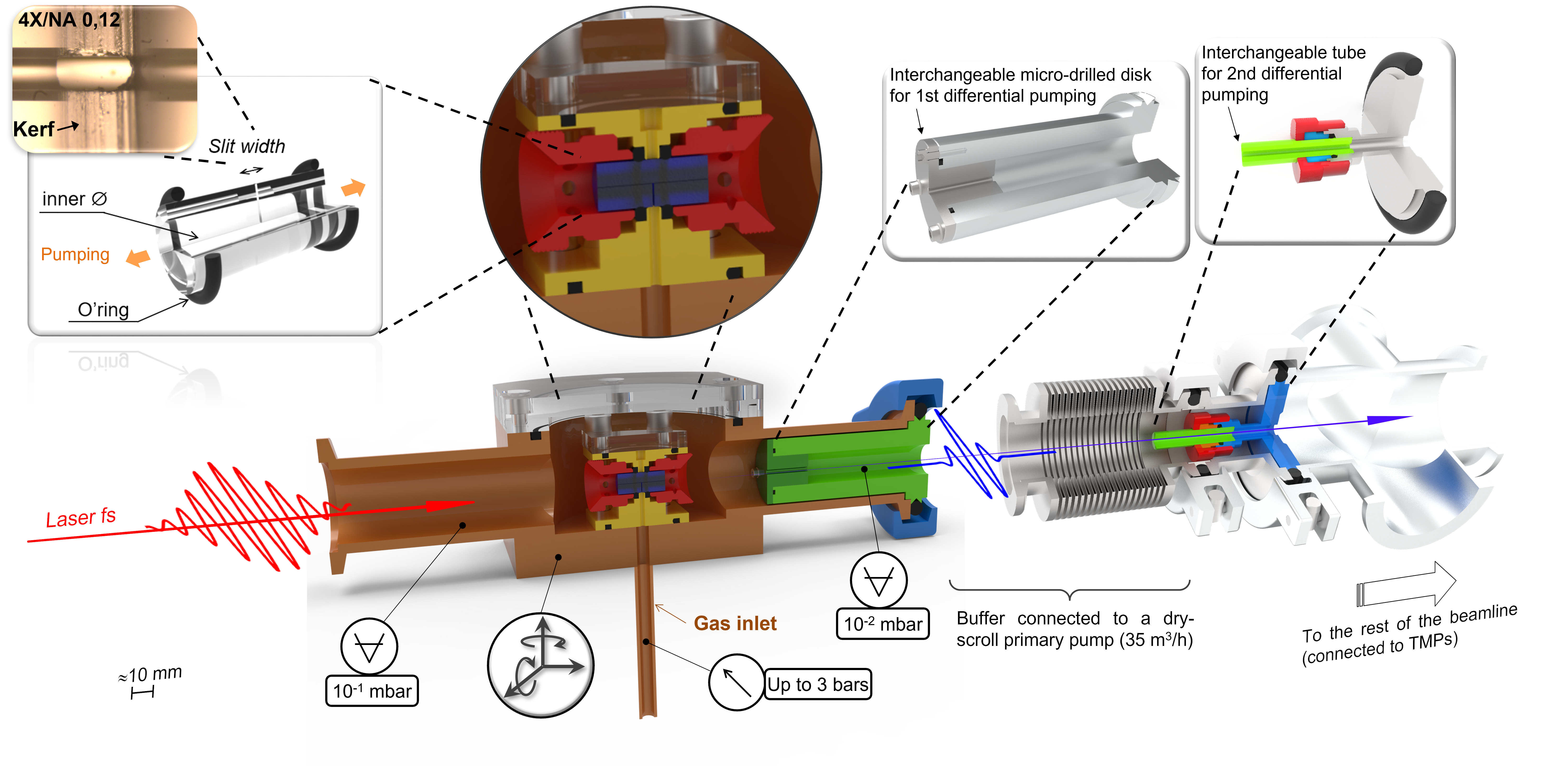}
\caption{Photorealistic cross-sectional (vertical median plane) view of the heart-source, highlighting the capillary (in blue at the center) inside the chambers, in false-colors and other constituent parts of the XUV source. Above are three magnified views of the gas cell assembly and of the waveguide. Artist view with the characteristic dimensions and the position of the two o'ring seals, separating the gas inlet volume from the vacuum, and constraining, compressing, the HCW to hold precisely and firmly in place when tightened up with Swagelok fitting. Microscope magnification of the kerf to create the slit. Gas load is delivered in the first chamber (yellow) from the gas inlet tube, directed towards the notch cut (of half of the outer diameter, and about $200$ µm width, corresponding to the thickness of the saw blade) in the HCW wall. Brown chamber is under vacuum. Red are Swagelok fittings which both maintain the HCW in accurate position and press on the o'rings for good gas-vacuum sealing. Green is the differential pumping component (the disk is a spare component with micro-drilled hole of different diameters), and it has also the ability to block a part of the residual seeding laser. Top is made of transparent PMMA for visual convenience during experiments. Blue component on the right side is a KF clamp. Laser (red pulse) comes in from the left. Generated XUV radiation (blue pulse) comes out on the right side. The approximate scale of the drawings is indicated on the bottom-left corner. Note that here the indicated $10$ mm scale is approximate, since the proportions of the drawn components are not fully respected, but this does not hinder the understanding of this pictorial representation.}
\label{fig:figCapillary}
\end{figure}
%
\subsubsection{Waveguide assembly and alignment protocol: the XUV source heart}
\label{sec_HCW_Descritpion}
One of the technical challenges one has to face up in such a development is the differential pumping. Indeed, XUV pulses must propagate from one chamber to another, windowless. Even though the photon-atom interaction volume is micrometric (even sub-micrometric) in the HCW, the vacuum actuators in the backstage must contribute to the elimination of the residual gas, promptly and efficiently. The whole beamline is presented in Fig. \ref{fig:fig4}, where one can see the XUV generation block-function at the beginning (left-side) of the beamline. Regarding especially the HCW, in that part, a deeper-to-deeper view of the HH generation process is sketched, and the map of electron trajectories has been obtained from computation of the TDSE on Matlab (see Supplemental Material) for a few-cycles laser pulse. A top-view of the whole setup is also displayed in Fig. \ref{fig:fig4_b}, showing the cascade of mean vacuum pressures that we have measured up the spectrometer, under normal operation conditions. In addition, the locations of the flow conductance reduction are shown (the CR acronym).
\begin{figure}[ht!]
\centering\includegraphics[width=13.9cm]{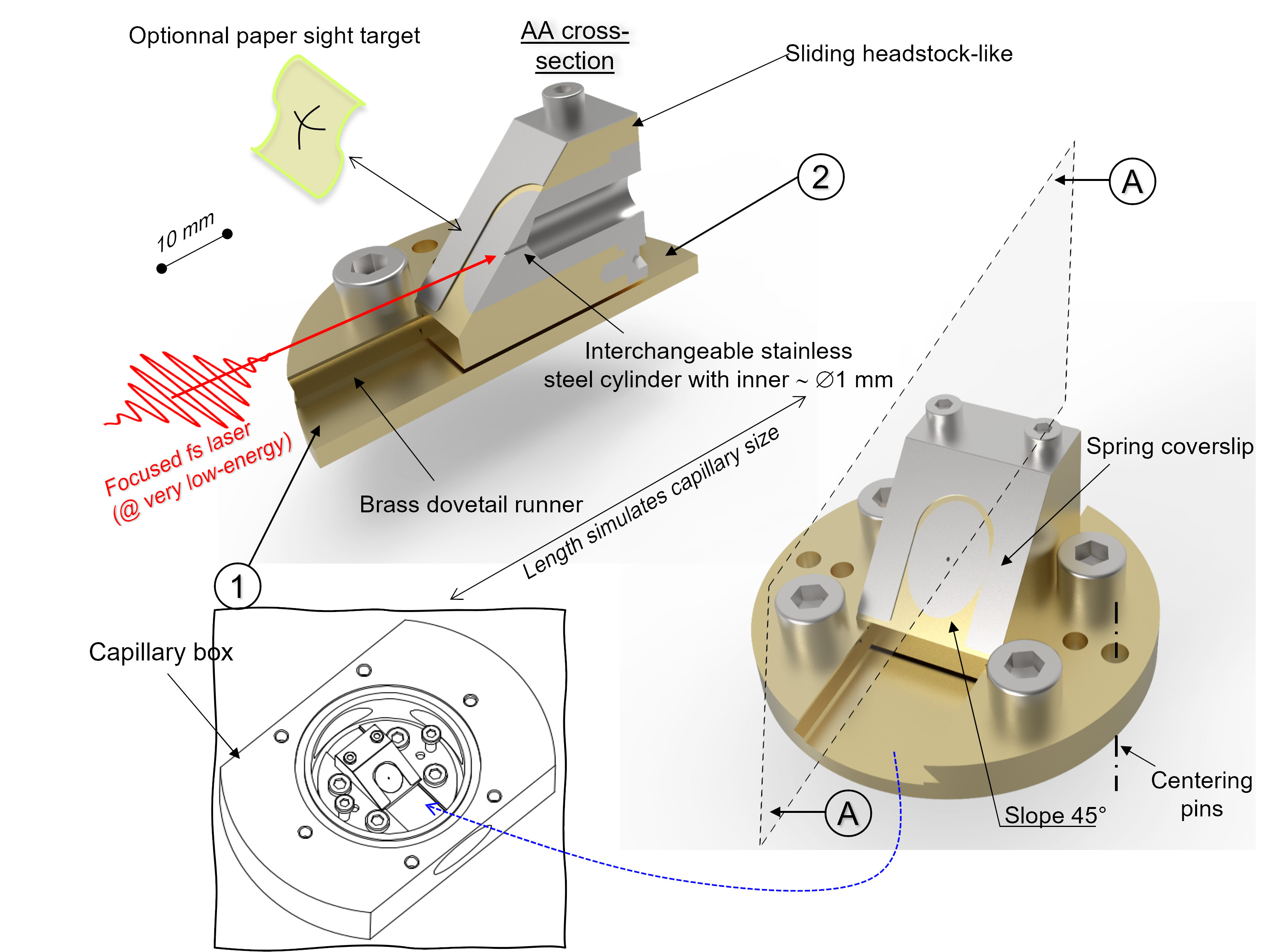}
\caption{Close-up rendering of the simple component use for pre-alignment of the laser path with the capillary, at atmospheric pressure, as we move the sliding headstock-like on a segment simulating the capillary length. The centering pin allow to have the same very accurate position of the body that receives the capillary waveguide. Once pre-alignment is coarsely carried out (at very gentle laser energy), this component is removed and capillary is installed, and finely aligned, before being pumped in vacuum. The $45$ degrees slope is made for visual convenience (but with a non-negligible surface roughness to decrease full specular reflection in distributing optical energy around it, for safety compliance requirement). The centering stainless steel cylinder is interchangeable allowing to use different hole diameters ($1$ mm, $500$ µm, ...) for iterating down towards more and more precise alignment until focused laser laps the internal edges of the micro-hole (bore). The brass dovetail runner can be dry-lubricated with graphite powder if some comfort is requisite. The bottom-left insert shows this component placed into the capillary box for aligning on the whole capillary length. }
\label{fig:fig4_AlignmentCapi}
\end{figure}
\\
Since we have concentrated our efforts on the versatility and modularity of the device, in terms of capillary diameters, lengths, and notch positions, we have opted for a double-structured HHG chamber, made of several building blocks, which is a custom designing and machining\footnote{Note that all small components have followed an ultrasonic cleaning process, in an ethanol bath, or with wipes soaked of isopropyl alcohol for cleaning of larger parts such as inner vacuum chambers.} and essentially made of simple and standard mechanics parts that are easily found in the gas and vacuum components of the professional market. It has been thought to be readily integrated and to handily receive HCWs of lengths $\sim 8 < L_{\textrm{med}} < \sim 70$ mm. It consists in a first inox vessel that allows the gas load in the HCW, placed and sealed inside a second buffer inox chamber that ensures vacuum for the swift expulsion of the gas at the capillary openings, see Fig. \ref{fig:figCapillary}. The first vessel has been made in multiple specimens. Its top cap is made of transparent PMMA, sealed with a rubber gasket. Transparency allows fine-tuning the laser alignment, empirically, in visualizing the shape and behavior of the plasma plume inside the capillary waveguide, while staying under vacuum. As one needs degrees of freedom for a proper alignment between the cell and the driving laser, the whole assembly is mounted on a five-axis positioning stage (XYZ and pitch-yaw rotating axes, that we denominate \textit{cradle}) for alignment along with the laser beam path, see technical blueprint in Fig. \ref{fig:Drawings_HeartSource}. Two welded bellows, connected on both opposite sides of the vessel containing the capillary and its nested box, allows for relative freedom in its movements. Once it is aligned with the laser beam (done at the atmosphere), the capillary vessels are curbed in order not to move when pumping volumes down to vacuum is realized. 
Note that we have chosen to use $304$ Stainless steel instead of Aluminium, to ensure a rigid body
\begin{figure}[ht!]
\centering\includegraphics[width=15.4cm]{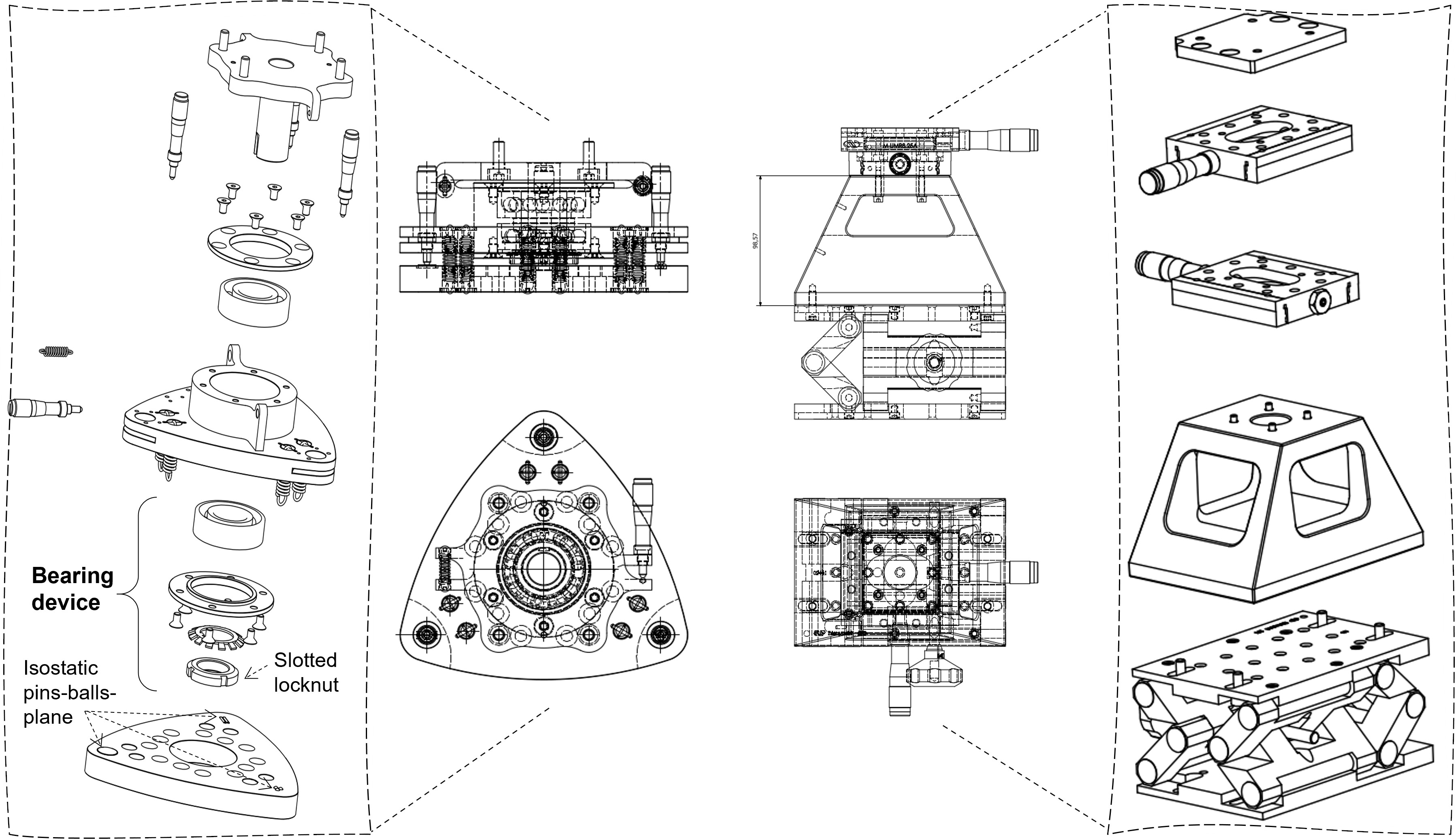}
\caption{Examples of ensemble drawings of the degrees (translation and roll, pitch and yaw) of freedom regarding the XUV source part. Top- and bottom- left figures are the side and top views of the cradle that ensures the rotation and tip-tilt of the capillary box. Right column figures display side and top views of the movements of translation of the capillary box. The pantograph at the basis of the device authorizes for a coarse adjustment in the \textit{z}-direction, perpendicular to the optical table. Sides: respective exploded views.}
\label{fig:Drawings_HeartSource}
\end{figure}
\begin{figure}[ht!]
\centering\includegraphics[width=15.1 cm]{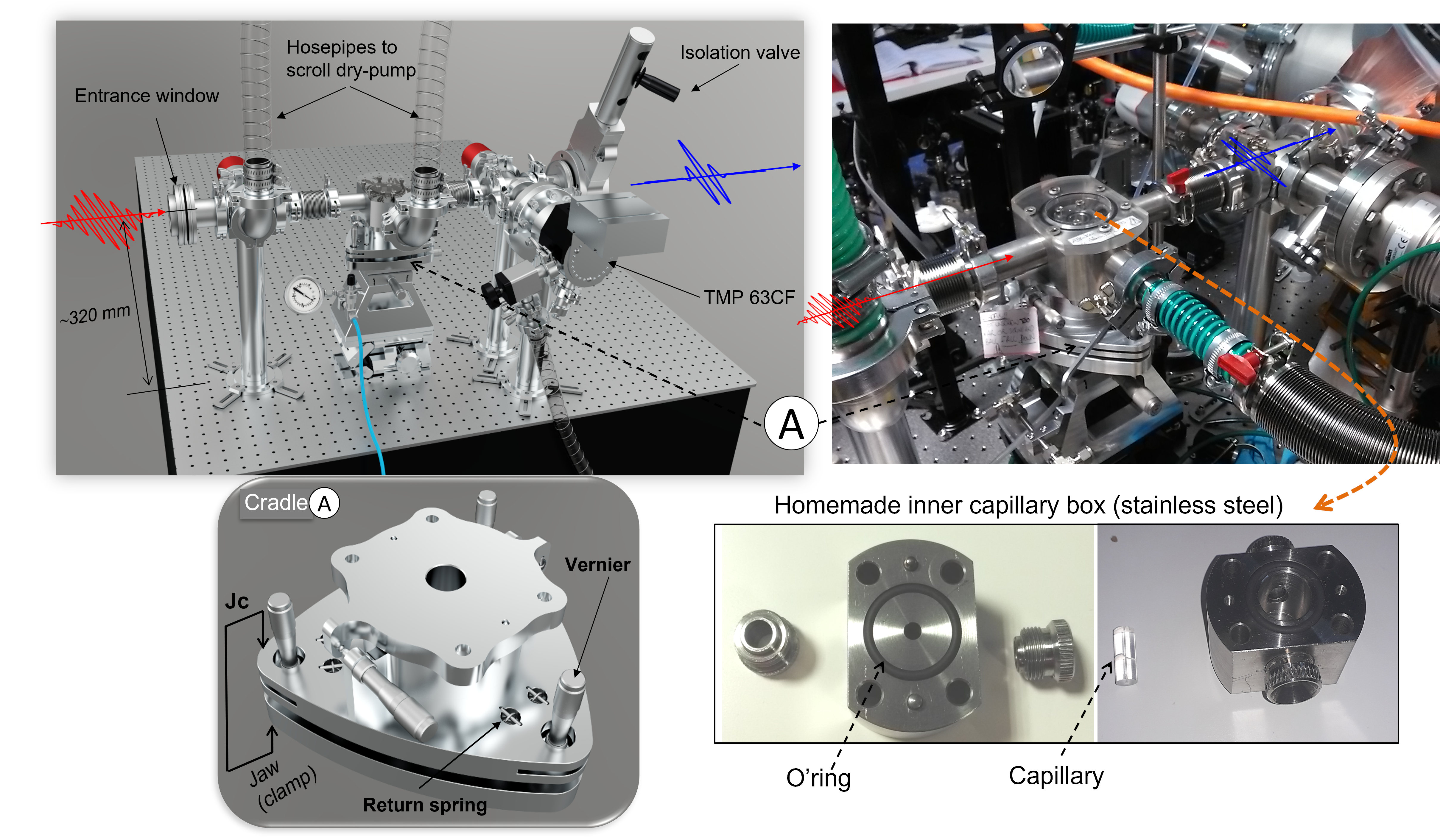}
\caption{The heart-part of the XUV source with the different degrees of freedom, highlighted. The cradle rests on a trapezoidal, monolithic and hollowed-out frame-block for a good mechanical stability and rigidity. A dual-pantograph lab jacks allows for a coarse lift or drop of the XUV source in the $y$-direction. The cradle is at last mounted on a three-axis translation linear stage, possesses $6$ repelling springs, and lays on a kinematic mount made of segments (roller) and balls bearing (line-point-plane mechanical linkage system), on which are positioned the micrometric heads (vernier). The latter have $10$ µm resolution per revolution. The upper plateau which receives the capillary box has centering rods. \textit{Jc} is a (helical) punctual/contact-point load (x$3$) for clamping while pumping down to vacuum, just in case (but based on our experience, is not mandatory \textit{a posteriori}). Finally, the whole has five-axis of motion. Plane of incidence is $\{x,z\}$, the horizontal plane given by the optical table. Right: photographies that show the XUV generation part, and the capillary with its box (called inner small box) made of $304$ Stainless steel. This is this part that is modular, containing o'rings, and that can be adapted to different values of HCW length. Laser comes from the left. }
\label{fig:fig_XUV_Heart}
\end{figure}
Furthermore, since differential pumping must play a cardinal role, as has been shown in the test bench described earlier, a specific component (green in Fig. \ref{fig:figCapillary}) has been designed to reduce the gas pipe conductance at the HCW exit. As the focused laser beam passes through the latter, where it is divergent at this distance (after focus), the length of this component has to be tested, because, the periphery of the beam should preferably be clipped (depending of the inner diameter) as it is a first step in blocking\footnote{Here, it is important to state that, while we have chosen to give the differential pumping component two functions, to gain space, the full blocking of the residual driving laser is carried out in the filtering chamber (Fig. \ref{fig:fig4}), because it is a very high thermal load (where ablation takes place, and discussed in the supplemental material, subsection \ref{sec:Thermal_MechanicStress_Optics}) that is necessary to move away, and to make in several steps, to preserve as long as possible the physical integrity of components in contact with it. } this residual driving laser which has vitiated all usefulness after the interaction region into the capillary, but also because, first and foremost, the XUV beam must propagate through this component without being diminished. This component has the form of an interchangeable disk (a wear part, is said), with a $500$ µm ($1$ mm has also been tested) drilled in its center, and about $2$ to $3$ mm thick. Then, downstream the micro-drilled disk of differential pumping, a flange connects the following small vessel (buffer) to a $35$ m$^3$/h scroll primary pump, and another differential pumping, an interchangeable tube attached into a Swagelok connector, as the ones studied in Section \ref{sec:Test_Bench}, is added tens of centimeters further, both come and complete this decisive function of reducing the pipe conductance, as shown in Fig. \ref{fig:figCapillary}. At the end, this ensemble allows us to reach efficiently up to $3$ atmospheres in backing pressure, while maintaining a safe vacuum of $10^{-6}$ mbar in the spectrometer, at the end of the beamline. This prevent damages to the XUV detector which must work below $10^{-4}$ mbar according to the manufacturer. 
The capillary waveguide is made of borosilicate glass (transparent), provided by Wilmad-LabGlass. One slit is cut perpendicularly at half the length of the capillary (the inner diameter needle) to ensure gas inlet (one-jet device configuration), as was made by \cite{PhysRevLett.82.1668}. With one slit, a gradient gas density appears from the center of the fiber to the tips, whereas with two slits, an interaction region with constant gas density is established in-between the slits, which should favor phase-matching. But, it is worth noticing that we have generated harmonics from a two-slits capillary (as sketched in Fig. \ref{fig:Capillary_TwoSlits}), as used in \cite{Popmintchev:08}, and very similar results were observed between the one slit or two slits configuration, but to the benefit of the one slit configuration. Moreover, we observed, as also reported in \cite{Goh2015}, that part of the guided light is scattered at various slits, so we chose to work with one slit to minimize the losses. To avoid gas leakage, the HCW is sealed from vacuum by two FKM O'rings rubber gaskets at each of its ends, pressed against the outer cladding (in being crushed by the Swagelok fitting), see Fig. \ref{fig:figCapillary} and insert in Fig. \ref{fig:fig4}. There are four variable parameters for the capillary waveguide: length, inner diameter, slit width and position of the slit. The outer diameter ranges from about $5$ mm to $6.5$ mm, depending on the inner capillary diameter, and is a limitation set by the manufacturer. Thus, we have realized a homemade set of fittings to accept all diameters, and to create a vacuum-tight seal around the outer diameter of the waveguide.
\begin{figure}[ht!]
\centering\includegraphics[width=15.5cm]{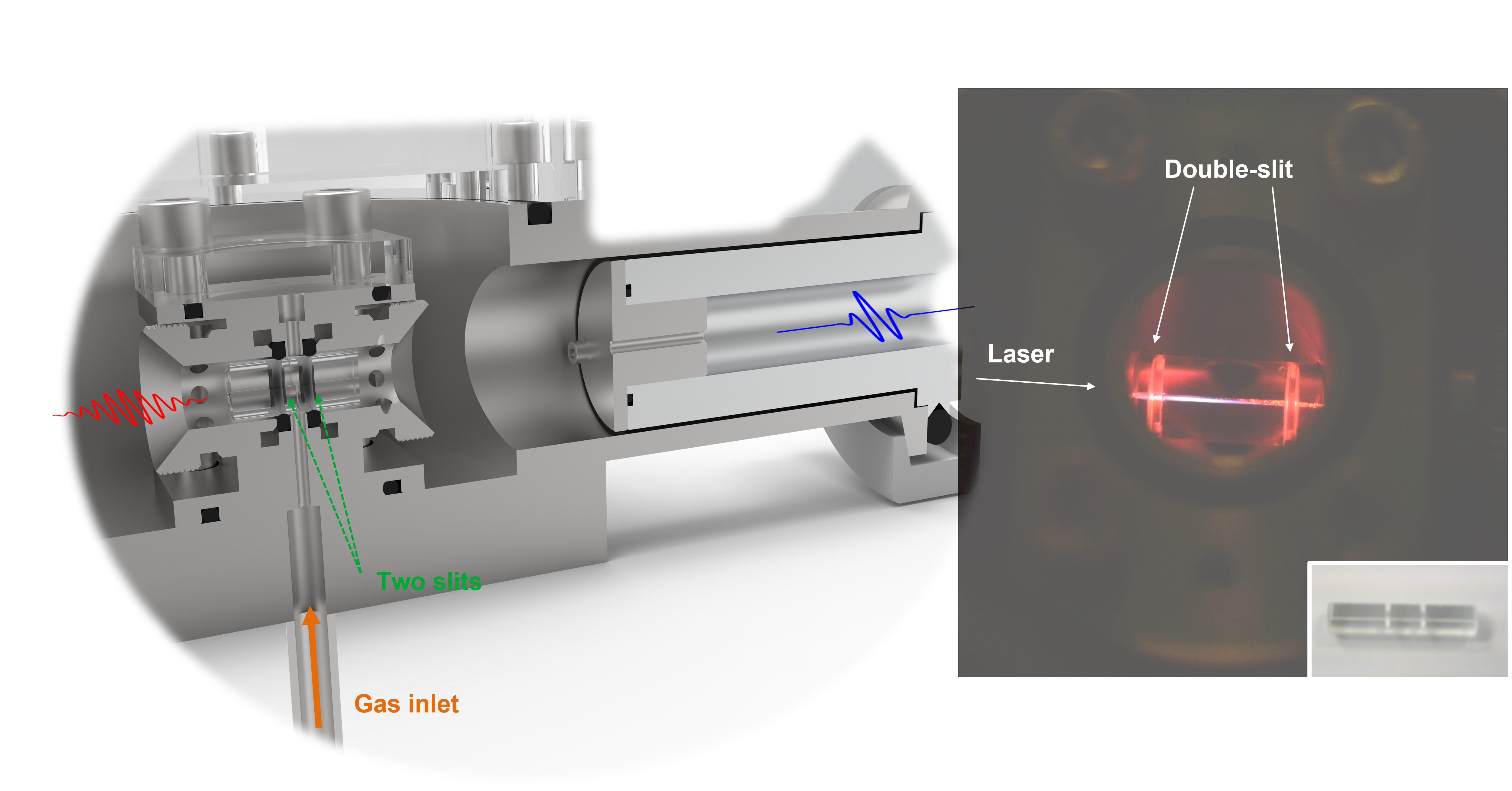}
\caption{Left: close-view rendering of the capillary, with two slits, for a constant gas volume in between the two slits. Right: snapshot of the capillary with double-slits, while experiment in progress, showing plasma inside (laser pulses come from the left). Insert shows one capillary taken alone.}
\label{fig:Capillary_TwoSlits}
\end{figure}
The HCW is then filled with inert gas (target gas is supplied by flexible polyurethane -polyethylene- hoses, and manual dosing valve with a shut-off system, from Balzers) from a vertical inox gas duct at a static pressure and continuous gas flow of Argon or Helium (both M$20$ Bottle AlphaGaz1 high purity), and accurate gas pressure diagnostic is realized at $50$ cm from the HCW inlet with a capacitive gauge Diavac-N from Leybold and a Bourdon gauge in complement for a double reading, especially regarding the higher pressure, $> 1$ bar. The vacuum pressure is measured with Wide/Full-Range and Pirani gauges. A focusing lens is mounted on XYZ driver to adjust and optimize the focus position of the laser pulses into the capillary. The gas is exhausted\footnote{One end is facing a Primary Pump flange, the other is facing a differential pumping component, a conductance reducer, and itself facing a Turbomolecular pump. Furthermore, sometimes the slit for gas inlet is not centered in the capillary length. Consequently, as will be seen in the Supplemental Material, it breaks the symmetry in the shape of the gas exhaust from the two opposite ends of the capillary, because the boundary conditions are asymmetric.} at the both opposite extremities of the HCW. 
To prevent damage at the entrance of the HCW while aligning the optical path of the laser beam, even though at very-low intensity, we have design a simple component (Fig. \ref{fig:fig4_AlignmentCapi}) that simulates the physical position of the capillary and which does not dread the laser intensity, especially because it comprises an interchangeable wear part that faces the laser beam. 
\\One of the key advantages of our setup is that by simply closing the isolation valve (see Fig. \ref{fig:fig_XUV_Heart}), as the buffer volume is small, atmospheric pressure is quickly reached, opening the box is fast ($4$xM$5$ CHC stainless steel screws), and, after the process of removing the former one then replacing the new HCW, vacuum down is also rapidly achieved. The laser alignment does not change, even though the pressure drop can be abrupt, if ill-handled, because we installed clamps (jaws) on the cradle (which lays on a point-line-plane isostatic framework, see Fig. \ref{fig:Drawings_HeartSource}). We estimate, by analyzing the torsor of the static mechanic actions of the exerted external forces reduced to the spring contacts (ball joint and pivot) with rest of the body, that springs AUT $10$-$30$ from Misumi are the most suitable. A micrometer dial indicator gauge placed on the buffer, whilst pumping down to vacuum, has shown that the ensemble moves of less that tens of microns (thus, $<5$ per cent of the inner diameter) only when pumping (by means of the bellows, and of the return springs (of repelling) with the appropriate stiffness, as defined earlier, see Fig. \ref{fig:Drawings_HeartSource} and Fig. \ref{fig:fig_XUV_Heart}). This means that it is reproducible and unimpeded by consecutive changes of the capillary (no need to re-align at every time). By the end, we have made two boxes to use capillary lengths from $10$ to $70$ mm. This can be seen in Fig. \ref{fig:fig_XUV_Heart} where the inner small box is shown together with one capillary waveguide, amongst a set of tens of capillary lengths. 
\\
Precision is a combination of mechanical tolerance and play. Regarding the angular precision of the ensemble (Fig. \ref{fig:Mircometer_Angular_precision}), we have attempted to evaluate it with AutodeskInventor. In considering one micrometer head BM$11.16$ set free to move, while the other two are fixed, one then takes the distances between the vertical axis of each of the micrometer heads at their application point of their ending ball on the pins (two metal rods or needles) or plane or balls (three coplanar) constituting the line-point-plane, for isostatism compliance. 
\begin{figure}[ht!]
\centering\includegraphics[width=12.4cm]{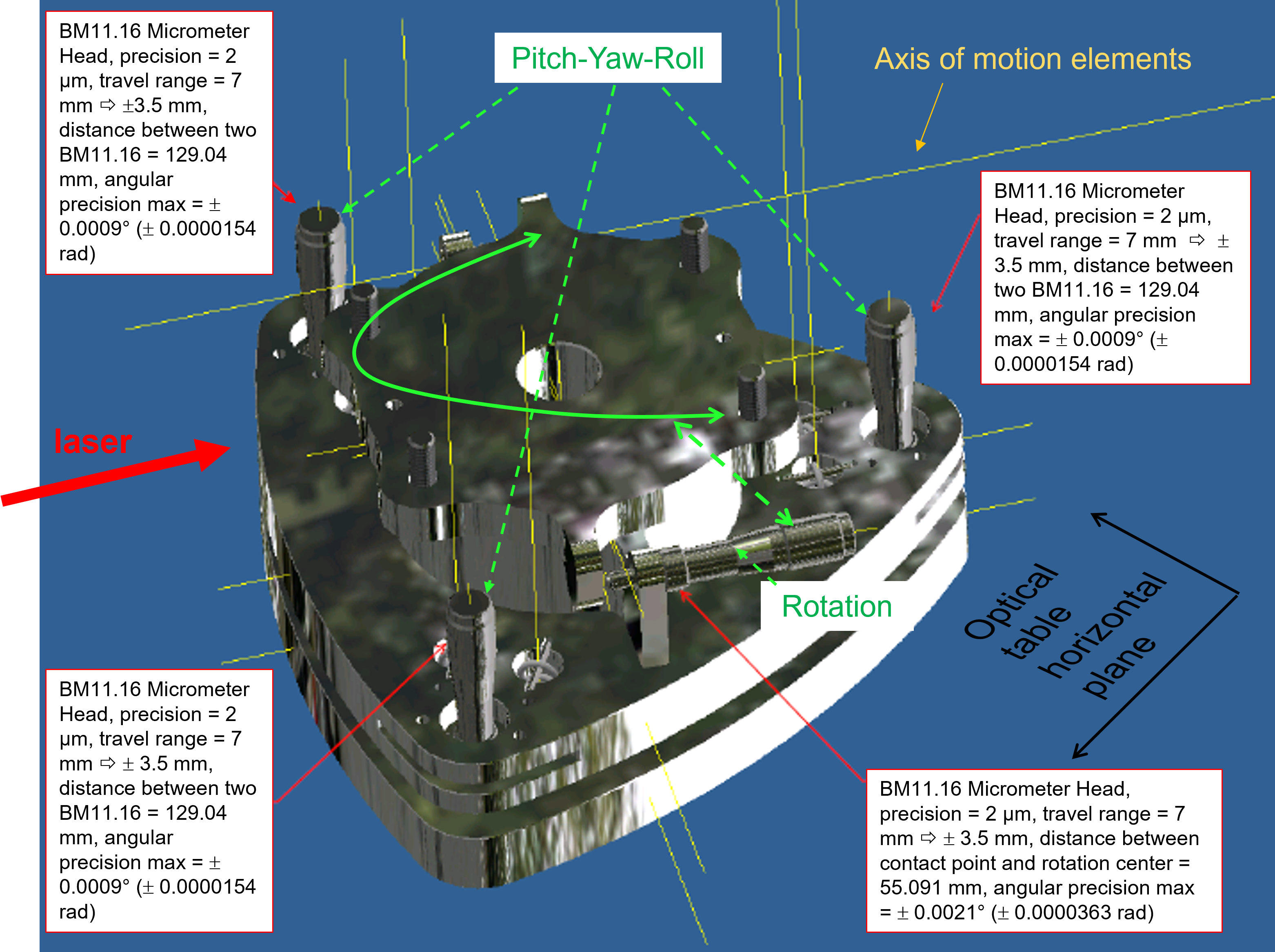}
\caption{Calculus of the angular precision of the cradle for capillary positionning with respect to the laser beam. Yellow lines are the axis of interest for yaw-pitch-roll, in order to calculate distances between elements. The equivalent lever arm at the focal distance of $400$ mm is therefore less than $1$ µm in one direction. Laser polarization is in the optical table plane. }
\label{fig:Mircometer_Angular_precision}
\end{figure}
Also, the stiffness of the springs must be taken into account. They are pre-wound mounted in the device, come by pairs for each axis, and have been selected to support the constraint imposed by the micrometer head, with an admissible strength of $40$ N. The bearings are in an angular contact ball arrangement and are wound by a slotted locknut, see Fig. \ref{fig:Drawings_HeartSource}, thus reducing or canceling the backlash. All these features led us to estimate the angular precision as shown in Fig. \ref{fig:Mircometer_Angular_precision}.
\\
In the following subsections, each vacuum chamber is described. The beamline was designed to assign to each vacuum chamber a unique function. Each block can be vacuum-insulated from the others by gate valves, for maintenance operation, or for laser foil filters substitution, for instance. Cylindrical and cross-shaped chambers, visible in Fig. \ref{fig:fig4} and Fig. \ref{fig:fig4_b}, have been chosen to satisfy the space saving in our lab and the cost savings by purchasing manufacturing standards in the boilermaking factories or suppliers. The HHG chambers are connected to one $35$ m$^3$/h ($35$-XDS) scroll dry pump from Edwards, upstream (left side in Fig. \ref{fig:figCapillary}) and one Turbomolecular pump (TMP) SL$80$H from Leybold coupled at its exhaust with a $15$-nXDi scroll dry pump (Edwards), downstream (right side in Fig. \ref{fig:figCapillary}). The global alignment through the beamline is then assessed using the zeroth order diffraction of the Spectrometer. 
\subsubsection{Filtering chamber}
\begin{figure}[ht!]
\centering\includegraphics[width=14.3cm]{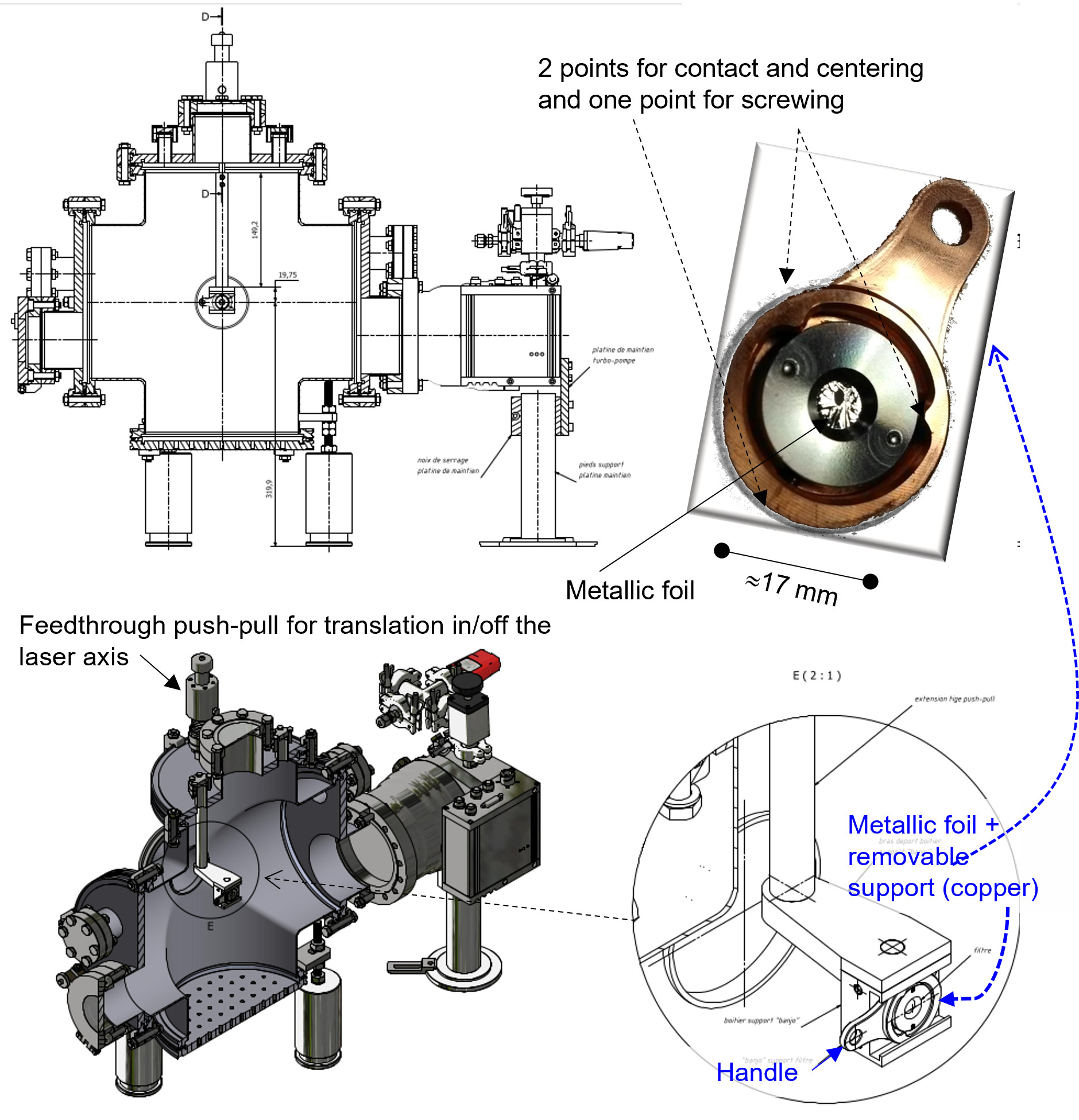}
\caption{Top-Left: drawings of the overall technical plans of the filtering chamber. Bottom: the cross-section shows the interior of the chamber, and the turbomolecular pump with the system of valve and gauge at its exhaust towards the primary scroll pump. The close-up view in the inset exhibits the metallic foil with its frame, both from Lebow Co., and its homemade support, handled with gloves, and attached to the push-pull arm from MDC Caburn. A flange on the cover is fitted out with a fast-opening and closing transparent door. Top-right: the filter foil (Lebow L0.5 rings) and its copper homemade support, for easy handling.}
\label{fig:Filtering_Chamber_Draw}
\end{figure}

\begin{figure}[ht!]
\centering\includegraphics[width=13.3cm]{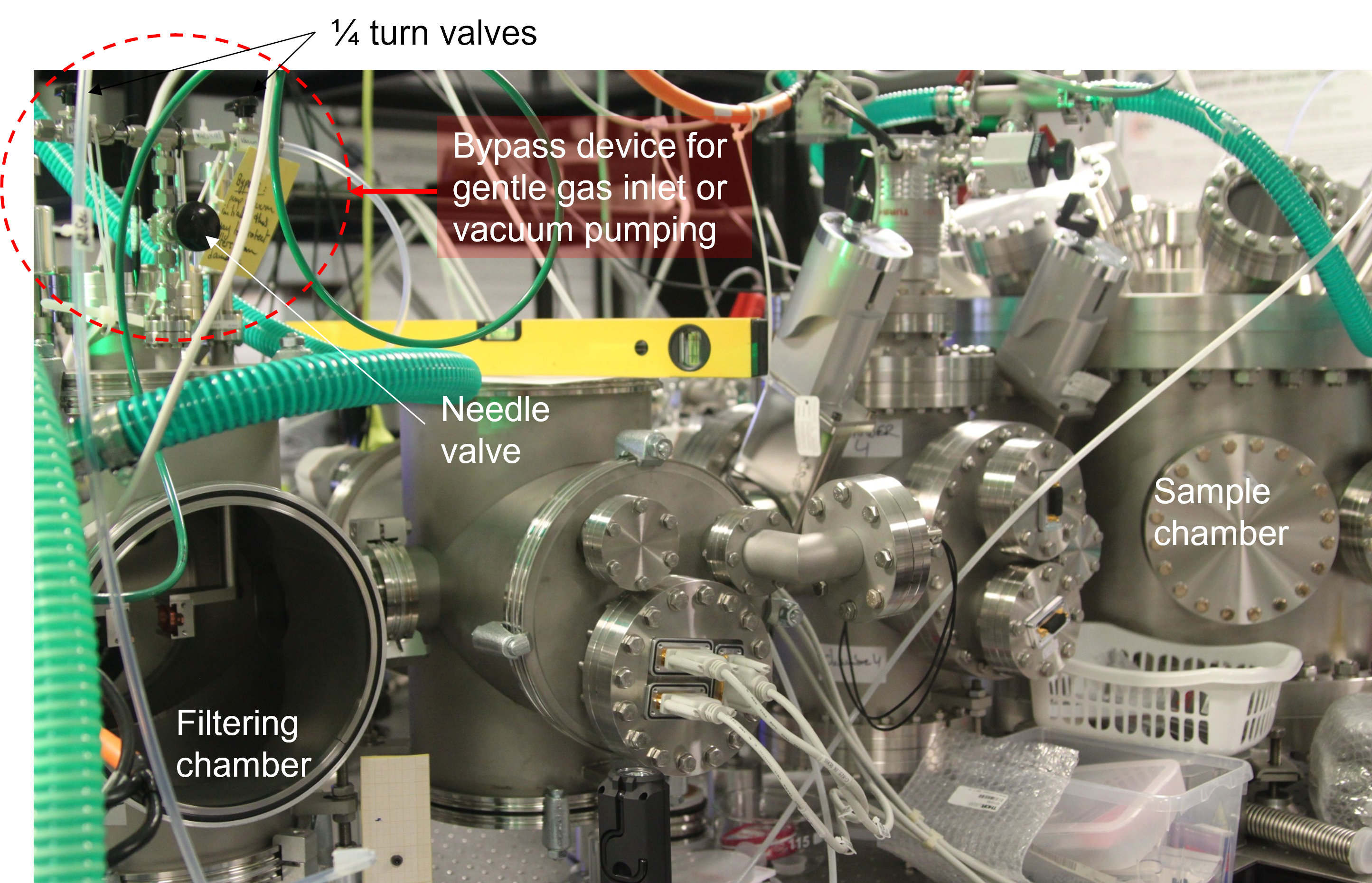}
\caption{Photography showing the vacuum chambers, and the gas and vacuum bypass device, for gentle flow regarding the integrity of the thin foils placed inside the filtering chamber (showed opened).}
\label{fig:Gas_Bypass}
\end{figure}

The laser beam was tightly focused with a $f=400$ mm N-BK7 lens at the entrance of the HCW with a $f$-number $f\# =f/D \approx 40$. The beam enters the XUV beamline through a $8$-mm thick anti-reflection (AR)-coated UV fused silica window. The lens is placed close to the entrance window to avoid damages arising from self-phase modulation. %
The vacuum is ensured with cascading $4$ turbomolecular pumps, allowing a descending pressure down to the last chamber containing the MCP and the XUV spectrometer.  These pumps are $3$ x Turbovac 350iX based on a hybrid rotor suspension, made of magnetic bearings, in DN100CF and DN160CF flanges. Note that the choice for turbo pumps with magnetic suspension (thus contactless) is justified because of mechanical vibrations arising from turbo pumps which could be communicated through their bearings to sensitive components inside the chambers, especially optical parts, which must not be put out of alignment, regarding the experiment integrity.
After generation, the fundamental of the laser is suppressed with thin metal foils, held on a movable and interchangeable support, slided in the beam axis thanks to a feedthrough telescopic arm (based on two push-pull actuators), manually activated from the outside of the chamber (Fig. \ref{fig:Filtering_Chamber_Draw}). These foils have two main purposes: they protect from optical damages the toroidal mirror (designed for accepting XUV intensities) and, above all, the XUV detector, which is very sensitive to intense NIR laser light, and, they remove this residual driving laser that would perturb the region under study at the sample place, which needs to be only excited (triggered to non-equilibrium) by the laser pump arm only, definitely not by the residual laser used for XUV generation. This chamber is mounted with a bypass on its top, consisting of a first butterfly $1/4$-turn valve to allow smooth ramp of gas (nitrogen) inlet when vacuum has to be broken, and smooth ramp of pumping down from the other $1/4$-turn valve, both together with a needle (accurate) valve so as to preserve thin foils integrity, see Fig. \ref{fig:Gas_Bypass}. Once a reasonable vacuum or gas pressure level is reached, the pumping or gas inlet can be safely amplified. 
\begin{figure}[ht!]
\centering\includegraphics[width=14.75cm]{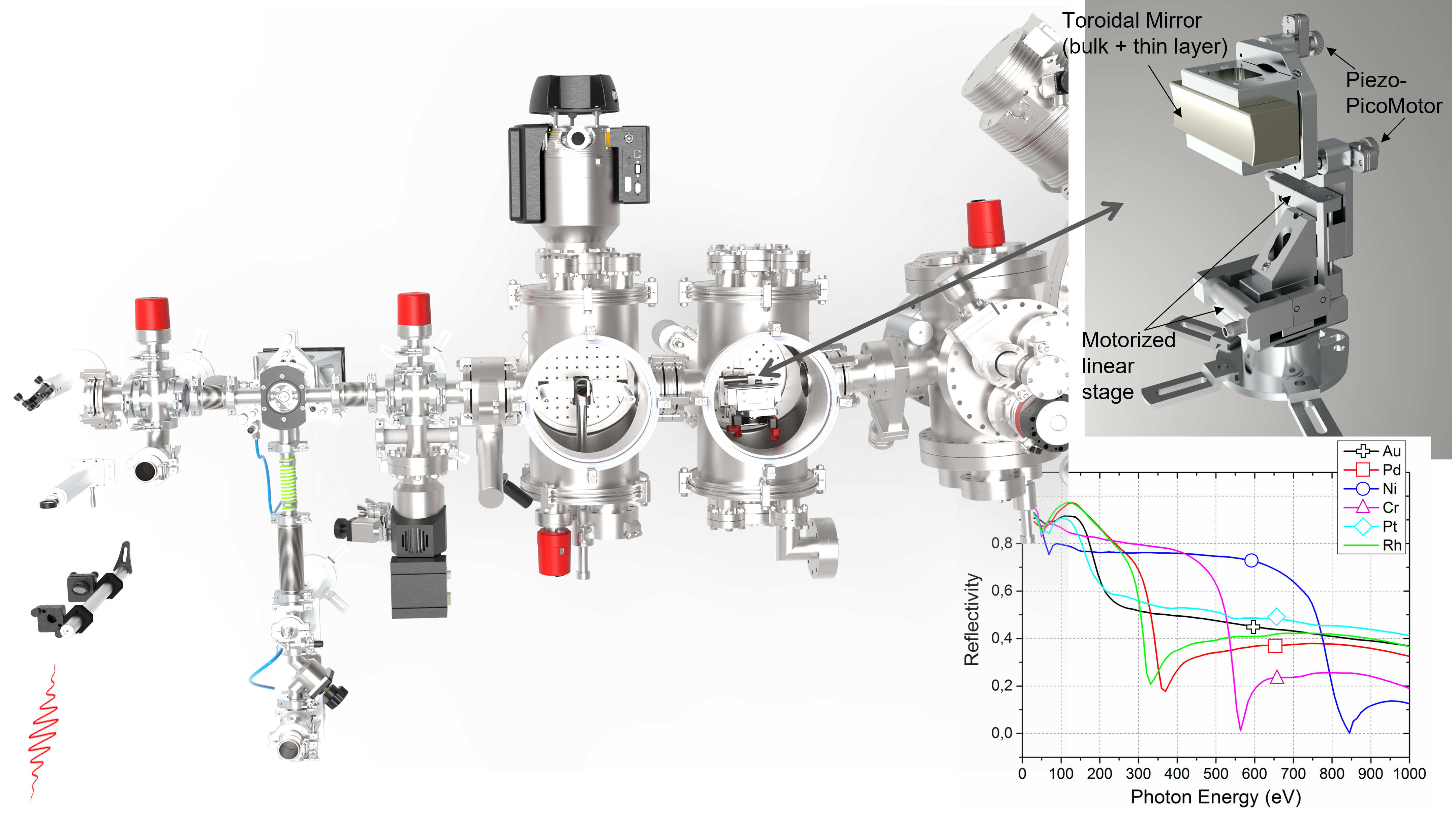}
\caption{A closer top-view of the beamline entrance (left) up to the TM chamber, and a zoom of the TM, showing also its support and its degrees of freedom. Chambers are represented opened, without their caps. Reflectivity of a few thin metallic layers, for a grazing angle of $87$ degrees, proposed for dressing the surface of the toroidal mirror, for which the bulk is in silicium (Si). Acronyms Au, Pd, Ni, Cr, Pt, and Rh stand for gold, palladium, nickel, chromium, platinum and rhodium, respectively. Data from CXRO \cite{CXRO}.}
\label{fig:reflectivity_CXRO}
\end{figure}
\subsubsection{Toroidal mirror chamber}
After its generation, the XUV beam is accurately micro-focused in the center of the interaction chamber, the future targeted region, with the magnetic sample to be investigated in pump-probe experiments. This function is handled by a Seso Thales toroidal mirror (TM), designed for a $3$ degree grazing incidence angle and with a magnification of $\sim1.6$ (both sagittal and tangential foci coincide). For a precise alignment, it is mounted on compact XYZ linear stage of the M11x series (2-phase stepper motor or open-loop DC gear motor, and backlash-compensated threaded spindle or ball screw), and with two PiezoMike piezoelectric linear actuators for tip-tilt axis, both from Physik Intrumente (PiMicos). The whole is remotely steered thanks to vacuum sealed electrical feedthroughs from the XAVAC series. The metallic surface for XUV reflection has been chosen to be platinum (Pt), as it has a relatively good feature over a wide photon energies range, see Fig. \ref{fig:reflectivity_CXRO}. The TM mount ensemble is home-built, and, as it is overhanging, machined parts are hollowed out to lighten their weight, as it can be perceived in the top-right insert of Fig. \ref{fig:reflectivity_CXRO}, both for stability and regarding the permissible torque in the three directions, indicated in the stepper motors datasheet. Finally, the whole device for XUV focusing described here is placed in a chamber that basically has the same architecture as the one shown in Fig. \ref{fig:Filtering_Chamber_Draw}.
\\
We have also carried out an experiment with a sample of a platinum-coated mirror, to study the effect of a possible optical damage that could arise on our TM, after the filtering chamber in case of filter foils absence. After $5$ hours of laser irradiation at the usual pulse energy, no damage has been observed on the sample. Therefore, our toroidal mirror can be used safely.
\subsubsection{Recombination chamber}
\begin{figure}[ht!]
\centering\includegraphics[width=13.4cm]{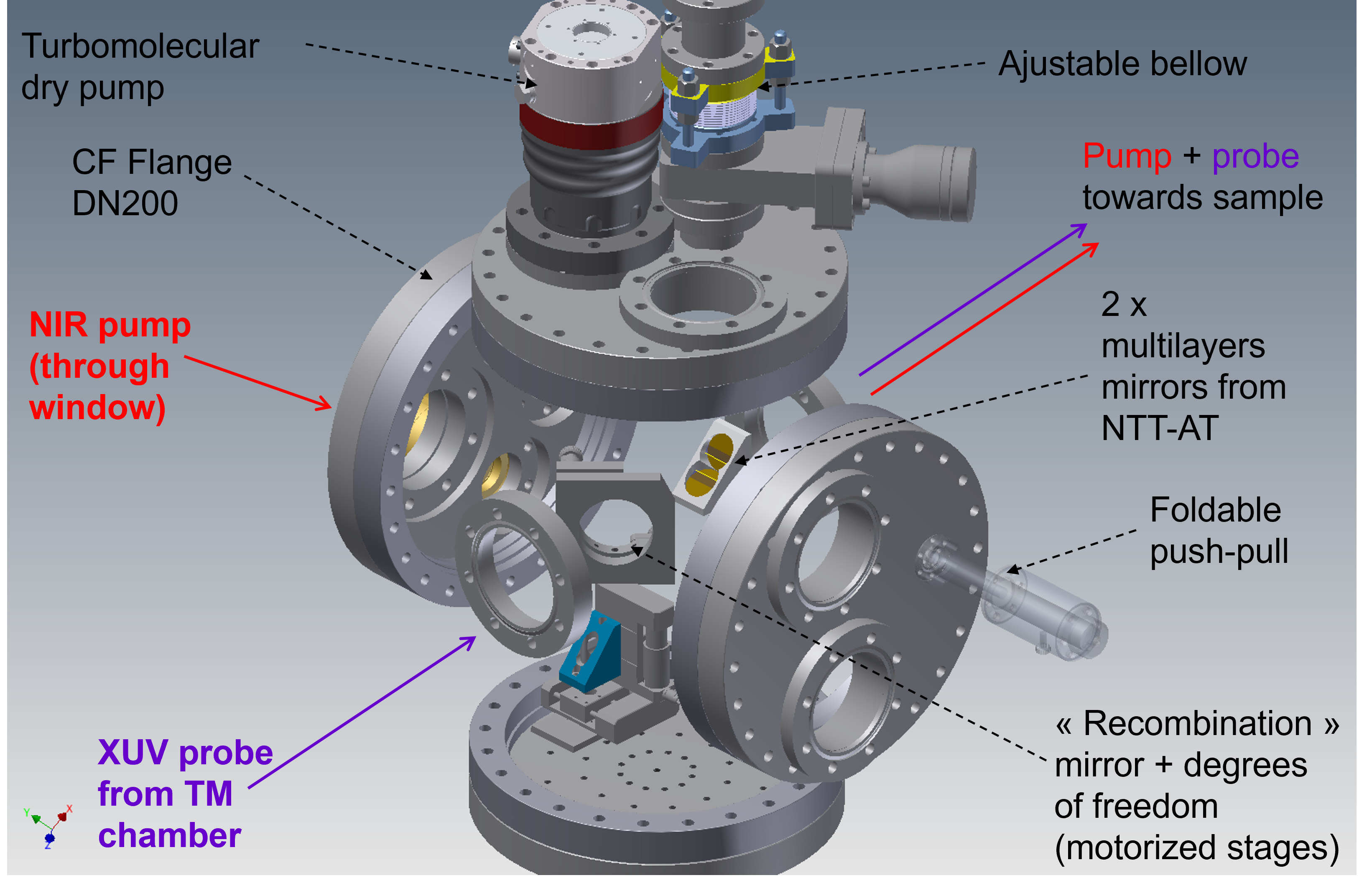}
\caption{View of the recombination chamber, taken alone.}
\label{fig:Recombination_Chamber}
\end{figure}
In the optical point of view, the laser pump, after passing into a controlled and motorized delay line, is focused before entering the beamline through an entrance window into the recombination chamber, see Fig. \ref{fig:fig4} and Fig. \ref{fig:Recombination_Chamber}. It is reflected with a close to $90$ degrees angle, either by a coated D-shape (half-mirror) mirror or a coated mirror drilled at its center, whence it begins to co-propagate with the XUV beam, up to the sample, then the pump is blocked with a filter foil, while the XUV probe, transporting the relevant signal, ends up in the spectrometer. The set of two multilayer mirrors from NTT-AT are mounted on a foldable push-pull and can be slided in the XUV axis of propagation to send the beam into a XUV CCD camera placed on the lid with an adjustable bellow to observe for the micro-focused spot of the XUV (with the distance [multilayer mirrors-XUV camera] being equivalent to the one [multilayer mirrors-sample plane], to exactly reproduce the image of the XUV focusing) and to measure its FWHM diameter.
\subsubsection{The sample interaction chamber}
\label{sec:Interaction_Sample}
\begin{figure}[ht!]
\centering\includegraphics[width=16.1cm]{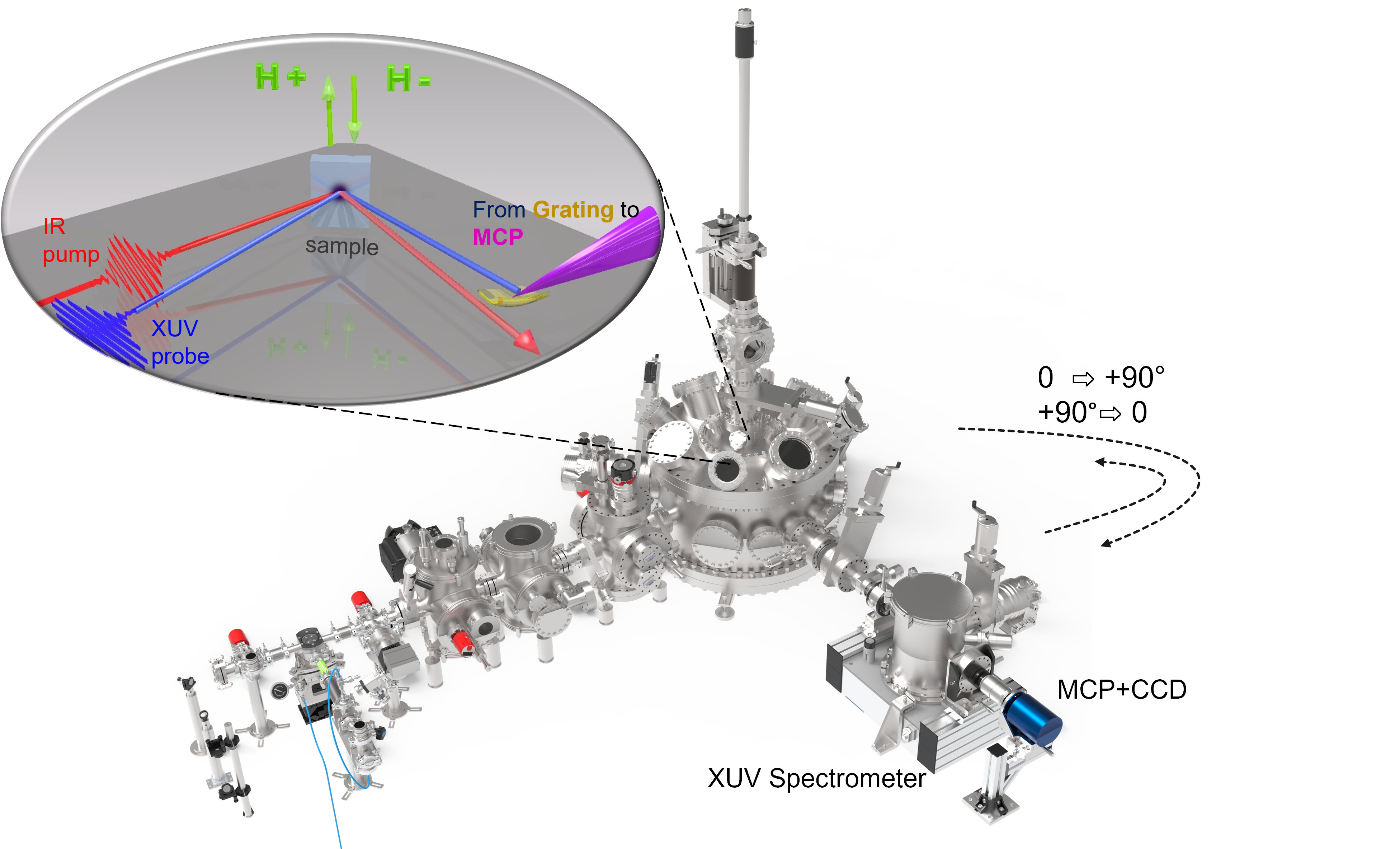}
\caption{3D rendering of the setup in the configuration for time-resolved Kerr Magnetic measurement (spectrometer sets at $90$ degrees, where magnetic contrast on the sample is optimized, in reflection, from the two reversed directions of the magnetic field $\mathbf{H}$, see inset). }.
\label{Kerr_Configuration}
\end{figure}

\begin{figure}[ht!]
\centering\includegraphics[width=14.5cm]{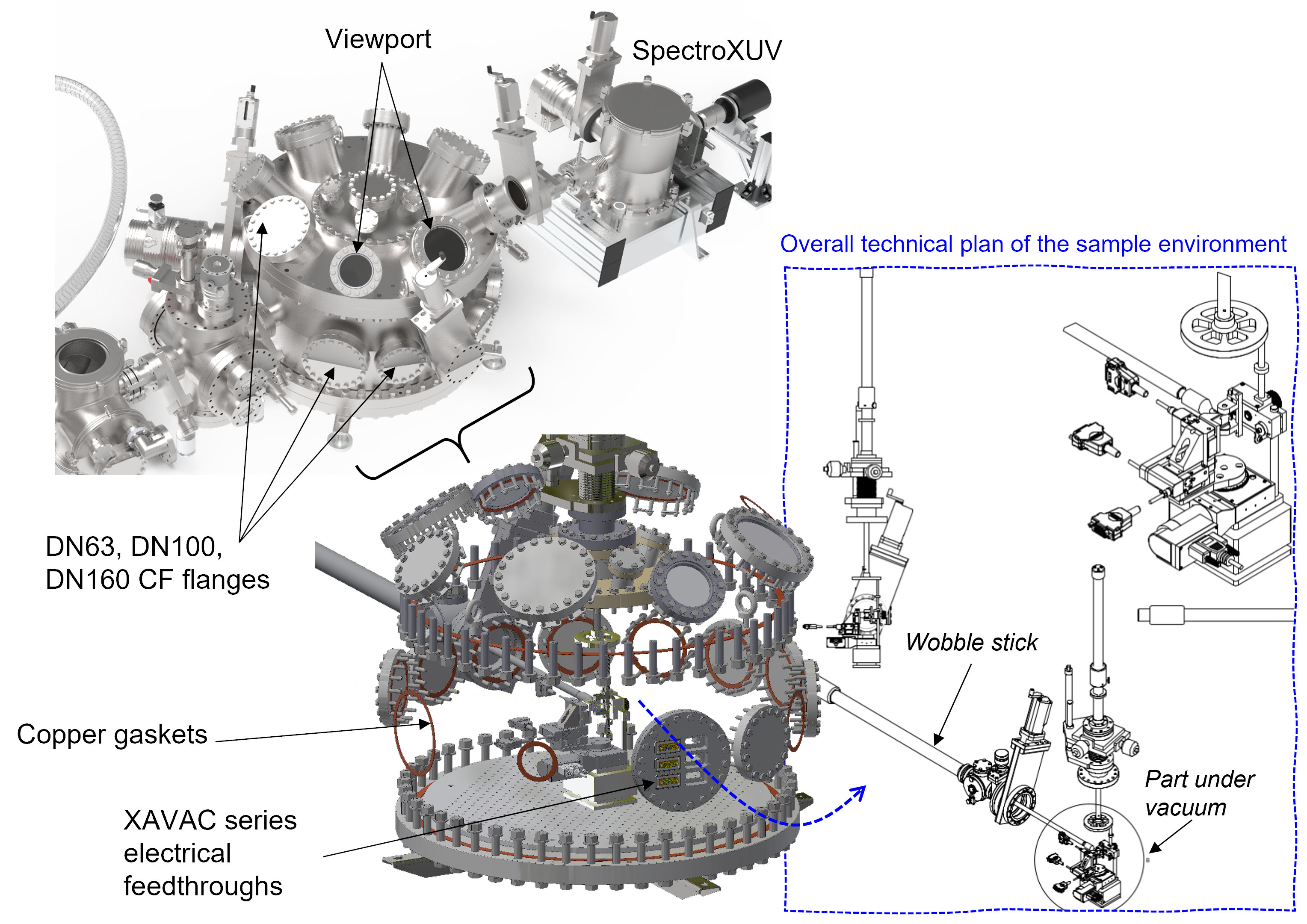}
\caption{View of the sample chamber and its environment. The inner diameter of the sample chamber is about $700$ mm, thus, the resulting needed thickness of the partitions and walls, to warrant rigidity, inevitably leads to a metric half-ton weight, at the very least. The optical table, in its entirety, on which sit on the laser and the rest of the beamline, is mounted on a pneumatic device that compensates for the drop due this heaviness. Top-left is a part of the XUV beamline. At the center, it is the sample chamber in transparency (partitions -or walls- are removed). At the right side, overall technical plans of the sample environment are shown.}
\label{fig:Sample_Environment_Chamber}
\end{figure}

\begin{figure}[ht!]
\centering\includegraphics[width=14.0cm]{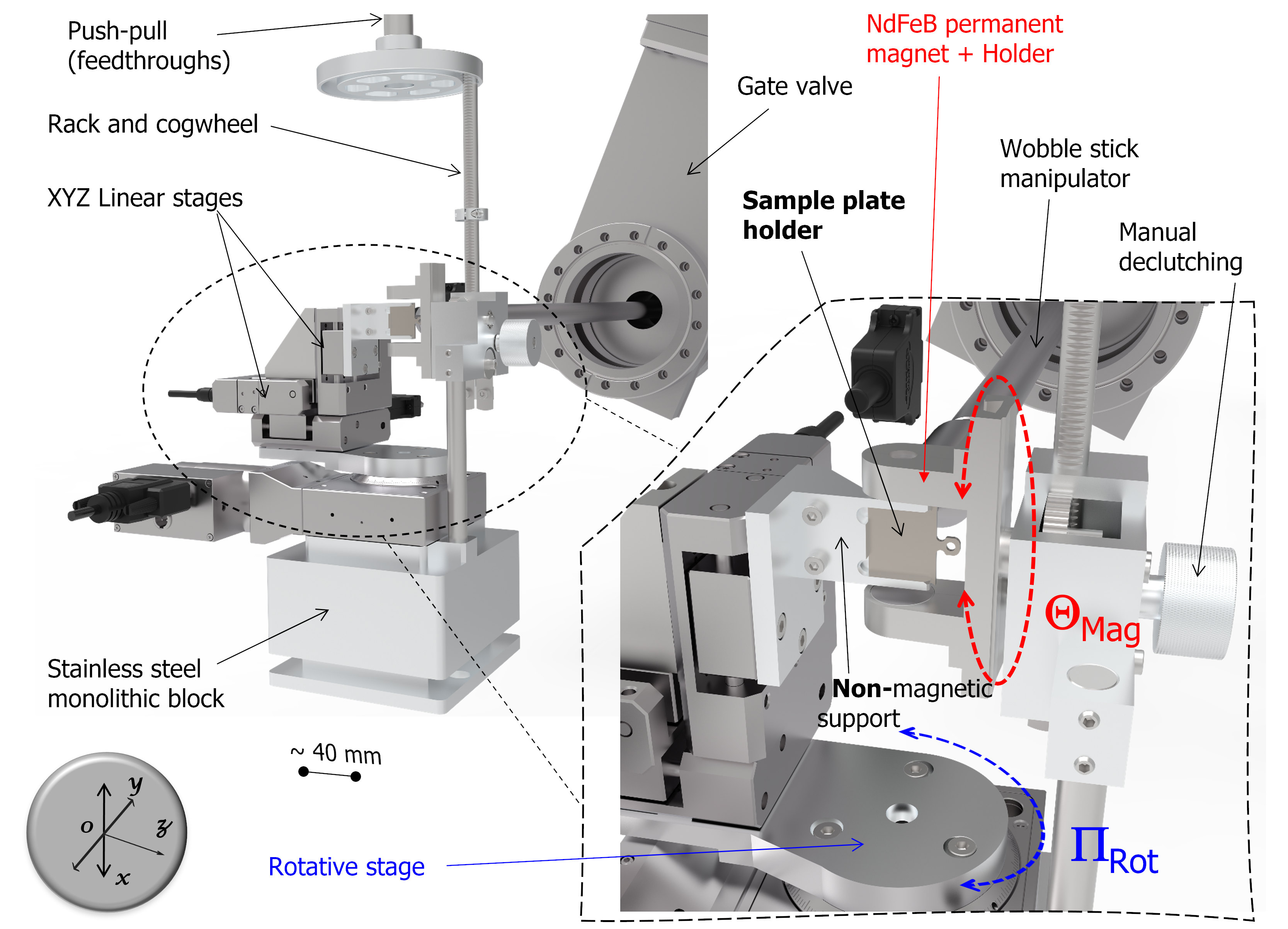}
\caption{A 3D-CAD of the complete device for managing the sample, inside the chamber depicted in Fig. \ref{fig:Sample_Environment_Chamber}. Right insert is a zoom view showing  the $\Theta_{\textrm{Mag}}$ rotation that allows the two H+/H- magnetization directions, parallel to the sample. The rotation $\Pi_{\textrm{Rot}}$ allows to transfer the system [sample-plate holder] in a position, locked on a non-magnetic support, then to present one of its faces towards the pump-probe beams.}
\label{fig:Sample_Environment}
\end{figure}

The sample interaction chamber is equipped with CF flanges periodically arranged around its perimeter. As a matter of fact, it is possible to perform experiments in the so-called time-resolved Kerr magnetic configuration, in the reflection configuration on the sample, as shown in Fig. \ref{Kerr_Configuration}, to be compared with the transmission configuration in Fig. \ref{fig:fig4}.
At the center of the interaction chamber, see Fig. \ref{fig:fig4} or Fig. \ref{fig:fig4_b}, the sample and all its environment have to be yet built and installed. However, the concept has been thought and drawn, the materials and components bought, and the whole is presented in Fig. \ref{fig:Sample_Environment_Chamber} and Fig. \ref{fig:Sample_Environment}. All motorized stages are piloted from outside, with electrical feedthroughs. The sample chamber is shown in more detail in Fig. \ref{fig:Sample_Environment_Chamber}. A set of ports (fixed and rotatable) are regularly distributed at the outer periphery for multi-purpose setup, and the choice of copper gaskets with CF flanges is driven by the necessity to reach at least the $1$E$-8$ mbar decade in this interaction chamber, specifically because rare-earths doping the heterostructures under study can migrate under moisture. The material is stainless steel grade $316$LN (mu-metal shield being too much expensive), for non-magnetic consideration, with a leak test $< 10$E$-9$ mbar.l/s guaranteed. Also, as a backplane-like, the inner bottom of the chamber is dressed with a breadboard of hundreds of M6 tappings, for which through holes have been drilled to avoid cavities (known as micro-voids) that could trap impurities and then degas, where miscellaneous measuring equipment can be easily installed. At last, when pumping over one night this chamber only, empty of any equipment, with all connecting valves that are being closed to isolate it from the other chambers, and after having rinsed the inner volume a couple of times and flooding it with dry nitrogen, we are able to reach about $1.5$E$-8$ mbar. 
Before pump-probe experiments and in order to load the sample of interest into the interaction chamber, the protocol is as follows. First, the sample to study is placed on a plate holder, at the end of a wobble stick, inside a transfer airlock (a small vessel with a fast opening door, not shown on Fig. \ref{fig:Sample_Environment}, or also called a fast entry load lock from Ferrovac GMBH) at atmospheric pressure, which is then pumped under vacuum with a vacuum turbo pumping station. Then, the transfer of the sample into the chamber is carried out with the wobble stick (PowerProbe series from Kurt J. Lesker). The rotative stage is turned in the $\{y,z\}$ plane, according to $\Pi_{\textrm{Rot}}$, to orient the nonmagnetic support of the sample towards the wobble stick axis, containing the sample, placed beforehand on the sample holder plate, and clamped on the nonmagnetic support. Afterwards, the rotative stage retrieved its initial position, as shown in the zoom of Fig. \ref{fig:Sample_Environment}, where the sample is facing the laser pump and XUV probe beams. The permanent magnets (Neodymium tablets of tens of mT) are placed on either side of the sample, and north-south poles are inverted by activating the rack and pinion from the outside of the chamber. The position of the sample, regarding the focus position of XUV beam from the toroidal mirror, is accurately set with the help of XYZ linear stages, and by making use of a YAG:Ce XUV-scintillator from the supplier Crytur. 
\begin{figure}[ht!]
\centering\includegraphics[width=14.0cm]{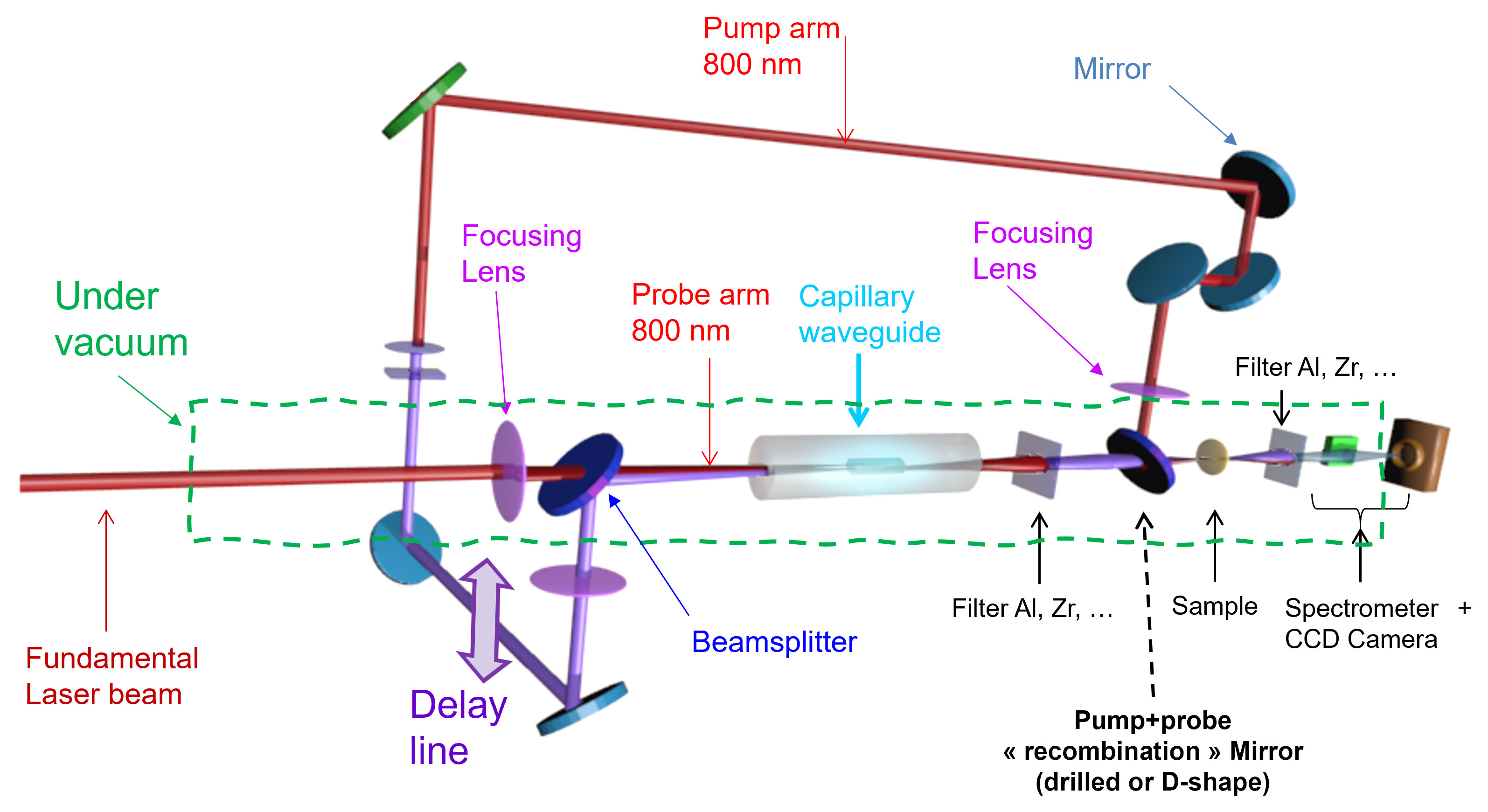}
\caption{The whole optical setup, isolated from the rest of the beamline: optical components and optical paths (not to scale). The fundamental is split into two beams, through a beamsplitter. The NIR pump arms and delay line are at ambient air. }
\label{fig:Optical_Setup}
\end{figure}
In Fig. \ref{fig:Optical_Setup}, the complete time-resolved pump-probe optical setup is shown. For now, the XUV probe arm, viz. the part under vacuum, is built. In such developments, it is the toughest segment to design. 
\subsubsection{Detection device}
To analyze for HH radiation, the grazing incidence flat-field X-ray spectrometer is a PGM$200$ from Horiba-Jobin Yvon, and possesses one toroidal mirror and two foldable aberration-corrected rotating plane gratings ($1800$ and $450$ grooves/mm holographic gratings with variable groove (line) spacing (VLS)). Then, a set of solar-blind dual scintillator (CsI-coated) microchannel plates (MCP), in chevron stack configuration, coupled to a fiberoptic P$46$ (Y$_3$Al$_5$O$_{12}$:Ce) phosphor screen from Photonis is mounted on the CF$100$ flange of the spectrometer output. MCP and phosphor screen are biased with high-voltages $-2.05$ kV and $+4.2$ kV, respectively. At last, a $1024$x$256$ pixels $16$ bits thermo-electrically cooled-CCD camera from Princeton Instruments (PIXIS256) coupled to a device of two top-to-tail Nikon $50$ mm$ f/1.4$ objectives are used for remotely monitoring and recording the horizontally spatially-resolved harmonic spectrum arising from the P$46$ screen, using IgorPro (WaveMetrics, Inc.) IDE environment. Black lens hoods are self-tailor-made with a 3D printer to tightly wrap the remaining free-space between the MCP screen-objective lens $1$ and between the CCD head-objective lens $2$ to prevent from ambient parasite light. Both objectives are tuned to the largest opening of the diaphragm and focused on $\infty$. Steady-state pressure vacuum into the spectrometer is ensured with a $350$ L/s turbomolecular pump from Leybold and its exhaust is connected to a $15$ m$^3$/h roughing Scroll pump from Edwards. The MCP detector being a costly and very sensitive component, one has to take care and watch when the inlet gas pressure is applied while high voltage is applied to it. We remarkably observed that, under normal operating conditions (even up to $3$ bars of He in the capillary), the vacuum pressure inside the spectrometer chamber is on the order of $10^{-6}$ mbar. This is allowed thanks to the cascade of differential pumping devices and also to the distance from the HCW to the spectrometer slit, being $\sim 2.5$ m, because many different conductances and volumes are crossed. The four legs of the spectrometer rest on PTFE (Teflon) pads (supporting the compression induced by the spectrometer weight) so as to move it effortlessly from the $0^{\circ}$ position to the $90^{\circ}$ one, with respect to the laser beam axis, as shown in Fig. \ref{Kerr_Configuration}. 
\\
We can say a few words on the experimental protocol. In T-MOKE experiments, the reflectivity for \textit{p}-polarized light depends on the (nonmagnetic) Fresnel coefficients and on the magnetic part via the dielectric magneto-optical constant $\epsilon_{xy}$ (the magnetization-dependent off-diagonal element of the dielectric tensor), which is linearly proportional to the magnetization component perpendicular to the plane of incidence\footnote{For a conventional pump-probe experiment at ambient, in visible wavelengths, a Wollaston, a monochannel detection (photodiode) and a lock-in-amplifier (LIA) is added, connected with a chopper, that modulates one optical arm. The LIA extracts the frequency component correlated to the chopped optical arm. The remaining components are rejected, as being some noise.}. XUV spectra are acquired for each reversal of the magnetic field \textbf{+H} et \textbf{-H} (the externally applied magnetic field), as shown in Fig. \ref{fig:Sample_Environment}.
Then, to extract the magnetic signal from the reflected intensity of the \textit{p}-polarized light, one calculates the asymmetry parameter (usually called A), which is the difference in reflected intensity from the sample for both magnetization reversals, normalized by the sum of the intensities \cite{PhysRevX.2.011005}.
\subsection{Synthesis on the complete approach and alignment procedure}
The complete optical alignment is carried out at atmospheric pressure (opened chambers) at the spectrometer zeroth-order at the end of the beamline. We used a simple back-propagation method, by installing a mirror in site and place of the toroidal mirror, as well as diaphragm (3D printed) between chambers to allow the laser path at the center of each chamber, without the presence of the capillary waveguide, at first.
\begin{figure}[ht!]
\centering\includegraphics[width=14.9cm]{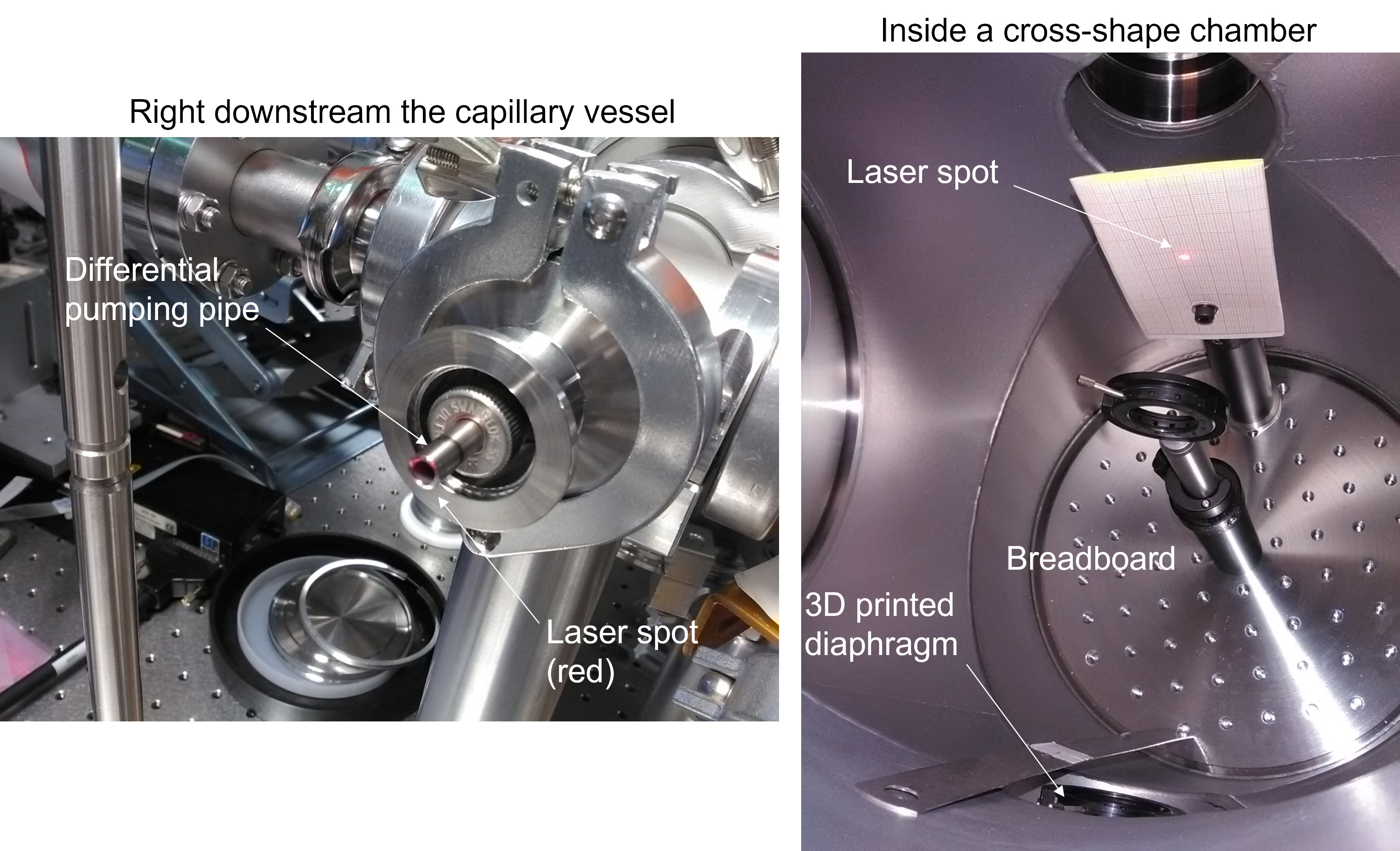}
\caption{Photographs showing some sites where occurs the femtosecond laser alignment. Left: the tube for differential pumping coming right after the capillary vessel. Its inner diameter is about $4$ mm, and XUV beam propagates through it. Right: an example of alignment in the center of a vessel chamber, using irises (or said diaphragm). Laser beam power is very weak, but still visible.  }
\label{fig:Alignment_Photos}
\end{figure}
Then, in order to accurately adjust the toroidal mirror, a BeamGage CCD camera (known as laser beam profiler, from Ophir Optronics) together with a set of optical densities (OD) in front of it, has been installed at the center of the interaction chamber, see Fig. \ref{fig:fig4}, which will be the position of the sample under study, and corresponding to the focal point of the toroidal mirror. Finally, to mimic the position of the HCW, the homemade component seen in Fig. \ref{fig:fig4_AlignmentCapi} is exploited. If we could make a self-criticism, in order to ameliorate our methodology, then the most elegant and efficient way for aligning, in back-propagating the (very weak) laser beam, would be to mount the toroidal mirror (and its support with the translation stages and rotating picomotors) on a goniometer base, in the azimuth plane, so as to make combine both forward and backward propagation. That could be considered in the future.  
\subsection{Risk assessment}
To complete this section, let us deal with the potential events which may unfavorably impact on individuals (the operators), and the machine environment. We identified four main risks, itemized as follows: 
\begin{itemize}
    \item Bad handling of a femtosecond laser could be hazardous and detrimental to the struck eye, if a lack of adapted protective goggles is observed,
    \item Costly optical components may be easily harmed if bad handling of the laser power, 
    \item Physical harm by falling of heavy pieces such as stainless steel flanges,
    \item Manipulation of high-pressures of noble gasses, if lack of awareness regarding the hazards warning with pictogram labels. 
\end{itemize}
The evaluation of risks allows us to decide on adapted precautions.
\section{Main results and discussion}
\label{sec:Results}
First, we consider that the duty cycle ($\alpha_{\textrm{dc}} = t_{\textrm{pulse}}/T_{\textrm{period}}\Leftrightarrow \sim 10^{-15}$ sec/$1$ kHz) of such laser is so tiny that the system under study, after being excited by a laser pulse, recovers back its equilibrium between two consecutive laser pulses. Second, to concentrate ourselves on the essential, in proving the functionality of our design, we arbitrarily fixed several experimental parameters: the laser energy and the laser pulse duration. In addition, for this first proof of the functioning of our development, the results that will be considered here are restricted to a capillary with a $300$ µm diameter. There are two reasons for that choice. First, we empirically observed that, in our case, the HH signal is of better quality and stronger. Second, a larger capillary diameter has shown that clipping and intensity clamping \cite{PhysRevLett.112.223902} are reduced, thus overheating is lower, preventing the aperture edges from being damage, in this very high thermal load environment (see the supplemental material). This is mainly because we have decided to fix the focal length $f_{\textrm{lens}}$ of the lens, in the tightly focusing configuration. The numerical aperture $f_{\textrm{lens}}/D_{\textrm{beam}}$, with $D_{\textrm{beam}}$ being the laser beam diameter at FWHM, is elevated to such an extent that the cone of light, leaving the lens, licks the edges of the capillary entrance, and ablates them, creating unwanted dust and glass debris \cite{YiLiu_2013}. In the supplemental material, a simulation illustrates this phenomenon in showing the stress at stake from the laser radiation pressure. With larger focal length (in the loosely focusing configuration, for instance), we can expect having a better HH signal with $150$ or $200$ µm capillary diameter, that would also have the advantages of an easier laser alignment than currently, a better output divergence of the XUV, and an improvement of the vacuum level in the following vessels, because of a lower volumetric gas flow rate. Therefore, capillary length and diameter have to be adapted according to the targeted application, and depending on the input experimental parameters, such as the lens focal length.%
\subsection{Some observations}
As mentioned earlier, the laser intensity at the focus, into the waveguide, is estimated to be $\sim 3.3$ x $10^{14}$ W/cm$^2$. With such an intensity, $n_\textrm{cr}$ is exceeded for Argon, while it is not for Helium \cite{Popmintchev10516}. Moreover, the saturation intensity where the atomic Coulomb Barrier Suppression regime occurs is slightly overtaken for Argon \cite{Gibson2004}. Nonetheless, working at such an intensity level for Argon, \textit{i.e.} above saturation, is not singular, since it has already been demonstrated in HCW by \cite{Cassou:14}, \cite{Tro__2022} or \cite{Gibson2004}, for instance. Even more, authors in \cite{Johnson2019} dubbed this high-density plasma regime \textit{overdriven} on which, strikingly, almost all demonstrations of HHG in SXR have been carried out so far. In that regime, the driving laser field is strongly reshaped by the gas plasma, leading to a decrease in the field intensity, and a shift of the envelope peak to the leading edge \cite{ZongyuanFu2022}. Actually, for harmonics generated at the ionization level $>\eta_{cr}$, the phase mismatch $\Delta k$ due to uncompensated plasma dispersion (giving rise to nonlinearities, such as plasma-induced defocusing (lensing) \cite{doi:10.1126/sciadv.aar3761}, induced by optical Kerr effect, and responsible for ionization loss, and is also known to drastically drop the laser intensity along the direction of propagation) reduces the coherent build-up length L$_{\textrm{coh}}$, thus decreasing the harmonics yield \cite{popmintchev2010}. Regarding our experimental conditions, it has been reported in section \ref{sec:Theoretical considerations} that experimental ionization fraction in argon (about half-ionized on the leading edge of the pulse, according to ADK calculation\footnote{Let us notice that the correction factor for barrier suppression ionization BSI, where ADK overestimates the exact ionization rate, is not used here.} derived in Section \ref{subsec:Theoretical considerations_SFA}.) exceeds the critical ionization limit $\eta_{cr}$. In that case, the HH generation can in principle not be fully phase matched, and consequently a phase-matched fraction of the pulse will be located in a time window before the peak. Indeed, when the fraction of ionized medium is higher than the critical ionization, it results in a transient phase matching regime, where only a small fraction of the rising edge of the laser pulse contributes to the, finally, usable harmonic signal \cite{Hadrich2014}, and the high gas density shifts the maximum HH emission at later times \cite{Heyl_2016}. Also, for Helium, the experimental weakly ionization fraction (below $0.1\%$ with ADK) is lower than the predicted one and thus is not optimal for phase-matching as well. Therefore, there is a further potential for scaling-up to higher photon flux and it will be considered in future experiments. As mentioned by \cite{Hadrich2014}, it is experimentally not trivial to find the good phase-matching conditions, because the corresponding terms (gas density gradient, focus size, focus position, pulse duration) in Eq. (\ref{eqn:eqn1}) change with time (it is meant that it evolves at femto-to-nanosecond scale), and we must rather deal with a $\Delta k(t)$. As a result, phase-matched generation is achieved transiently in time and within a part of the interaction volume. Therefore, phase-matched HH generation seldom obeys a locked and unique set of ingredients, but rather is versatile, sensitive, and can be achieved for a large range of generation parameters, as seen from Eq. (\ref{eqn:eqn1}) and as stated in \cite{Weissbilder:22}, \cite{Heyl_2016} for instance, which have to be adapted depending on the application requirements. See Supplement for supporting content.
\\
\subsection{XUV spectra}
\begin{figure}[ht!]
\centering\includegraphics[width=13.7cm]{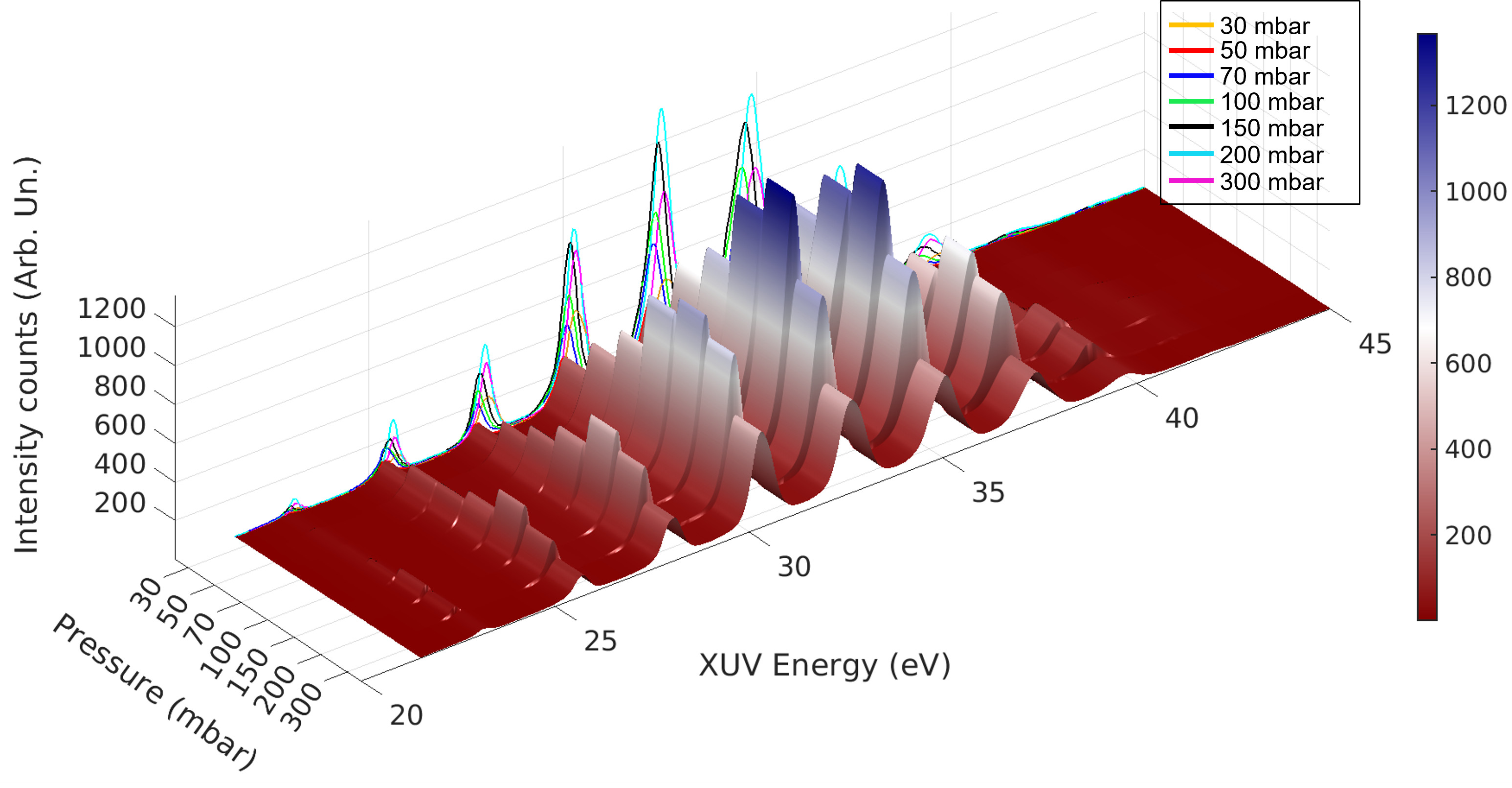}
\caption{Map of stacked HHG spectra for different pressures, in Argon, $2a=300$ µm, length = $31$ mm, one slit at the middle. Grating $450$ gr/mm. Each of the 3D pressure-dependent curves is projected on the background wall. Amplitudes are the raw signals, without correction.}
\label{fig:figmapAr}
\end{figure}

\begin{figure}[ht!]
\centering\includegraphics[width=13.7cm]{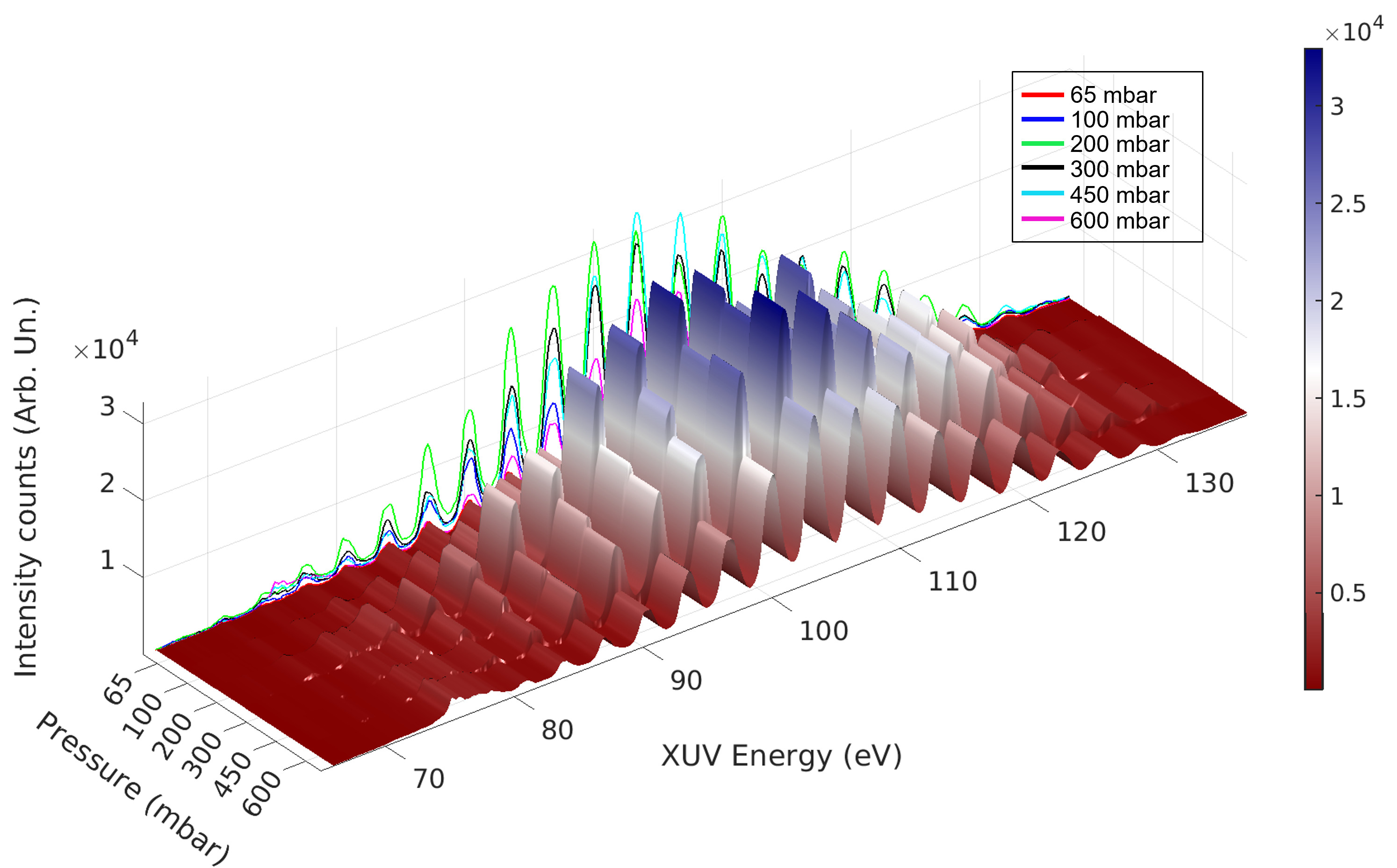}
\caption{Map of stacked HHG spectra for different pressures, in Helium, $2a=300$ µm, length = $13$ mm. Grating is $1800$ gr/mm. Spectrometer slit is opened at $\sim 500$ µm. Same as Fig. \ref{fig:figmapAr}. Amplitudes are the raw signals, without correction.}
\label{fig:figmapHe}
\end{figure}

\begin{figure}[ht!]
\centering\includegraphics[width=12.4cm]{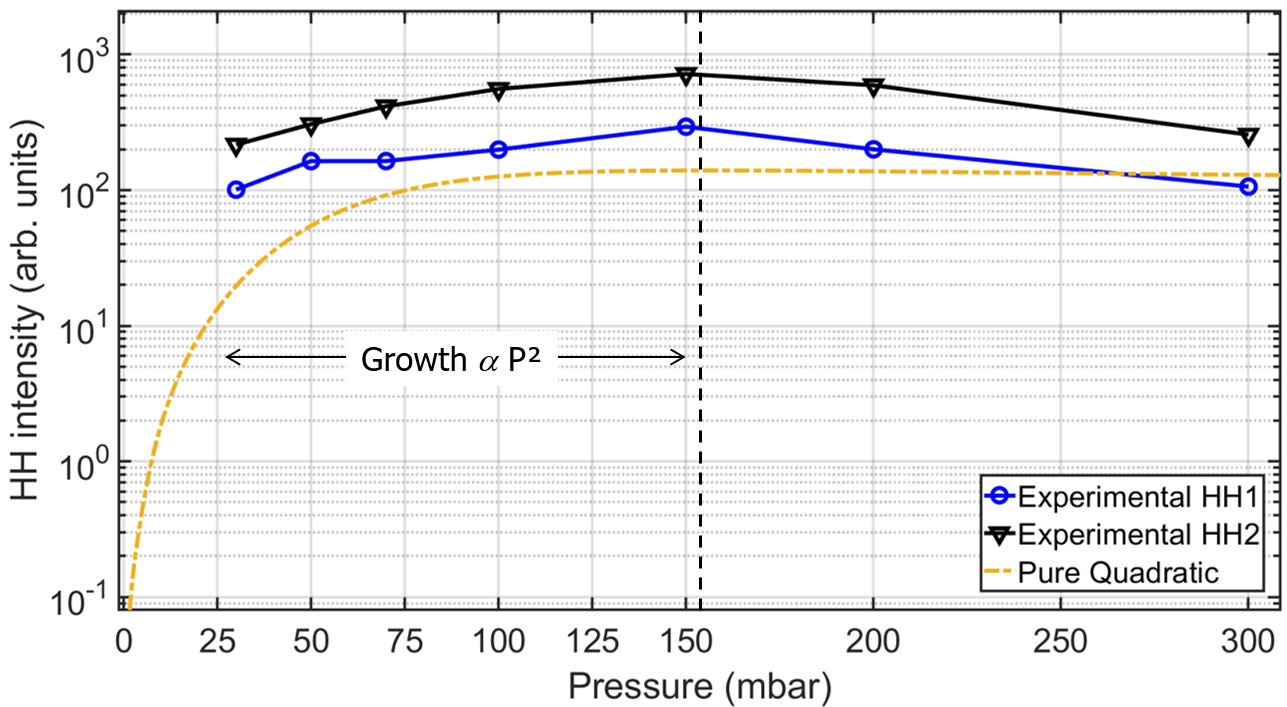}
\caption{Intensity evolution of two harmonic as a function of the Ar pressure $P$ as compared with a quadratic growth (guide for the eye) to the first experimental points, using Eq. (\ref{eqn:FluxPhoVsLcoh}). It shows a maximum at $150$ mbar, in agreement with Section \ref{sec:Theoretical considerations}.  HH are taken at the center of gravity from Fig. \ref{fig:figmapAr}, so based on the calibration, we can say at $\sim 30$ eV. The quadratic curve is the  trends that would be expected from our experimental conditions, in a phase matching regime. Here, we use $\Delta k$ derived from Fig. \ref{fig:fig1} which gives $L_{\textrm{coh}}$ (=$\pi/\Delta k$), and $L_{\textrm{abs}}$ varies with the pressure, according to Eq. (\ref{eqn:eqn2}). Note that amplitude of the quadratic curve is voluntary underestimated to facilitate the reading. }
\label{fig:Evo_P_Argon}
\end{figure}

\begin{figure}[ht!]
\centering\includegraphics[width=12.4cm]{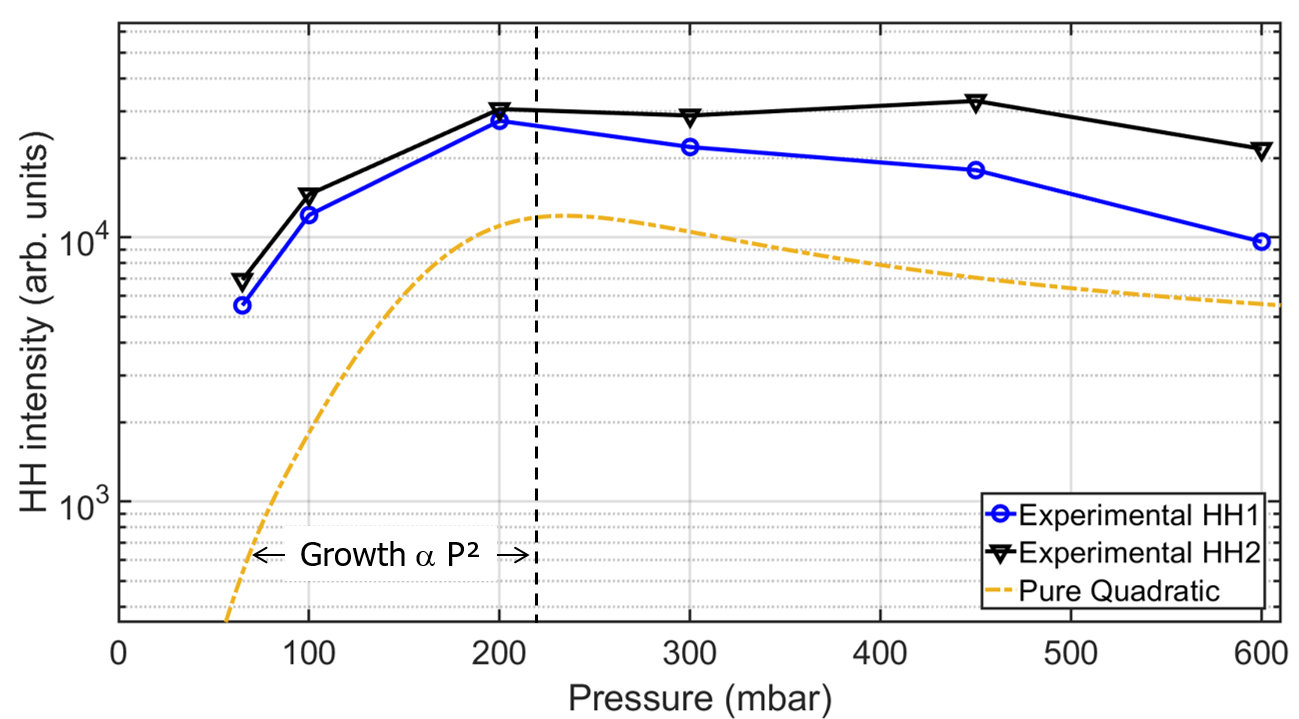}
\caption{Same as Fig. \ref{fig:Evo_P_Argon}, but for Helium. Maximum is in the range $200$ to $400$ mbar, in agreement with Section \ref{sec:Theoretical considerations}. HH are taken at the center of gravity from Fig. \ref{fig:figmapHe}, so based on the calibration, we can say at $\sim 35$ eV. The quadratic curve is the  trends that would be expected from our experimental conditions, in a phase matching regime. Here, we use $\Delta k$ derived from Fig. \ref{fig:fig2} which gives $L_{\textrm{coh}}$ (=$\pi/\Delta k$), and $L_{\textrm{abs}}$ varies with the pressure, according to Eq. (\ref{eqn:eqn2}). Note that amplitude of the quadratic curve is voluntary underestimated to facilitate the reading. }
\label{fig:Evo_P_Helium}
\end{figure}

In this section we show experimental results emanating from the overall beamline presented in Fig. \ref{fig:fig4}, that being, what is seen from the XUV point of view: $1$- the distance XUV virtual source-CCD detector is about $3$ m, $2$- the optical elements are the metallic foil filter, the toroidal mirror, the spectrometer (entrance slit, toroidal mirror, grating). 
The reader will find a study of several measurements in the next section.
\\First, an example of typical stacked spectra of the generated odd harmonics in the far-field are shown in Fig. \ref{fig:figmapAr} and Fig. \ref{fig:figmapHe}, for Argon and Helium, respectively. These maps correspond to the cumulative and vertical binning from the full-frame raw images on the CCD, without amplitude correction. The abscissa axis is expressed in eV, and the calibration in terms of CCD pixels to energy will be explained later. Each spectrum has been recorded over $40$ acquisitions and with an integration time of $500$ ms per acquisition, and is wall-projected. Thus far, harmonic spectra are recorded without any metallic filter to discard the fundamental laser beam, because in our configuration, MCP detector acts as a filter for removing the NIR laser \cite{Zhang2014}. This "sea" of harmonic peaks shows that HH generation can take place between a wide range of gas pressure, but obviously with distinct conversion efficiencies ($\lambda^0_{NIR} \rightarrow$ $\lambda^i_{XUV}$).
Fig. \ref{fig:Evo_P_Argon} and Fig. \ref{fig:Evo_P_Helium} show the evolution, as a function of backing pressure, of two harmonics for Argon and Helium, respectively, and extracted from Fig. \ref{fig:figmapAr} and Fig. \ref{fig:figmapHe}. Qualitatively, one can see that the HH intensity grows according to a quadratic law, based on Eq. (\ref{eqn:FluxPhoVsLcoh}) (to some simplifying assumption, where $A_q$ is taken here as a constant), with the pressure as predicted by \cite{Kazamias2011}. In accordance with \cite{Heyl2012}, this quadratic behavior resembles a fully phase-matched case with no major re-absorption. Eq. (\ref{eqn:FluxPhoVsLcoh}) indicates that, when approaching phase matching, \textit{i.e.} $\Delta k \rightarrow 0$, then $L_{\textrm{coh}}$ diverges to $\infty$ and prevails over the other characteristic lengths, $L_{\textrm{med}}$ and $L_{\textrm{abs}}$, and $\Psi_{\textrm{HHG}}$ tends to its maximum, as a result.
Let us remind that, as aforementioned for Argon, phase-matching occurs within a short time interval in the rising edge of the driving laser pulse, namely, before the peak \cite{Rothhardt_2014}, \textit{i.e.}, before the critical ionization is reached. For Argon, optimal backing pressure at $150$ mbar is comparable with experiments using similar parameters in \cite{Butcher2013} or \cite{PhysRevLett.82.1668}. Lastly, for both gases, after their optimal pressure is attained ($\sim 150$ and $\sim 300$ mbar, for Argon and Helium, respectively), the phase mismatch $\Delta k$ begins to increase which leads to the decline of the harmonics intensity, as it is visible in Fig. \ref{fig:Evo_P_Argon} and Fig. \ref{fig:Evo_P_Helium}, on the right side of the vertical dashed line.
%
%
%
The newly generated frequencies are integer multiples $n_{\textrm{HHG}}$ of the fundamental laser frequency $\hbar\omega_L$, up to some cutoff order $q_{\textrm{max}}$, and are emitted in the forward direction. As the interaction medium for the XUV generation is a monoatomic gas, it is only possible to produce odd harmonics orders because of centrosymmetry, where the lowest-order nonlinear response of the material system is the third-order contribution to the polarization (dipole moment per unit volume, and depending on the electric strength), described by a $\chi^{(3)}$ electrical susceptibility \cite{RBoyd_NLO}. 
\begin{figure}[ht!]
\centering\includegraphics[width=14.9cm]{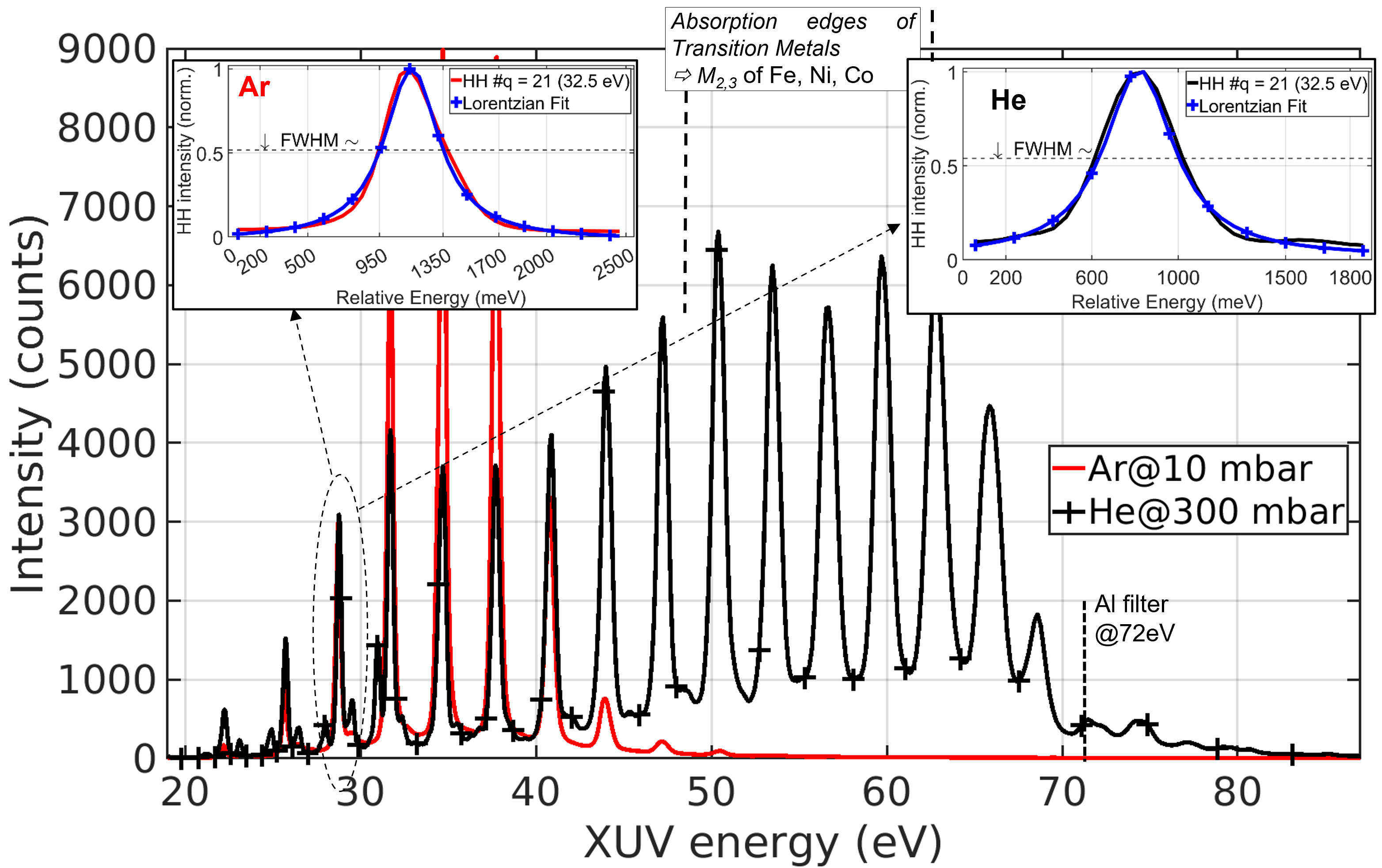}
\caption{Calibrated spectra of the HHG for phase-matched pressures, in Argon and Helium, from the the grating $450$ gr/mm, averaged over $20$ acquisitions with $500$ ms integration time. The Al filter L$_{2,3}$ absorption edge is also shown. $2a=300$ µm, length L$_\textrm{med}= 11$ mm, one slit in the middle of the HCW. Some transition metals of interests are also shown. The two inserts display the Lorentzian fit of one specific harmonic, for both gases.}
\label{fig:figSpecLow}
\end{figure}
\begin{figure}[ht!]
\centering\includegraphics[width=14.9cm]{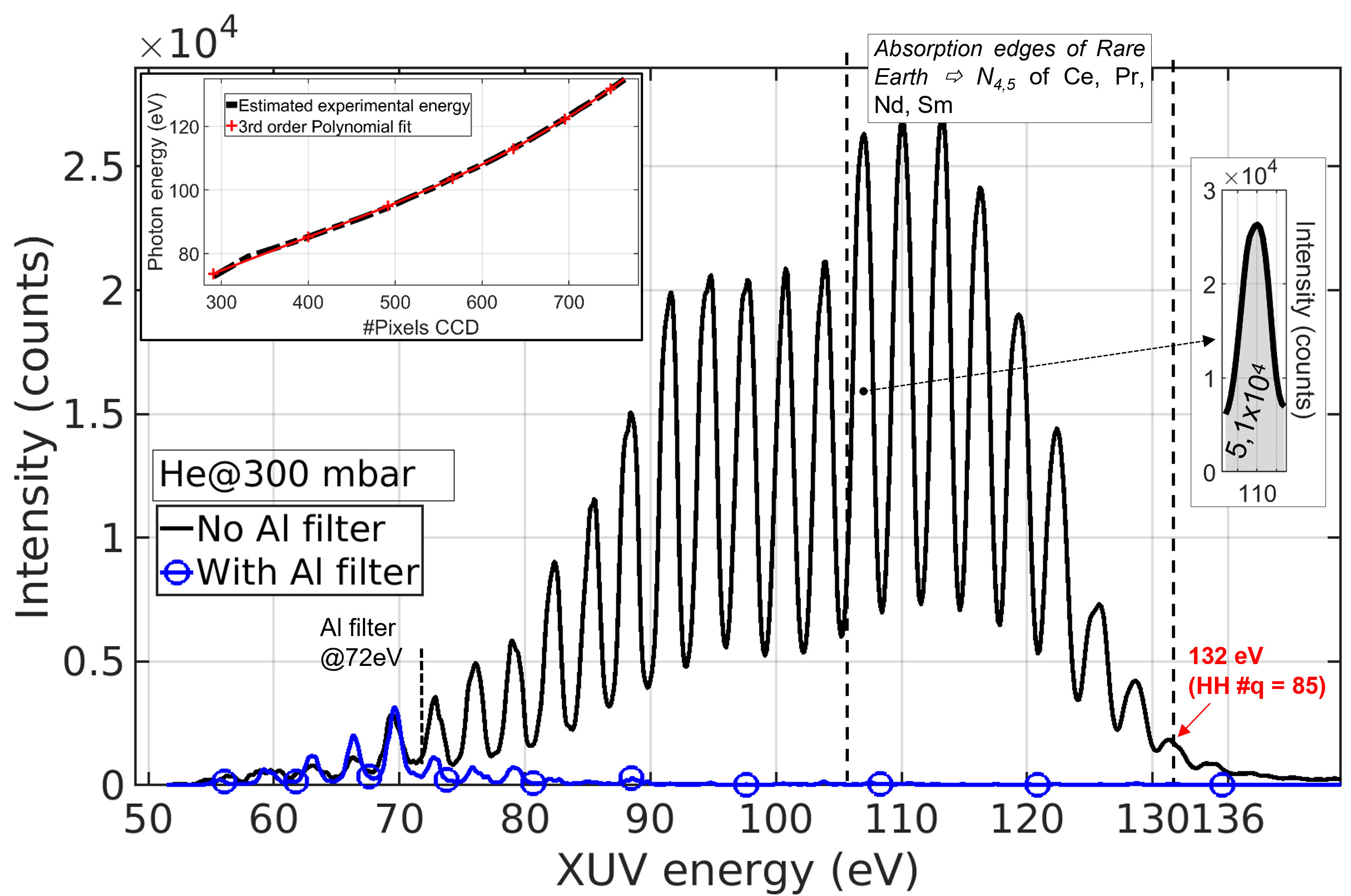}
\caption{Calibrated spectra of the HHG for phase-matched pressures, in Helium, from the grating $1800$ gr/mm, averaged over $20$ acquisitions with $500$ ms integration time. The Al filter L$_{2,3}$ absorption edge is also shown. $2a=300$ µm, length L$_\textrm{med}= 11$ mm, one slit in the middle of the HCW. Spectral region of some rare earth elements of interests is also delimited. Left-insert represents the law of dispersion of our optical detection device, and used for $x$-axis calibration. Right-insert indicate the distributed number of counts for the harmonics $q=71$ (shaded area under curve).}
\label{fig:figSpecHigh}
\end{figure}
For Helium, it is known from section \ref{sec:Theoretical considerations} that phase matching is optimal at a pressure $\sim 400$ mbar (for energies of about $110$ eV), for $2a=300$ µm and L$_{\textrm{med}}=13$ mm. Fig. \ref{fig:Evo_P_Helium} shows a maximum backing pressure between $200$ and $400$ mbar, corresponding to an average pressure of $100$ to $200$ mbar into the capillary waveguide, as predicted by numerical simulations, see the Supplemental Material. 
Let us add, that, at such a pressure, there is a strong gas ejection visible through the shape of a plasma plume at the entrance of the HCW, meaning that part of the laser energy (in the tail part of the pulse) is screened before entering the HCW. Therefore, the focus position was adjusted in order to mitigate the size of the plume and optimize for the best HH intensity. 
\\
As a general rule (which has been applied for Fig. \ref{fig:figSpecLow} and Fig. \ref{fig:figSpecHigh}), photon energies axis have been calibrated from raw CCD pixels using first the L$_{2,3}$ absorption edge at $72$ eV of an Aluminium (Al) foil thin films of $0.2$ µm thick from Lebow Company. Then, a third order polynomial fit is used to describe the dispersion law of the energy-to-CCD-pixel relationship (assuming that there is a period of $2$x$1.55$ eV between each harmonic peak, and that the steep slope on the right side of the spectrum corresponds to $72$ eV) in the form $\hat{E}_{XUV}$ (eV) $ = \Sigma_{i=0}^{j}c_i.p_E^i$, where $\hat{E}_{XUV}$ is the estimated energy, $j \in \{1..3\}$ is the order of the polynomial, $p_E^i$ the pixel index of energy $\hat{E}_{XUV}$, and $c_i$ the polynomial fit coefficients. This handling is done for both gratings. The results are shown in Fig. \ref{fig:figSpecLow} and Fig. \ref{fig:figSpecHigh}.
\\
Note that in Fig. \ref{fig:figSpecLow}, Argon background-substracted spectrum is plotted for a HCW of $L_{\textrm{med}}=11$ mm, instead of $31$ mm because of experimental convenience with the Helium spectrum. Thus, phase-matching here is not optimized in Argon. It is shown in Fig. \ref{fig:figSpecHigh} that cut-off is $\sim 132$ eV, corresponding to the $q = 85$th harmonic of the fundamental $800$ nm ($1.55$ eV). Cut-off at $\sim132$ eV considering a driving at $800$ nm in Helium is in good agreement with prior experimental observations \cite{Popmintchev:08} and consistent with Fig. \ref{fig:Cutoff_E} and other theoretical predictions \cite{Popmintchev10516}, reinforcing our study as well as our experiments. 
For future pump-probe experimental conditions, a set of metallic foil filters will be considered (Indium, Boron, Silver, Titanium and Zirconium).
\\For Argon and Helium, from Fig. \ref{fig:figSpecLow}, a best-fit analysis of the Lorentzian curve \cite{Zurch2015} gives that the high-contrast harmonic peak $q = 21\equiv32.5$ eV is spectrally narrow, approximately $400$ meV at half-maximum (comparable to \cite{Heyl2012}). The relative bandwidth at the harmonic $q = 21\equiv 32.5$ eV is $\Delta E/E\approx1/81$. Note that here, the opening size of the entrance slit of the spectrometer was set to $\sim 500$ µm (the opening range of the slit $\in [10; 2000]$ µm). Thus, there is a margin to taper off the spectral width by closing a bit the entrance slit, but this would be done to the detriment of HH intensity (since, at the moment, there is no toroidal mirror re-focusing the XUV beam in the plane of the spectrometer entrance slit). This trade-off has to be taken into account depending on what is expected from the experiments.
\\
\subsection{Quantitative analysis of HH flux}
Now, in order to roughly (since, usually, a XUV calibrated photodiode is used, which we do not have yet) estimate the absolute XUV photon yield of the generated HH directly after the HCW, from the measured number of counts/s with the CCD, one has to take into account the overall optical elements efficiency. In the direction of propagation, they are: the filter-foil, the focusing toroidal mirror, both the toroidal mirror and the grating of the spectrometer, the MCP, the phosphor screen, the set of objectives and the CCD sensor. The final HHG flux in photon/s for the specific largest distributed harmonic $q=71$ ($\sim 110$ eV, see Fig. \ref{fig:figSpecHigh}) is obtained from the following \cite{Fan14206}, \cite{Hadrich2014}, \cite{Sayrac2017}
\begin{equation}
N_{\textrm{ph}} = \big(S_{\textrm{CCD}}.\sigma_{\textrm{CCD}}\big).\big(A_\textrm{CsI}.\textrm{QE}_{\textrm{MCP}}.(E_{\textrm{ph}}/1.12).g_{\textrm{MCP}}.\textrm{QE}_{\textrm{CCD}}.T_f.r_{\textrm{TM}}^2.\varphi_{\textrm{sr}}\big)^{-1}
\label{eqn:eqnPhotonYield}
\end{equation}
where $T_f$ stands for the transmittance of all optical elements
\begin{equation*}
T_f =t_{\textrm{filter}}.t_{\textrm{spec}}.t_{\textrm{ps}}.t_{\textrm{obj}}.t_{\textrm{exp}}.\phi_{\textrm{slit}}
\label{eqn:eqnPhotonYield_Tf}
\end{equation*}
%
%
%
%
The denominator in Eq. (\ref{eqn:eqnPhotonYield}) is the optical power budget of the beamline hence $S_{\textrm{CCD}}$ is the signal in counts, $\sigma_{\textrm{CCD}}$ is the CCD sensitivity and is $6$e-/counts, $A_\textrm{CsI}$ is the absolute quantum efficiency of the CsI coating combined with the Open Area Ratio of the MCP, the acronym QE corresponds to the quantum efficiency of the component in its subscript, $g_{\textrm{MCP}}$ is the MCP gain, (E$_{\textrm{ph}}/1.12$) is the number of electron-hole pairs created per photon from the P46 at a band gap energy of $1.12$ eV, where $1.12$ eV is the band gap of the silicum, $r_{\textrm{TM}}$ is the polarization dependent reflectivity of the toroidal mirror at the considered energy, $t_{\textrm{ps}}$ is the conversion efficiency of the phosphor screen (Fiber-Optic Plate is taken into account \cite{Izumi_2012} where angular distribution is more directional than that predicted by Lambert's cosine law \cite{KANDARAKIS2006508}) at the peak wavelength $\sim 550$ nm, $t_{\textrm{filter}}$ is the filter foil transmission (as well as accounting for the thin surface layer of oxide that is known to appear upon contact with air, typically a few nm thickness on both sides of the foil, and responsible for reducing transmission by $\sim$x$2$) and is wavelength-dependent, $t_{\textrm{exp}}$ is the exposure time, $t_{\textrm{obj}}$ is the objectives transmission, $t_{\textrm{spec}}$ is the spectrometer transmission. A portion of the XUV photons shower is lost $\phi_{\textrm{slit}}$ because it is blocked at the entrance slit (because, in the current configuration, the XUV beam is focused in the center of the sample chamber, $\sim 1$ m before the slit, and there is not yet another TM to re-focus the beam into the slit). Losses due to the collection of the solid angle of the P46 screen are included in $\varphi_{\textrm{sr}}$. The CCD quantities were all provided by the supplier.
Finally, the transmission from the HCW exit through the vacuum system is considered here unity. The estimated photon flux exiting the HCW at the most intense particular harmonic $q$ (see insert in Fig. \ref{fig:figSpecHigh}) is thus  $N_{\textrm{ph}} \simeq 4.9$ x $10^{7}$ photons.s$^{-1}/0.3$ nm bandwidth, so, $E = 1.6$x$10^{-19}$x$N_{\textrm{ph}} \approx 7.83$ pJ, and is comparable to other recent XUV beamlines \cite{Heyl_2016}, \cite{Aurore_Consortium}, where, for the latter, it has been measured with an avalanche photodiode (APD). A bandwith of $\Delta \lambda = 0.3$ nm at $q=71 \approx 106$ eV $\approx \ 11.7$ nm corresponds to an energy bandwith $\Delta E \approx E_0\Delta \lambda/\lambda_0 \approx 2.7$ eV.  The energy calibration factor of our detection is $\sim 1.42$x $10^{-1}$ femtojoule/count, and is precisely consistent with \cite{Goh2015}, thus strengthening our work. Afterwards, the total energy in the High Harmonics beam is derived by summing over the whole comb-like of harmonics. 
\\Somehow, let us notice that this estimation concerns the photon flux at the \textit{output} of the HCW\footnote{As already mentioned earlier, and as Eq. \ref{eqn:eqnPhotonYield} demonstrates it, it is in fact the measurement at the end of the beamline, and stems from the spectrometer combined with the MCP imaging detector.}. In reality, the XUV source being created \textit{into} the HCW, so situated downstream of the HCW output, we can argue that our $N_{\textrm{ph}}$ might be underestimated due to a partial re-absorption by the residual (ionized or neutral) gas in expansion at the HCW exit, on the way to the detector. It is due to the fact that at the capillary output, the atomic number density is difficult to stringently evaluate, because, amongst others, is wavelength dependent. In other words, it is the available XUV intensity for the user, but not the total generated intensity, down into the heart of the HCW.
\subsection{Some routinely vacuum levels as a function of gas pressures}
\label{Multi_Atmo}
\begin{table}[ht]
\centering
\begin{tabular}{||c c||} 
 \hline
 Argon@ $P_{\textrm{in}}$ (mbar) & Vacuum pressure in the \\
  & IR filtering chamber (mbar)\\ [0.5ex] 
 \hline\hline
  $5$ & $8.10^{-5}$  \\
 $10$ & $9.10^{-5}$ \\  
   $15$ & $1.10^{-4}$  \\
    $20$ & $1.2.10^{-4}$ \\
   $30$ & $1.6.10^{-4}$  \\
   $50$ & $2.5.10^{-4}$  \\
   $70$ & $4.5.10^{-4}$  \\
   $100$ & $8.3.10^{-4}$  \\
     $150$ & $1.1.10^{-3}$  \\
   $200$ & $1.1.10^{-3}$ \\
  $300$ & $1.2.10^{-3}$\\ [1ex] 
 \hline
\end{tabular}
\caption{Example of measured vacuum pressures in the IR filtering chamber Fig. \ref{fig:fig4} as a function of the Argon backing pressure sent into the HCW. The HCW is $L_{\textrm{med}} = 39$ mm, $2a=400$ µm, slit at $18$ mm from the left extremity (side of the entrance fs laser). }
\label{table:3}
\end{table}
In Table \ref{table:3} the performance of vacuum levels under the gas input of Argon is presented, using a $400$ µm inner diameter (the worst diameter in terms of gas flow). One can see that even when injecting $300$ mbar of Argon in the HCW, an acceptable (in the sense of the design operating state regarding the turbomolecular pump) vacuum pressure of $1.2$x$10^{-3}$ mbar is reached into the IR filtering chamber. The vacuum pressures downstream this first (crucial) chamber are then naturally lower, and the MCP detector can thus be high-voltage biased, securely. Obviously, decreasing the capillary inner diameter, at constant length, compare with the one presented in Table \ref{table:3}, improves the vacuum quality in the following chambers, and the IR filtering chamber reaches down the $10^{-4}$ mbar decade $@P_{\textrm{in}} = 300 $ mbar. At the contrary, diminishing the capillary length, at a constant inner diameter, degrades the vacuum in the IR filtering chamber. For Helium, those values are pretty much the same, although the vacuum level is more favorable with Argon due to the efficiency of evacuating gas with higher atomic scale, linked to the leaking (and outgassing) rate (through gaskets for instance) of Helium that is higher than the Argon counterpart (the device is not absolutely vacuum-tighten), because of the smaller diameter of the helium atoms. Our device can then seamlessly extend to multi-atmosphere gas inlet. 
\subsection{A summary of the HCWs used in this work}
\label{Capillary_SumUp}
Let us insist on the fact that many capillaries have been tested with our setup. Not all of them are discussed. Therefore, Table \ref{table:CapillarySumUp} recapitulates the main attributes that have been observed for each. The inner diameter $300$ µm is the most commonly used in our experiments. 
\begin{table}[ht!]
\centering
\begin{tabular}{||c c c c||}
 \hline
 \rowcolor{orange!20}
 \multirow{2}{2.0cm}{$2a$ (µm)} & \multirow{2}{1.5cm}{L$_\textrm{med}$(mm)}  & \multirow{2}{2.5cm}{Slit(s) number and relative position} & \multirow{2}{4.5cm}{Observations}  \\
 [6.0ex] 
 \hline\hline
  \multirow{5}{*}{$150$} & $ 25$ & two,  & \textcolor{purple!80! magenta!60!}{The hardest $\O$ to align /} \\
  & & $8$ mm left and & \\
  & &  $8$ mm right & \\
  & & & Weaker HH signal \\
  & & & \textcolor{green!80! blue!80!}{Less gas consumption} \\
  \hline
 \multirow{4}{*}{$200$} & $13$ & one, & \\
 & $23$ & one, $5$ mm right & \\
 & $31$ & one, $18$ mm left & No significant difference with $2a =150$ µm\\ 
 & $39$ & one, $18$ mm left & \\
 & $39$ & one, $10$ mm right & \\
 \hline
    $250$ & $10$ & &  \\ 
     \hline
     \multirow{10}{*}{$300$} & $11$ & one, in middle & \textcolor{green!80! blue!80!}{Good compromise HH signal/ease of alignment} \\
     & & & \textcolor{blue!90!}{$11$ and $13$ mm: our best choices } \\
     & $13$ & one, in middle & Easy to align / \\
     & $16$ & one, in middle & \\
     & $23$ & one, $5$ mm right & \\
     & & & \textcolor{green!80! blue!80!}{Nice-looking plasma column into HCW} \\
    & $23$ & one, $5$ mm left & \\
     & $31$ & one, $18$ mm left & Less easy to align than $23$ and lower / \\
        & & & \textcolor{green!80! blue!80!}{Nice-looking plasma column into HCW} \\
     & $35$ & &  \\
    & $39$ & one, $18$ mm left & Even less easy to align than $23$ and lower / \\
     & & & Nice-looking plasma in the HCW \\
     & $45$ & two, centered, and & \\
       & & $\sim 10$ mm center-to- & \\
       & & center distance & \\
    \hline
   \multirow{8}{*}{$400$} & $11$ & in middle & \textcolor{purple!90! magenta!60!} {More gas consumption /}\\
   & $23$ & one, $5$ mm right & \textcolor{green!80! blue!80!}{Easier to align }\\
   & $25$ & two, centered, and & \\
  & & $\sim 10$ mm center-to- & \\
  & & center distance & \\
    & $31$ & one, $18$ mm left & No benefit earned compared with\ \\
    & $39$ & one, $18$ mm left & same lengths for $2a = 300 $ µm \\
    [1ex] 
 \hline
 \hline
\end{tabular}
\caption{Summary of the capillaries (HCWs) tested in this work. Width of each slit is about $200$ µm, which is the thickness of the diamond saw blade. With Argon input. Note that alignment ability is to put in regard with our tight focusing configuration, compared with the (loosely focusing) meter-scale focal length commonly used in hollow-core fiber compression of femtosecond laser pulses (see for instance \cite{Goncalves_2016_ScientificReports}). The slit position said left or right means taken from the laser entrance in the HCW or from to the exit, respectively, in the sense of laser propagation. Colors: purple is for relevant major drawback, green for advantage.}
\label{table:CapillarySumUp}
\end{table}
%
\section{A collection of measured Spectra as a function of input experimental parameters}
We now present a set of interesting recorded spectra, obtained by tuning the input experimental parameters of our XUV platform, to complete this work. As was formerly said, we have tested several lengths of the HCWs, but, also, a few positions of the slit (for gas entrance): middle, right, and left. We chose not to hold forth on these results because we have not distinguished any relevant variation between them, at least in our case. 
\begin{figure}[ht!]
\centering\includegraphics[width=14.9cm]{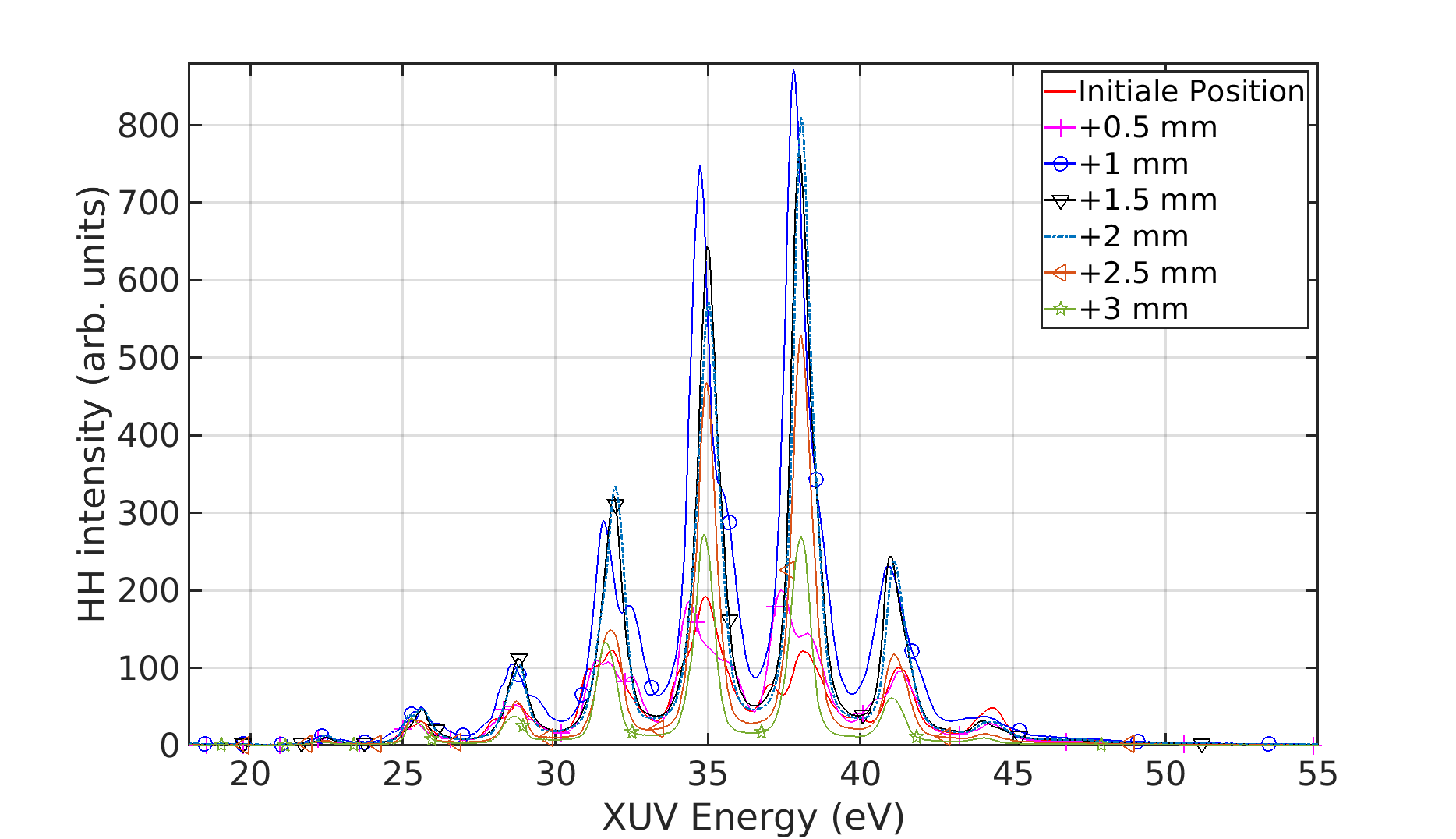}
\caption{A few (uncalibrated in amplitude) spectra in Ar at $130$ mbar as a function of positive lens positions (thus, focus position in the HCW) from an initial arbitrary lens position. Background is substracted. Capillary length $L_\textrm{{med}}= 39$ mm, inner diameter $2a = 200$ µm, slit position is $18$ mm from the left (thus from the laser entrance side), entrance slit of the spectrometer is opened at $1$ mm. }
\label{Fig_LensPosAr}
\end{figure}
In Fig. \ref{Fig_LensPosAr}, one can see the influence of the position of the focus of the NIR laser pulses in the gaseous medium. In \cite{PhysRevA.55.3204},the authors show that, depending on whether focus position is before or after the gas jet, on axis $r=0$, noncollinear or collinear phase matching is achieved, and efficiently or inefficiently building up the harmonic emission. It is tangible here that focus position entails an influence on the amplitude or on the shape of the harmonic. In tuning the lens position, one goes through an optimized position (here, about $+1$ mm), then the HH signal drops-off towards its prior value of the initial lens position. Also, a slight shift in the bluer frequencies appears (as well as a harmonic spectral broadening and shoulder peak -that could evolve in a double peak or a spectral splitting- structure, also visible, for $+0.5$ mm), that we could attribute to either chromatic aberrations exhibited from our focusing singlet lens or to filamentation phenomenon where the the driving laser field experiences spatio-temporal shaping during its nonlinear propagation in the active medium inside the capillary waveguide (see \cite{Holgado2016} and references therein). In other words, although intensity is confined into the hollowed capillary, many nonlinear effects are involved which affect the spatiotemporal profile of the laser pulse, and, consequently, one can suppose that the distribution of the intensity along the capillary length, so along $z$, is varying, thus complicating the nature of the calculation. The latter authors have shown numerically that the spectral structure can be complicated because of self-phase modulation arising due to an important Kerr effect, and the occurrence of a blue-shift coming from the presence of the plasma. The rear part competes with the front part of the pulse, where the latter first causes an expelling of the light off axis (induced-plasma defocusing after the ionization of the gas neutrals), then a spatial replenishment takes place because optical power is still intense enough to induce nonlinear effect that refocused the light conveyed in the rear part of the pulse. This can be equivalently found in moving the position, not of a lens, but of the active medium, as found in the gas jet configuration. In addition, we have observed that a longer capillary could potentially cut the upper part of the HH spectrum, and about $2$ or $3$ HH peaks could be lost, as can be seen in Figs. \ref{Fig_LensPosAr} and \ref{Fig_FourHHG_Spectra}. 
\begin{figure}[ht!]
\centering\includegraphics[width=13.5cm]{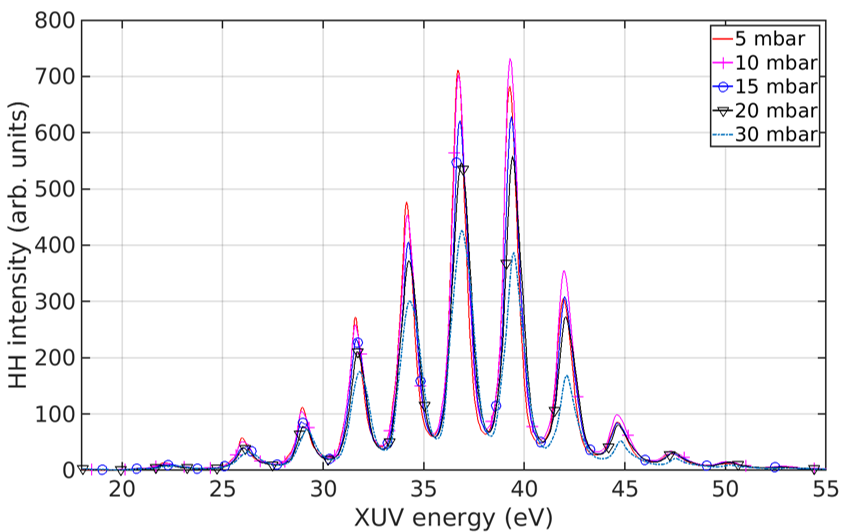}
\caption{A few (uncalibrated in amplitude) spectra in Ar as a function of low pressures. For the rest, same features as Fig. \ref{Fig_LensPosAr}. }
\label{Fig_ArLowPressures}
\end{figure}
Fig. \ref{Fig_ArLowPressures} gives five spectra recorded in the low-pressure range, for Argon. Here, the longer medium of interaction allows for harmonics generation at lower gas pressure, due to the gap from the gas injection spot to the site where it is drained away, in the directly surrounding vessel, under primary vacuum.  
\begin{figure}[ht!]
\centering\includegraphics[width=13.5cm]{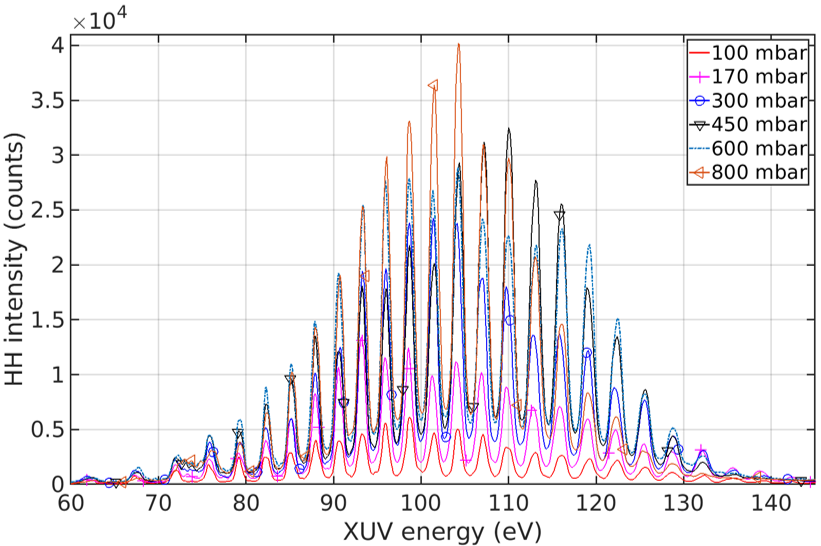}
\caption{Several (uncalibrated in amplitude) spectra in He as a function of higher pressures. Background is substracted. Capillary length $L_\textrm{{med}}= 11$ mm, inner diameter $2a = 300$ µm, slit position in the middle, entrance slit of the spectrometer is opened at $0.5$ mm, $30$ cumulative acquisitions, integration time is $600$ ms/acquisition/spectra. }
\label{Fig_HeHighPressures}
\end{figure}
In He, Fig. \ref{Fig_HeHighPressures}, the intensity of the HH signal is clearly greater (amplitude swells by a factor $\sim $ x $8$ between $100$ mbar and $800$ mbar, for the most intense HH at Pixels CCD $=549$, around $105$ eV) with pressures $>300$ mbar, especially in the first half-part (corresponding to the lower part of the whole XUV frequencies) of the spectrum. In He, again, Fig. \ref{Fig_Influence_LensTM}, shows the importance of an accurate alignment of the focusing optical elements, also, in the HH signal intensity growth. We first started with an initial, and arbitrary, position of both the TM and the focusing lens of the driving laser. Then, only the focus lens is tuned. Once the best position is found, it is the turn of the TM to be adjusted until HH signal is fully optimized. Therefore, it is an interplay between focusing optical elements.
\begin{figure}[ht!]
\centering\includegraphics[width=13.5cm]{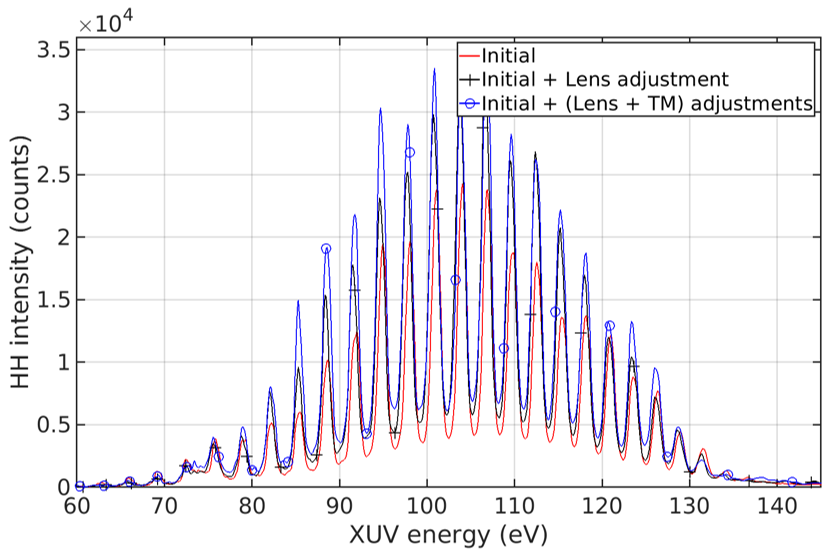}
\caption{A few (uncalibrated in amplitude) spectra in He at $300$ mbar showing the influences on finely tuning the focusing lens and the position of the toroidal mirror. For the rest, same features as Fig. \ref{Fig_HeHighPressures}. }
\label{Fig_Influence_LensTM}
\end{figure}
In Fig. \ref{Fig_FiveHHG_Spectra}, we show one example of HH spectra measured with a capillary for which the gas entrance slit is shifted $5$ mm upstream, in the sense of laser propagation (so, the capillary face at the opposite of the laser entrance). It means that a portion of the inlet gas is rapidly ejected at the nearest capillary end tip, and could deplete the opposite side of the capillary, where laser is focused. But as mentioned earlier, we did not discern any relevant differences between a shifted slit and a centered one. 
\begin{figure}[ht!]
\centering\includegraphics[width=13.5cm]{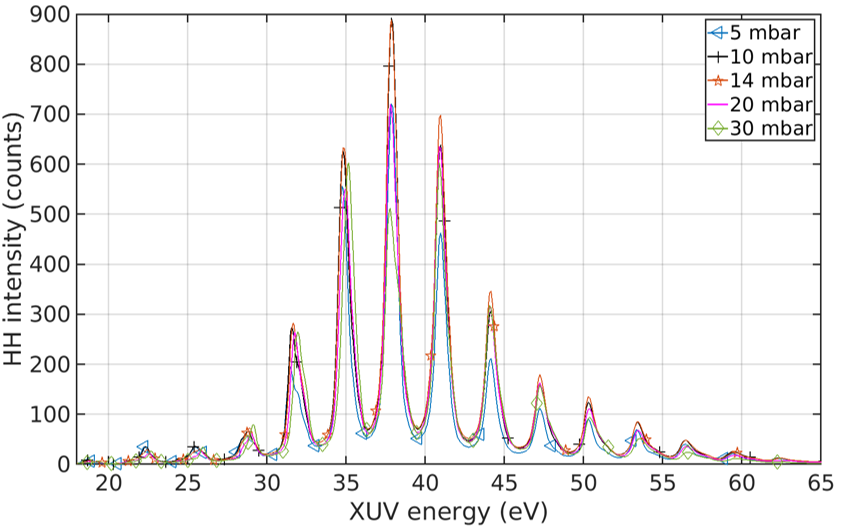}
\caption{Five HHG spectra, for different Ar inlet pressures. Entrance slit of the spectrometer is set $\sim 1000$ µm, capillary length $L_{\textrm{med}} =23$ mm, inner diameter $2a=400$ µm. Here, the slit, for gas inlet into the capillary, is not centered on the capillary but is set at $5$ mm to the right, in the direction of laser propagation.}
\label{Fig_FiveHHG_Spectra}
\end{figure}
In Fig. \ref{Fig_FourHHG_Spectra}, we achieved the shortest capillary length that we are able to test presently with our setup. This length is $L_{\textrm{med}} =10$ mm, and, here the inner diameter is $2a=250$ µm. Note that this kind of diameter is a bit more difficult to optically align than the $300$-$400$ µm versions, but it allows for a better management of the vacuum level, because, of course, the volumetric flow rate is smaller in its case, than larger diameters, considering the same length of capillary. In contrast, using capillary diameters of $150$ or $200$ µm is even more efficient in terms of vacuum, but to the detriment of the lower ease of optical alignment.
\begin{figure}[ht!]
\centering\includegraphics[width=13.5cm]{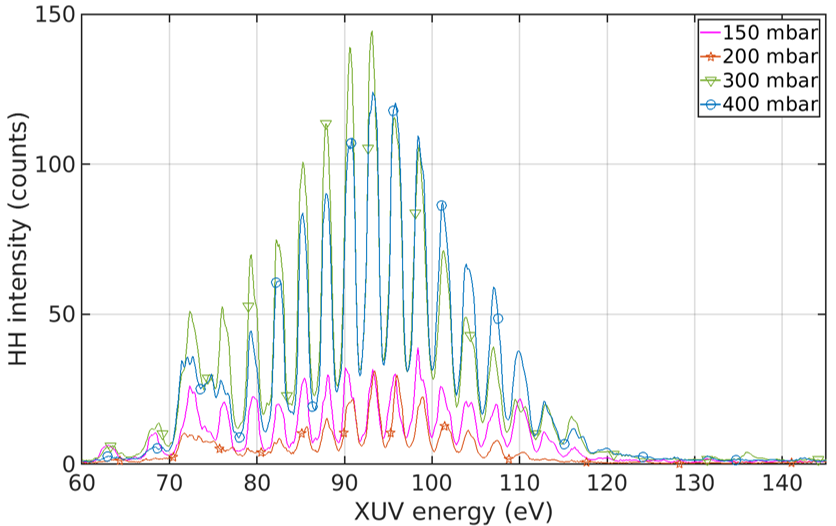}
\caption{Four raw HHG spectra, around expected phase-matching pressures of He. Capillary length $L_{\textrm{med}}=10$ mm, inner diameter $2a=250$ µm. Spectrometer resolution (entrance slit) is unoptimized. Compared with Fig. \ref{Fig_HeHighPressures}, one can see that the highest HH peaks have disappeared, showing the difficult task to perform a good alignment of the laser in the capillary, coupled to a stronger gas concentration in the inner duct (at constant gas flow speed). }
\label{Fig_FourHHG_Spectra}
\end{figure}
We have conducted experiments with a capillary length $L_{\textrm{med}}=23$ mm, an inner diameter $2a =300$ µm, an Ag filter, and have recorded raw spectra at the spectrometer zeroth order, with Helium as the interacting medium, and from pressures right above the atmospheric pressure, $1.6$, $2.5$ and $2.8$ bars, respectively. We have observed that there is clearly a strong enhancement for backing pressure of He $>2.5$ bars, by a factor of $5$. Of value, it has also been observed that, even with a backing pressure of $2.8$ bars, our differential pumping (described in section \ref{sec:Exp_Setup}) remains efficient, and $10^{-6}$ mbar are maintained at the end of the beamline (the spectrometer), position of the (fragile) MCP detector. 
The ensemble of spectra in Fig. \ref{Fig_Jitter_Helium} shows shot-to-shot (shot of $1$ s) fluctuations in a $30$ s time-frame. The inset is constructed according to the ISO 21748:2017(en) standard, titled \textit{Guidance for the use of repeatability, reproducibility, and trueness estimates in measurement uncertainty evaluation}, where the Uncertainty of Type A, related to the repeatability of a measure, is given by $s_x/\sqrt{N_m}$ where $s_x$ is the standard deviation of the set of measurement results and $N_m$ is the number of measurements. Note that the spectra are cut in their upper part, which is a testament to poorer laser alignment. However, the peaks-to-baseline jitter can be explained, in all likelihood, in terms of gas perturbations arising at the output of the capillary, as suggested in the supplementary material, relying on the Comsol Multiphysics (a software based on Finite Element Method) simulations, Section \ref{Gas_Fluidic_Comsol}, and as it is the only one source of turbulence regarding the rest of the beamline. It will be a target of improvement. Another source of fluctuations has already been reported in \cite{Goh2015} where the authors attribute them to ionization-induced nonlinear mode mixing during propagation of the drive laser pulse inside the HCW. However, the true output stability of HHG has to be observed over hours, which is a point that will be considered in future work. Another feature to determine is the degree of spatial coherence of our XUV beam, which could be accomplished with the famous double-slit Young's experiment \cite{Popmintchev:11}. 
\begin{figure}[ht!]
\centering\includegraphics[width=14.4cm]{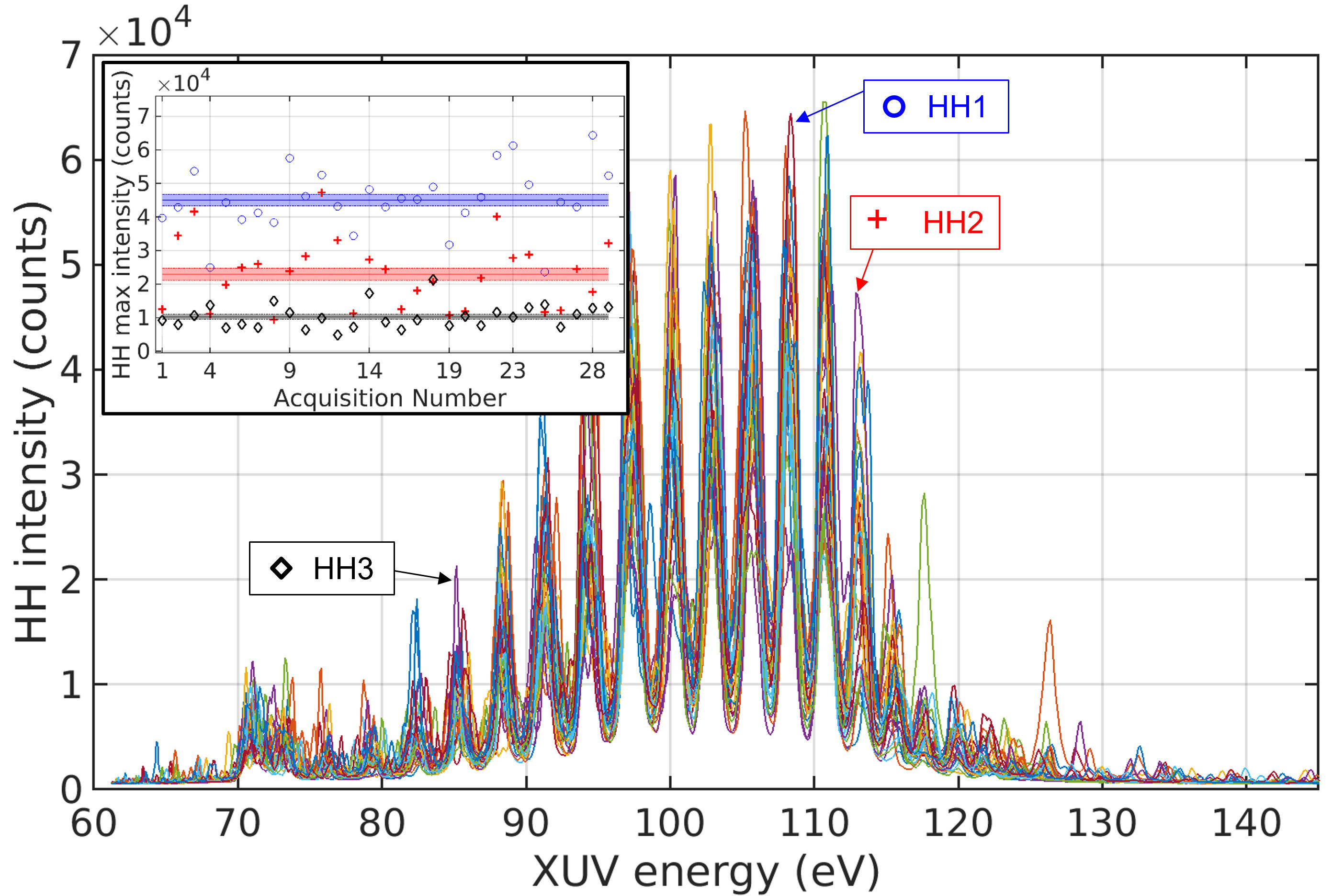}
\caption{A landscape of subsequent spectra recorded every $1$ sec. Entrance slit is $\sim 500$ µm, capillary length $L_{\textrm{med}}=13$ mm and inner diameter $=300$ µm, pressure about $300$ mbar of He. Grating is $1800$ gr/mm. Acquisition time $= 1$ sec over a total time of $30$ seconds, $N_m = 30$ spectra (a $100\%$ duty cycle). Insert shows the maximum amplitude of $3$ harmonics. Straight solid line is the statistical mean, and shaded area corresponds to the HH intensity standard Type A uncertainty, for each of the three chosen harmonics peak.}
\label{Fig_Jitter_Helium}
\end{figure} 
\\
To conclude this section, we have targeted a few levers of improvement regarding HH intensity as well as HH stability, especially in Helium, which, with a relatively little amount of effort, can be itemized as follows: the laser intensity control, a finer handling of the lens focus and its position relative to the center of the capillary, different focal lenses (for tightly and loosely focusing) and multi-slits waveguide. Regarding the latter, a modification must be brought to the capillary box, to insulate the gas inlet part from the gas exhaust part, itself separated in two sections \cite{Goh2015}: one for differential pumping with several slits evenly distributed at the end of the capillary, the other to drain the remaining gas into the capillary itself towards the surrounding vessel under vacuum. Also, the lens could be replaced by a concave mirror, to avoid dispersion or else aberrations produced into the singlet lens bulk, since here working in transmission. 
\section{Conclusion and mid-term outlook}
In summary, in this technical report, we have described the complete development of our table-top HHG beamline. We then have reported on the generation of XUV pulses in a compact, simple and relatively affordable setup from intense $800$ nm femtosecond laser pulses. In order to do so, we have developed a fully-fledged table-top apparatus and succeeded in generating high-harmonics, in Argon and Helium, for which not only the center-of-gravity of the frequency components is situated within the upper part of the XUV region, but also in the pedestal of the soft X-ray range, since it has been shown that spectra build up to $\sim132$ eV, in Helium. It is quantitatively and qualitatively in compliance with the presented basic theoretical predictions or corresponding peer works reported elsewhere. 
A set of capillary waveguides of different lengths and inner diameters has been successfully and easily tested, thanks to our modular design. In addition, two-slit capillaries have been employed, with no significant improvement observed compared to a single-slit. Experiments have been conducted by sweeping the inlet gas pressure, either with Argon or Helium, in order to find the optimal pressure value regarding the HH phase matching. 
Thus it is shown that the phase matching conditions we predicted in Section \ref{sec:Theoretical considerations} correspond to our experimental observations and that our beamline is able to sustain an inlet backing pressure of at least $3$ bars, compatible with the needs to generate Soft-X rays above $132$ eV, which requires multibar gas pressure. 
\\
An envisioned future development is the extension of the cutoff energy to the Soft-X range, including the N$_{2,3}$- absorption edges (or localized-4\textit{f} electron shells) of rare-earth (RE) elements ($150-300$ eV). In that case, it will be mainly devoted to the study of ultrafast demagnetization of transition metals and rare-earth in elements-specific time-resolved visible pump-XUV probe experiments. This extended range of energies will allow us to study RE magnetization dynamics in a pump-probe T-MOKE configuration \cite{maghraoui2023_PRB}. Soft X-ray generation will be achieved using a High-Energy commercial Optical Parametric Amplifier (HE-OPA) \cite{Colosimo:08} and with a few atmospheres of He in the HCW, for which, as said above, we have reported its performance and suitability at relatively high pressure of $\sim 3$ bars.
\section*{Funding}
The project has been supported by the French Agence Nationale de la Recherche (ANR) under the EquipEx n°ANR-10-EQPX-52.
\section*{Acknowledgments}
The authors gratefully acknowledge Jean-Yves Bigot for helpful discussions on scientific purposes, the French Comsol Support based in Grenoble, the IT support from LPC Caen, and, at IPCMS, Oleksandr Kovalenko, Gauthier Dekyndt, Olivier Crégut, Tom Ferté, Cédric Leuvrey, Bruno Michel and Michèle Albrecht for their invaluable help and their technical support throughout the Project.
\section*{Disclosures}
The authors declare no conflicts of interest.
%

%% file: M_Supplemental_Material.tex
%
\appendix
\section{Introduction}
In this supplemental material accompanying the main document titled "A complete and compact broadband high-harmonics beamline based on a modular hollow waveguide for photon generation centered on the upper region of the extreme ultraviolet spectral range - A thorough analysis", we aim to append theoretical framework that drives the HHG process, as these aspects are essential to comprehend the nature of the difficulties in building such a setup. First, we take another look at the ionization of a mono-atomic neutral gas under strong field excitation. Then, we concisely address the quantum origin of HHG by using the Time-Dependent Schrödinger Equation (TDSE) describing the dynamics of a single atom embedded in an intense laser field. Afterward, basic optics and a simple model of thermal stress that occur in the capillary waveguide are discussed from Comsol numerical computations. Finally, we deal with the microfluidics (at sub-millimetric scale) mechanisms in play inside the capillary waveguide submitted to a gas load in its center, and connected to vacuum at its open ends, and also the fluid flows outside the HCW, a field called CFD (Computational Fluid Dynamics). Comsol simulations are an attempt to understand and report on multiphysical events that occur in our XUV source. 
%
%
\section{Atoms in strong field ionization}
\label{Sec:theory_basics_AtomsHighfield}
This chapter is devoted to accompany the glimpse given in the section \ref{subsec:Theoretical considerations_SFA}.
\subsection{the ADK model}
\begin{figure}[ht!]
\centering\includegraphics[width=14.9cm]{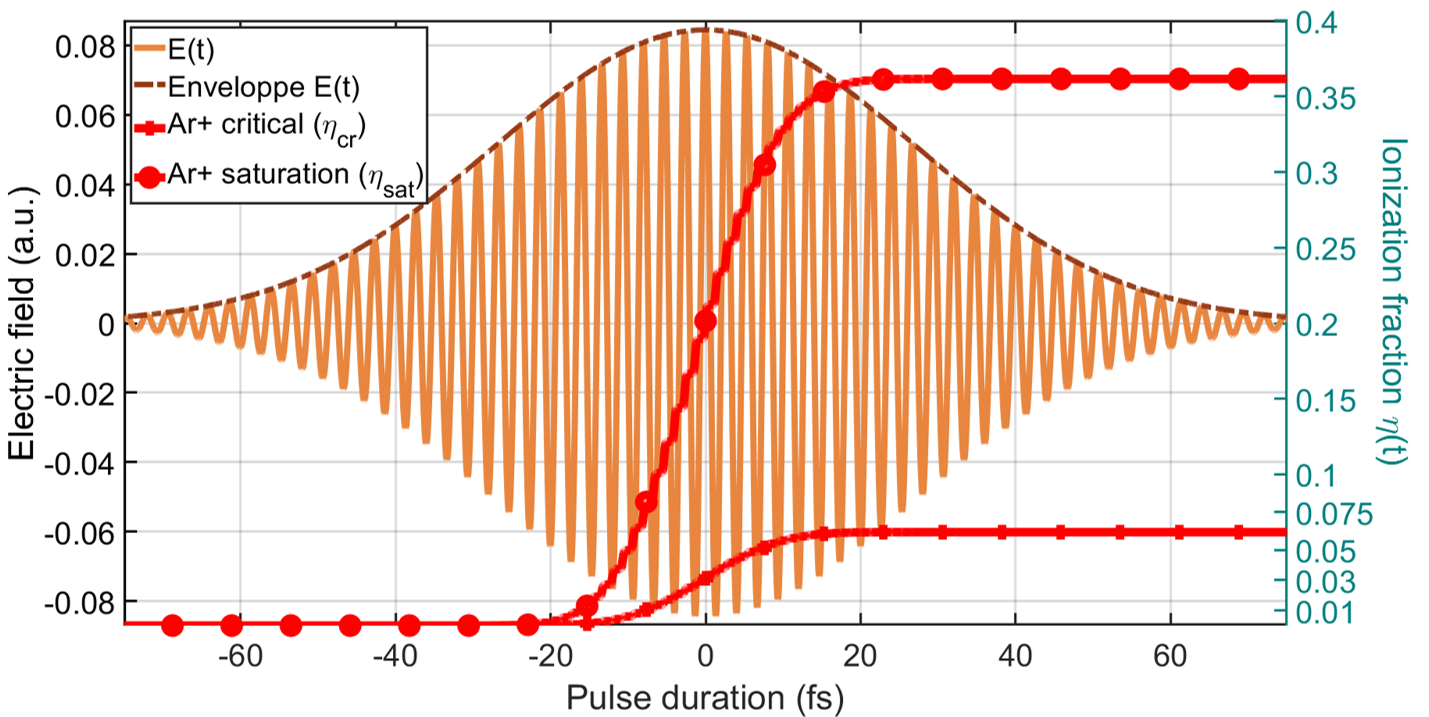}
\caption{Evolution of the parent ions (or alternatively, free-electrons) population for predicted laser intensity $I^c_0 \approx 1.8$x$10^{14}$ W/cm$^2$ evaluated at critical fraction of ionization (red curve marked +) for Argon and $t_p=45$ fs at FWHM, order of magnitude consistent with \cite{Popmintchev10516}, \cite{Weissbilder:22}. Ionization at saturation ($\eta_{\textrm{sat}}$) laser intensity (red curve marked o) $I_{\textrm{sat}}$ is also shown. Neutral population (not shown) is worth $1-\eta(t)$.}
\label{fig:Population_Ar}
\end{figure}

\begin{figure}[ht!]
\centering\includegraphics[width=14.9cm]{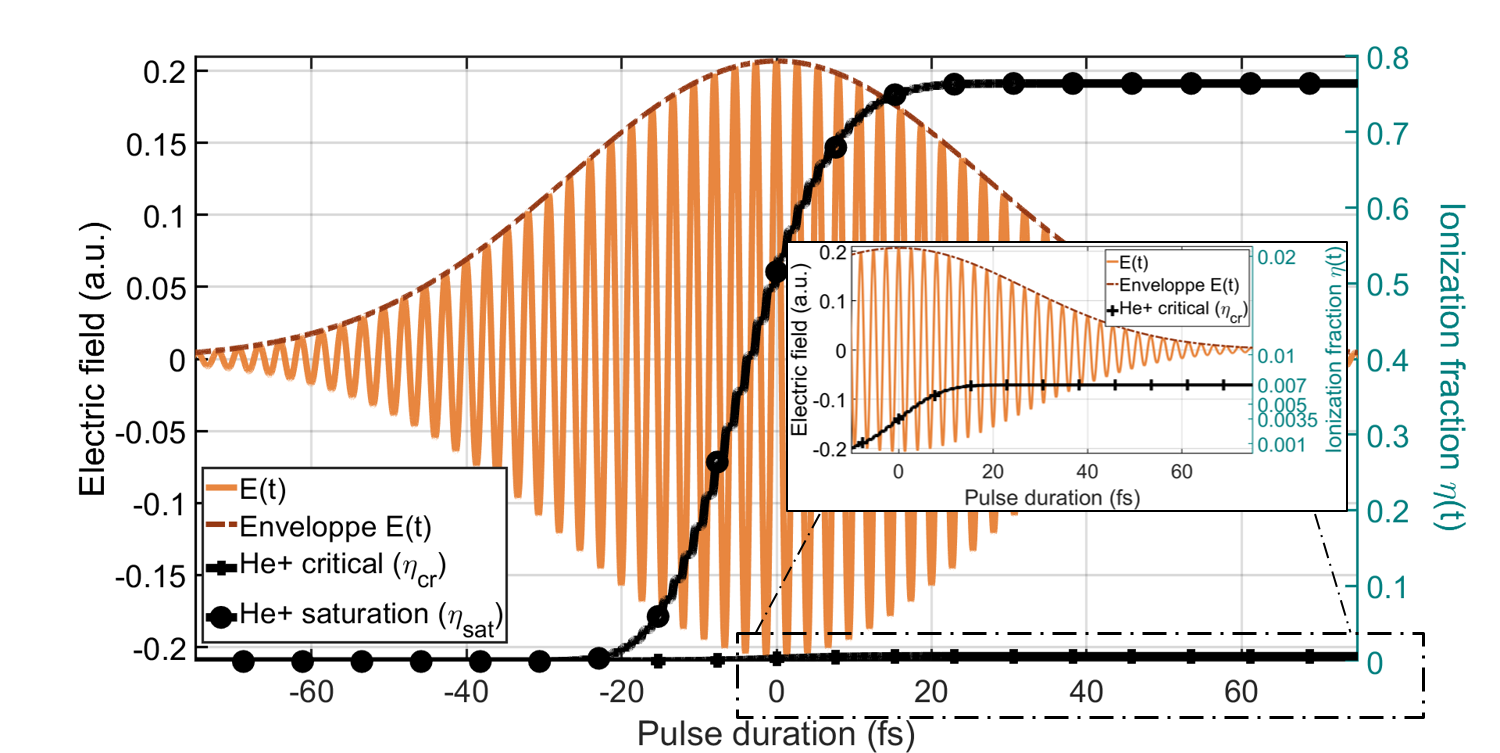}
\caption{Evolution of the parent ions (or alternatively, free-electrons) population for predicted laser intensity $I^c_0 \approx 5.5$x$10^{14}$ W/cm$^2$ evaluated at critical fraction of ionization for Helium and $t_p=45$ fs at FWHM, order of magnitude consistent with \cite{Popmintchev10516}. Ionization at saturation ($\eta_{\textrm{sat}}$) laser intensity (black curve marked o) $I_{\textrm{sat}}$ is also shown. Neutral population (not shown) is worth $1-\eta(t)$.}
\label{fig:Population_He}
\end{figure}

We shall restrict our interest and analyze to the adiabatic limit\footnote{For a (our) low-frequency laser rate, the corresponding time-varying electric field is slowly evolving and one can consider that the wave functions of bound electrons have enough time to adjust themselves to the varying potentials in such a way that the electron penetrates the partially suppressed (to be discussed later) potential barrier “horizontally,” i.e., without changing its total energy \cite{Kruchinin_2018}. } case and with the motion of the ionized electron (into the continuum) remaining non-relativistic.
The strong-field ionization considered here lies within the tunnelling-ionization regime, which is defined by the so-called Keldysh parameter $\gamma = \omega_0\sqrt{2I_p}/E_0$ where $I_p$ is the ionization potential, $E_0$ denotes the laser electric field amplitude, in atomic unit (Hartree atomic unit). The amplitude of the optical laser field $E_0$ is related to the intensity $I_0$, in SI units, by $I_0$[W/cm$^2$] = $|E_0$(V/cm)$|^2$/($2Z_0$) where $Z_0=377 \Omega$ (impedance of free space).
The Keldysh parameter $\gamma$ (also known as adiabaticity parameter \cite{Levy_23}), provides a quantitative understanding of the dominant photoionization mechanism by comparing the period of the oscillating electric field $T$ with the characteristic response time $\tau$ of the system \cite{T_Moon_24}. It is referred to as the borderline value between multifaceted photoionization regime. When $\gamma \ll 1$, tunneling ionization takes place, which also means that the perturbative character of the interaction prevails, while the multi-photon ionization process occurs in a non-perturbative interaction, for $\gamma \gg 1$. Let us bear in mind, that, for argon, for instance, the tunneling time can be as low as a few hundreds of attosecond \cite{Camus2017}.
To estimate the ionization fraction corresponding to ideal experiments, we use the Ammosov, Delone and Krainov (ADK) theory. 
The rate of optical tunnel ionization from the ground state is given by \cite{Spielmann98}
\begin{equation}
w(t) = \omega_p |C_{n^*}|^2 \bigg(\frac{4\omega_p}{\omega_t}\bigg)^{2n^*-1} \exp\bigg(\frac{-4\omega_p}{3\omega_t}\bigg)
\label{eqn:eqnTunnelRate}
\end{equation}
with
\begin{equation*}
\omega_p =\frac{I_p}{\hbar}
\end{equation*}
\begin{equation*}
\omega_t =\frac{eE(t)}{(2m_eI_p)^{1/2}}
\end{equation*}
\begin{equation*}
n^*=Z \bigg( \frac{I_{ph}}{I_p}\bigg)^{1/2}
\end{equation*}
\begin{equation*}
|C_{n^*}|^2= 2^{2n^*}[n^*\Gamma(n^*+1)\Gamma(n^*)]^{-1}
\end{equation*}
where $I_p$ is the ionization potential of the considered gas and is expressed in atomic unit (1 a.u. $\equiv 27.21$ eV) \cite{Spielmann98}, $\hbar$ is the reduced Planck's constant, E(t) $=E_0\cos(\omega_0 t)\exp(-t^2/t_p^2)$ is the electric field of the laser with $E_0$ expressed in a.u. (1 a.u. $\equiv$ $m_e^2e^5\hbar^{-4}\approx5.14$x$10^9$ V/cm) \cite{Kostyukov2018}, $\omega_0$ is the angular frequency of the laser (1 a.u. $\equiv 4.16$x$10^{16}$ rad/s), $t_p$ is the laser pulse duration at full width at half maximum (FWHM) (1 a.u. $\equiv 2.41$x$10^{-17}$ s), $Z$ is the ion charge number after tunnel ionization (1 a.u. $\equiv Z=1$), $m_e$ and $e$ are the mass of the electron and the electron charge, respectively, $I_{ph}\approx 13.6$ eV is the ionization potential of Hydrogen, $\Gamma$ is the Gamma function \cite{Abramowitz}. In the following, we use atomic units (1 a.u. $\equiv \hbar = m_e = e = 4\pi\epsilon_0 = 1 $), unless otherwise stated.
The total ionization fraction (or density of freed-electrons) is defined as \cite{Spielmann98}, \cite{Gibsonthesis},
\begin{equation}
\eta(t) = 1-\exp\big[ - \int_{-\infty}^t \textrm{d}t' w(t')\big],
\label{eqn:IonFrac}
\end{equation}
where $t'$ is a dummy variable that holds for integration. In Fig. \ref{fig:Population_Ar} is shown the time-dependent neutral atoms depletion in Argon for laser intensity equivalent to the one needed to reach the critical (this term is discussed in the main document) ionization $\eta_{cr} \approx 6.25\%$ ($4.8\%$ in \cite{Lytlethesis} for a given harmonic and a pulse duration $t_p = 25$ fs). In Helium, Fig. \ref{fig:Population_He}, the critical ionization is $\eta_{cr} \approx 0.7\%$ ($0.5\%$ in \cite{Lytlethesis} for a given harmonic and a pulse duration $t_p = 25$ fs). It is seen that about half the critical ionization is reached in the center (the peak at $t = 0$ fs) of the pulse. Note that Coulomb-Barrier Suppression\footnote{or, said over-the-barrier (OTB), occurring when $\gamma \ll 1$ and the field is gently stronger than the one for tunneling ionization, where both (OTB and tunneling) induce a distortion of the Coulomb potential, that yields a lowering of the barrier.} ionization regime for Ar and He atoms is reached for $I_0= I_{\textrm{sat}} \approx 2.5$x$10^{14}$ W/cm$^2$ \cite{Gibson2004} and for $I_0 = I_{\textrm{sat}} \approx 15$x$10^{14}$ W/cm$^2$ \cite{Spielmann98}, respectively, above which ADK theory is not anymore accurate \cite{Calvert2016}, but can be corrected as suggested in \cite{Tong_2005}. 
These results dictate the ultimate limit regarding the laser intensity to deliver for which the HH process can be efficiently generated, and is discussed in the main document. 
%
%
%
\subsection{The dipole phase}
Now, at this point of the discussion, it is customary to point out that the picture of the semi-classical model described previously, although giving an intuitive insight of the HHG mechanism and being sufficient to well agree with many observed experimental features, does not inform on the quantum phenomena occurring in the generation process from an atom exposed to an intense, linearly polarized, electromagnetic field. Upon this consideration, a more-realistic approach has been developed and is celebrated as the Strong-Field Approximation (SFA), where the contribution of the evolution of any bound states beyond the ground state $\ket{0}$ is neglected \cite{Lewenstein94}, \cite{Amini_2019}. The classical electron paths are then generalized to complex-valued functions, the quantum trajectories. These quantum paths are an extension of the classical electron trajectories depicted in the semi-classical three-step picture, and interpret the origin of quantum effects such as phase accumulation due to tunneling ionization and quantum diffusion of the electron wave packet during its propagation.
\\ 
Especially, in the next subsection we will deal with the necessary macroscopic response in high harmonics generation so that phase matching (a coherent signal building of the waves emitted by individual atoms, called the emitters, in a macroscopic medium) occurs, and will see that it depends upon five contributions. One of these concerns the dipole phase, and depends on the laser intensity and on the electron trajectories into the continuum. 
Those electron trajectories are sorted in two categories: the short and long ones (implicit in the last term of Eq. (\ref{eqn:eqn1})), and can be analytically derived using the saddle-point equations \cite{Haessler2012}, in the SFA framework, to obtain a closed-form expression, or from a purely numerical treatment of the TDSE, with the use of judicious mathematical maneuvers (see section \ref{sec:TDSE_CN}). 
\\
In what follows, we broach the subject in giving only very general outlines. Let us start with the time-dependent induced dipole moment 
\begin{equation}
\mathbf{D}(t)= \braket{\Psi(t)\vert\mathbf{\hat{D}}\vert\Psi(t)}
\label{eqn:Induced_Dipole_General}
\end{equation}
where the dipole operator $\mathbf{\hat{D}} = -\mathbf{\hat{r}}$ in the single atom case, and, according to the SFA assumptions, the electronic state $\ket{\Psi(t)} $ that stands for the time evolution of the atomic system is a coherent superposition of the ground $\ket{0}$ and the continuum $\ket{\mathbf{p}}$ states, and takes the following ansatz
\begin{equation}
\ket{\Psi(t)} = \textrm{e}^{iI_pt}\bigg( a_0(t) \ket{0} + \int d^3\mathbf{p} b(\mathbf{p},t)\ket{\mathbf{p}} \bigg)
\label{eqn:WaveFunctionSFA_General}
\end{equation}
where $a_0(t)$ represents the amplitude of the ground state and $b(\mathbf{p},t)$ denotes the transition amplitude to the continuum states. 
From the latter, and without entering into details, for the sake of conciseness, the time-dependent dipole moment can then be expressed as \cite{NAYAK20191}, \cite{Amini_2019}
\begin{equation}
\mathbf{D}(t) \simeq i \int_{-\infty}^{t} \textrm{d}t' \int \textrm{d}^3\mathbf{p} \underbrace{\textrm{e}^{-iI_pt}\mathbf{d^{\ast}}[\mathbf{p}+\mathbf{A}(t)]}_{\textrm{Recombination}}\underbrace{\textrm{e}^{-iS(\mathbf{p},t,t')}}_{\textrm{Propagation}}\underbrace{\textrm{e}^{-iI_pt'}\mathbf{E}(t')\mathbf{d}[\mathbf{p}+\mathbf{A}(t')]}_{\textrm{Ionization}},
\label{eqn:Induced_Dipole}
\end{equation}
where
$\mathbf{A}(t)$ is the vector potential associated with the electric field, defined as $\mathbf{E}(t) = -\partial\mathbf{A}(t)/\partial t $, $I_p$ is the atom's ionization potential, $\mathbf{p}$ is the electron canonical momentum, $\mathbf{d}(\mathbf{p}) = \braket{\mathbf{p}\vert\mathbf{r}\vert \Psi_0}$ is the dipole moment of the field-free ground-continuum transition between the ground state $\ket{\Psi_0}$ and the plane wave state $\ket{\mathbf{p}}$, and is expressed in terms of matrix elements, $\mathbf{E}(t)$ is the laser electric field, and
\begin{equation}
S(\mathbf{p},t,t')= - \int_{t'}^{t} \textrm{d}t'' \Bigg[ \frac{[\mathbf{p}+\mathbf{A}(t'')]^2}{2} + I_p(t-t')\Bigg]
\label{eqn:SaddlePoints_1}
\end{equation}
is referred to as the quasi-classical action. The fourfold integration in Eq. (\ref{eqn:Induced_Dipole}) is called the Lewenstein Integral, and can be interpreted as the continuous summation of probability amplitudes of the three distinct processes under the integral kernel \cite{PhysRevA.52.4747}. Indeed, reading Eq. (\ref{eqn:Induced_Dipole}) from right to left, each of the three terms corresponds successively to one step of the classical three-step model: the first term is the probability amplitude for one electron to make the transition (tunneling) to the continuum at time $t'$ with drift momentum $\mathbf{p}$; then, in the second term, it propagates in the continuum, freely moving in the laser field since influence of atomic potential is assumed to be small, and accumulates a phase given in Eq. (\ref{eqn:SaddlePoints_1}), until it recombines to the atomic ground state $\ket{0}$ at time $t$, the third term. The triple-integral over the $\mathbf{p}$ drift momentum space for the Propagation Term in the integral kernel of Eq. (\ref{eqn:Induced_Dipole}) stands for an infinite sum of drift momenta, accounting for all quantum paths, in the sense of Feynman path integral \cite{doi:10.1126/science.108836}. \footnote{The prior expression for the induced dipole $\mathbf{D}(t)$ can also be interpreted as Landau-Dykhne formula for transition probabilities applied to the evaluation of the observable $\mathbf{D}(t)$ in Eq. (\ref{eqn:Induced_Dipole_General}), with $\Psi(t)$ the time-dependent electronic wavefunction \cite{Lewenstein94}.}
\\
Turning into the spectral domain, we thereby obtain the dipole radiation spectrum derived from the Fourier Transform of the time-dependent dipole moment $\mathbf{D}(t)$
\begin{equation}
\mathbf{\tilde{D}}(\omega_q) \simeq i \int_{-\infty}^{\infty}  \textrm{d}t \int_{-\infty}^{t}  \textrm{d}t' \int \textrm{d}\mathbf{p} \,\mathbf{d^{\ast}}[\mathbf{p}+\mathbf{A}(t)] \mathbf{F}(t')\mathbf{d}[\mathbf{p}+\mathbf{A}(t')] \textrm{e}^{-iS(\mathbf{p},t,t')+i\omega_q t}
\label{eqn:Induced_Dipole_Spectrdomain}
\end{equation}
from which the harmonic spectrum I($\omega_q$)$ \, \varpropto \, \omega_q^4|\mathbf{\tilde{D}}(\omega_q)|^2$ is obtained. The integral can be solved in different ways, of which one mathematical technique is known as the Saddle-Point approximation, the other is a purely numerical integration (trapezoidal rule) but is not straightforward due to the highly oscillating integrand induced by the term $\exp[-iS(\mathbf{p},t,t')]$ \cite{MHoegner}. The solutions are stationary phase for which the action is minimum, stemming from the classical least action principle (Fermat's principle), satisfying $\nabla_{\mathbf{p}}S(\mathbf{p},t,t') = 0$ and $\delta_{t,t'}S(\mathbf{p},t,t') = 0$.
\\
Now, the phase of the dipole is given by the quasi-classical action $S(\mathbf{p},t,t')$ accumulated by the electron along all corresponding quantum pathways $j$ (or quantum orbits) under the only influence of the laser field \cite{PhysRevA.80.033817}. For each constituted harmonic, just a few quantum paths are mainly involved, whose intensity-dependent phase can be approximated as $\phi_q(I) \simeq -\alpha^j_q I_0$ where $\alpha^j_q$ is the reciprocal intensity, and is roughly proportional to electron excursion time $\tau^j_q$ \cite{PhysRevA.65.031406}. It is possible to derive the different quantum path contributions from the previous calculated atomic dipole moment $\mathbf{D}$ at a specific harmonic $q$, for several laser peak intensities. Indeed, to extract information about the phase components, the intensity dependent dipole phase is first multiplied by a Gaussian Window function, yielding an apodized dipole term $\mathfrak{d}_{q}^{\phi} (I)$, on which is then applied a wavelet-like sliding window Fourier transform \cite{Csizmadia_2021}, that can also be viewed as a Gàbor-type analysis,  
\begin{equation}
\mathcal{G}(\alpha_q,I') = \mathcal{F}\{\mathfrak{d}_{q}^{\phi} (I)\} = \int_{-\infty}^{\infty} \textrm{d}I \, \textrm{e}^{i\phi_q(I) } \textrm{e}^{-(a_w(I'-I))^2} \textrm{e}^{-i\alpha_q I},
\label{eqn:GaborTransform}
\end{equation}
where $\phi_q(I) = \textrm{arg}(D(q,I))$, $I'$ is the central intensity for a given apodization window and $a_w$ is the length of the apodization window. 
\begin{figure}[ht!]
\centering\includegraphics[width=14.8cm]{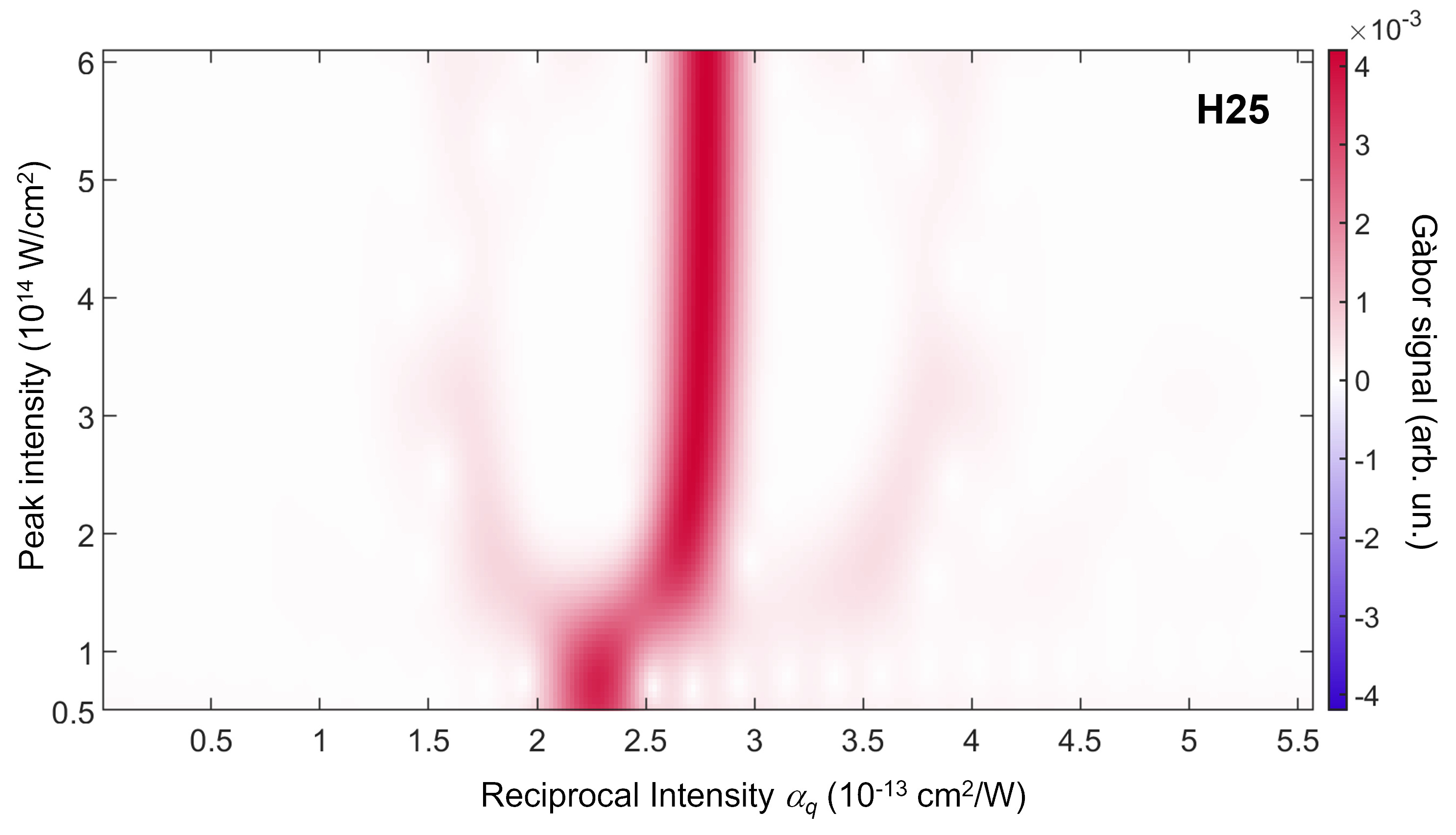}
\caption{First two quantum trajectories $j=1, 2$, for $25$th harmonic in Argon, $t_p=45$ fs, represented in the intensity ($I'$)-reciprocal intensity ($\alpha_q$) plane.}
\label{fig:QuantumTraj_Ar}
\end{figure}
The two distinguishable and main quantum paths correspond to families of electron trajectories that have spent different times in the continuum, begetting disentanglement of different quantum phases, see Fig. \ref{fig:QuantumTraj_Ar} and Fig. \ref{fig:QuantumTraj_He}. The two branches of the tuning fork-like structure are respectively called long $\mathscr{l}$ (@$\alpha_q \sim 30$x$10^{-14}$cm$^{2}.$W$^{-1}$) and short $\mathscr{s}$ (@$\alpha_q \sim 15$x$10^{-14}$cm$^{2}.$W$^{-1}$) trajectories. This pair of trajectories tends to come closer and merging at a point of bifurcation as a single class of trajectory corresponding to the so-called energy \textit{cut-off} in the harmonics spectrum. Note that while the electron wave packet can undergo multiple visits (higher-order returns) of the parent ion by accelerating in the fundamental laser field, they have been neglected (therefore only the contributions of trajectories shorter than one optical cycle have been considered) because they have a longer return time, thus suffering more quantum diffusion \cite{Balcou1999}. Quantum paths with $j>2$ exist but are very faint. Therefore, the two first quantum paths $j=1, 2$ are then said to be the dominant ones. Short and long trajectories descended from the ionized electron wave packet are also known to produce quantum interferences patterns with the portion of the remaining bound electron wave packet when returning to the origin (the parent ion). These quantum interferences, signature of quantum beat between two coherently excited quantum states, result in an oscillating dipole and emitted light \cite{Bengtsson_PRA_2023}, and play a foremost role in the formation and shape of the high harmonic spectrum \cite{Csizmadia_2021}. Finally, the value for $\alpha_q$ that we will retain for the following, is $\alpha_q^{Ar} \simeq 22.5$x$10^{-14}$ and $\alpha_q^{He} \simeq 23.8$x$10^{-14}$cm$^{2}.$W$^{-1}=$ cste $\forall$ $q$, for the sake of ease.
\begin{figure}[ht!]
\centering\includegraphics[width=14.8cm]{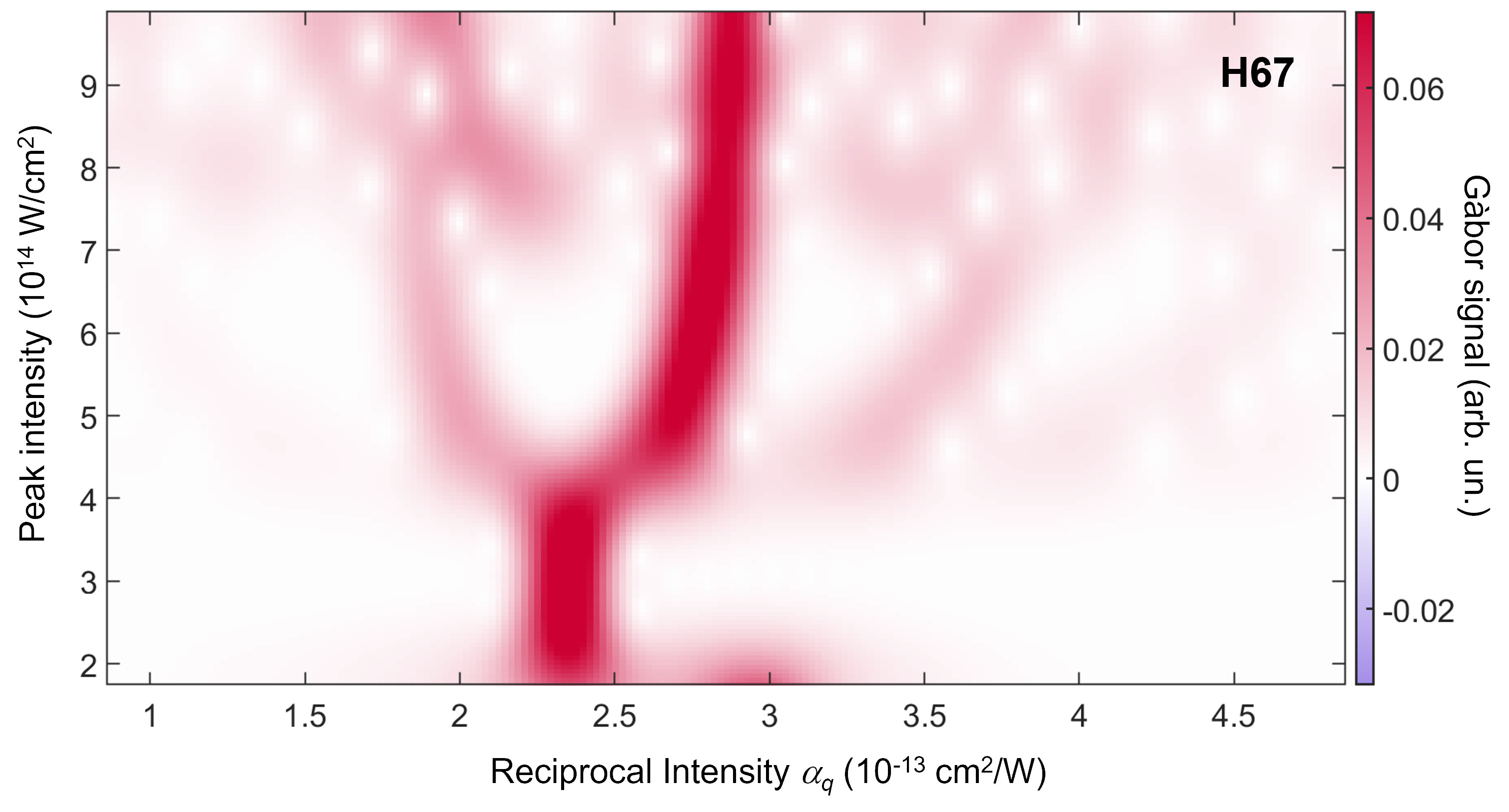}
\caption{First two quantum trajectories for $67$th harmonic in Helium, $t_p=45$ fs, represented in the intensity ($I'$)-reciprocal intensity ($\alpha_q$) plane.}
\label{fig:QuantumTraj_He}
\end{figure}
%
\section{Brief overview of the Crank-Nicolson scheme for the TDSE}
\label{sec:TDSE_CN}
\begin{figure}[ht!]
\centering\includegraphics[width=14.9cm]{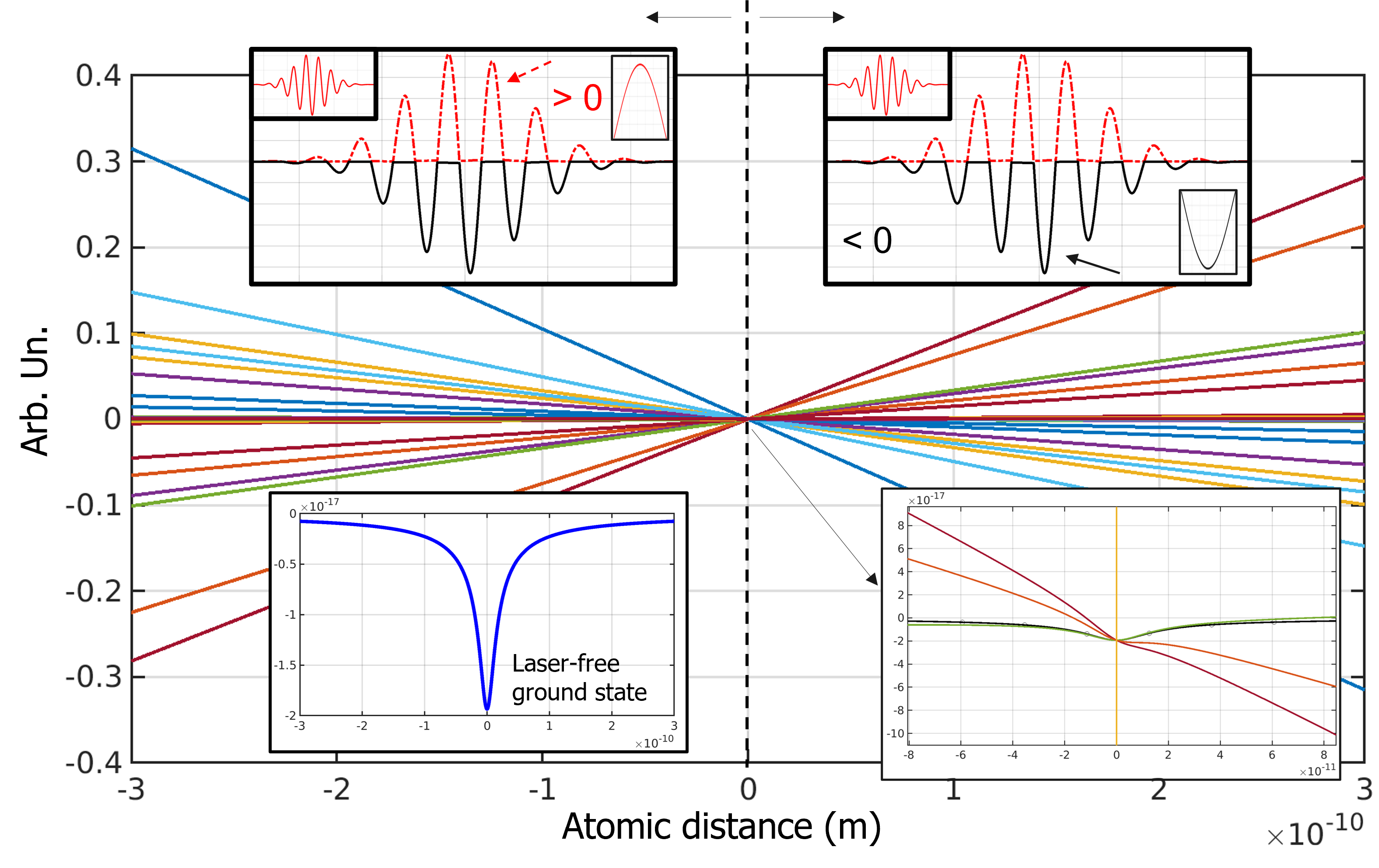}
\caption{A few deformations of the effective potential $V_0 (x) + x E(t)$ under the influence of the sign of the oscillating laser field. When laser field amplitude is negative (black portion of the Gaussian pulse in the top-right insert), potential inclines towards positive distance (bottom-right corner of the graphic), and when laser field amplitude is positive  (dashed-red portion of the Gaussian pulse in the top-left insert), potential bends towards negative distance (bottom-left corner of the graphic). Bottom-left insert is the unperturbed atom potential. Bottom-right insert shows a zoom of a few laser-perturbed shapes of the atomic potential for negative laser electric field. }
\label{Fig_Atom_Potential}
\end{figure}

In this subsection, we simply derive the outlines of the numerical implementation of the TDSE, which depicts the quantum origin of the HHG phenomenon, as briefly discussed in section \ref{subsec:Theoretical considerations_SFA}.
\\
To take into account \textit{ab initio} quantum description of the HHG mechanism, we introduce the Time-Dependent Schrödinger Equation which describes the evolution of the electronic wavefunction over time and space
\begin{equation}
i\hbar\frac{\partial}{\partial t} \ket{\Psi(x,t)}  = \hat{H}\ket{\Psi(x,t)}
\label{eqn:TDSE}
\end{equation}
where $\hat{H}$ is the sum of the field-free time-independent Hamiltonian $\hat{H}_0$ and the interaction Hamiltonian $\hat{H}_L$, in presence of the laser field (which is on the other hand \textit{classically} treated), as a disturbance term
\begin{equation}
\hat{H} = \underbrace{- \frac{\hbar}{2m_e}\frac{\partial^2}{\partial x^2} - \frac{e^2}{4\pi \epsilon_0} V_0(x)}_{\hat{H}_0} + \underbrace{xeE(t)}_{\hat{H}_L}
\label{Eq:Hamiltonian}
\end{equation}
The Hamiltonian $\hat{H}$ has a finite size, thus has a finite number of eigenstates, due to the discretization of the space grid.
\\
The one-dimensional model consists of a single electron, originally trapped in an atomic potential well $V_0(x)$, is expressed in the soft-core Coulomb potential, and written as 
\begin{equation}
V_0(x) =- \frac{1}{\sqrt{x^2+b^2}}
\label{SoftCoulombPotential}
\end{equation}
with $b = 1.1892$ is the regularization parameter, chosen to adjust the ionization potential $I_p$ of the atom. In Fig. \ref{Fig_Atom_Potential} a few potential tilts at various times $t_i$ in the laser field sine wave are sketched, which acts as a perturbation, and corresponding to the two last $x$-varying terms in Eq. (\ref{Eq:Hamiltonian}). It is explained because the high laser field strength, that the laser CPA scheme allows, is of the order of the inner atomic potential. The result of this atomic strength leaning down is such that a potential barrier exits, where an electron can \textit{horizontally} tunnel through, and constitutes the first step of the ionization and recombination process responsible for HHG. 
To numerically compute the solution of the TDSE Eq. (\ref{eqn:TDSE}), one can use the Crank-Nicolson (CN) method which is an implicit scheme that ensued from the Backward- and Forward-Euler methods. 
\begin{figure}[ht!]
\centering\includegraphics[width=11.1cm]{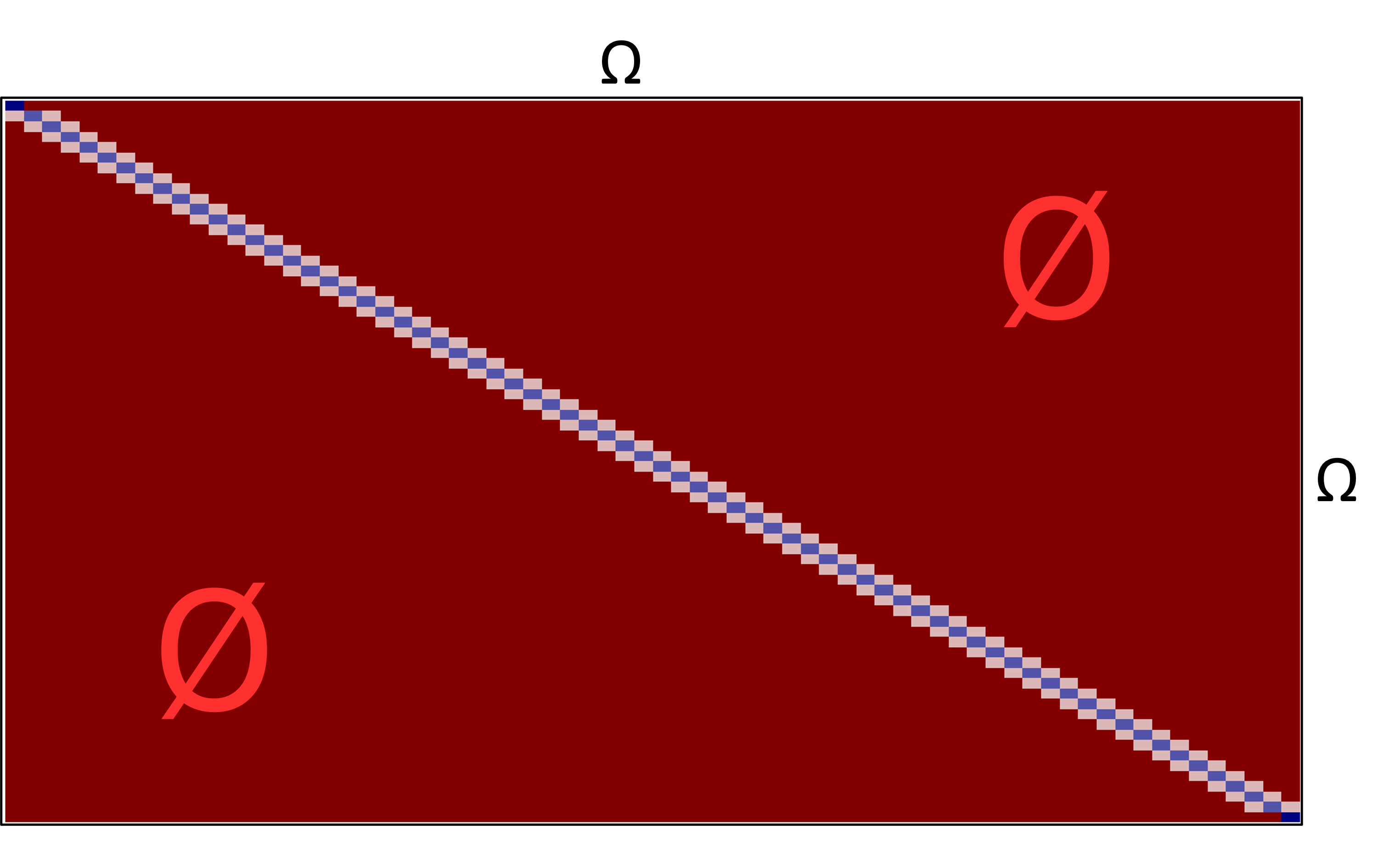}
\caption{Illustration of a typical tridiagonal matrix $|\hat{H}|$ used to solve for the TDSE, where sub-super and main diagonals are visible. The domain is bounded at $\Omega$.}
\label{Fig_TridiagMatrix}
\end{figure}
The time evolution operator, also called time propagator, given in Eq. (\ref{Eq:Hamiltonian}), used to propagate the wavefunction forward in time, after spatial and time discretization, is then deduced
%
\begin{equation}
a_1 \Breve{\Psi}_{i-1}^{k+1} + b_1 \Breve{\Psi}_{i}^{k+1} +c_1 \Breve{\Psi}_{i+1}^{k+1}  = a_2 \Psi_{i-1}^{k} + b_2 \Psi_{i}^{k} +c_2 \Psi_{i+1}^{k},
\label{eq:CN_Recursive}
\end{equation}
where the accent $\Breve{}$ dresses the unknown vector, the superscript $k$ and the subscript $i$ stand for the time and the spatial nodes, respectively, with $a_1=c_1=-\frac{i\hbar\Delta t}{4m\Delta x^2}$, $a_2 = c_2 = +\frac{i\hbar\Delta t}{4m\Delta x^2}$  and $b_1= 1 + \frac{i\hbar\Delta t}{2m\Delta x^2} + \frac{V_i^k\Delta_t}{2i \hbar}$, and $b_2 = 1 + \frac{i\hbar\Delta t}{2m\Delta x^2} - \frac{V_i^k \Delta_t}{2i \hbar} $  which can be written in the following form
\begin{equation}
\hat{U}(t_{n+1},t_n) = \bigg \lbrack \hat{\mathbb{I}} + i\hat{H}\frac{\Delta t}{2} \bigg \rbrack^{-1} . \bigg \lbrack \hat{\mathbb{I}} - i\hat{H}\frac{\Delta t}{2} \bigg \rbrack,
\label{eq:CN_Propagator}
\end{equation}
where $\hat{\mathbb{I}}$ indicates the N-by-N Identity matrix. Eq. (\ref{eq:CN_Propagator}) is another way to write the so-called Cayley's form. Finally, the wave function is propagated from one time step to the next one, which, in the compact form using the quantum nomenclature, formally yields
\begin{equation}
\braket{x_i\vert\Psi(t_{n+1})} = \braket{x_i\vert\hat{U}(t_{n+1},t_n)\vert\Psi(t_{n})}
\label{eq:CN_GLobal_TDSE}
\end{equation}
where the subscript $i$ ranges from $1$ to N, and stems from the spatial grid in the $x$-dimension of the grid. 
\\
Rearranging terms with the use of basic matrix operations such as $A^{-1}A = AA^{-1} =\hat{\mathbb{I}} $ and $A\hat{\mathbb{I}} = \hat{\mathbb{I}}A = A$, we have, with Eq. (\ref{eq:CN_Propagator}), \cite{garcia2000numerical}
\begin{multline}
\Psi^{k+1} = \bigg \lbrack \hat{\mathbb{I}} + i\hat{H}\frac{\Delta t}{2} \bigg \rbrack^{-1} . \bigg \lbrack \hat{\mathbb{I}} - i\hat{H}\frac{\Delta t}{2} \bigg \rbrack \Psi^{k}\\
= \bigg \lbrack \hat{\mathbb{I}} + i\hat{H}\frac{\Delta t}{2} \bigg \rbrack^{-1} . \bigg \lbrack 2\hat{\mathbb{I}} - \bigg( \hat{\mathbb{I}}+ i\hat{H}\frac{\Delta t}{2} \bigg) \bigg \rbrack \Psi^{k}\\
= \bigg \lbrack 2 \bigg(\hat{\mathbb{I}} + i\hat{H}\frac{\Delta t}{2} \bigg)^{-1}  -\hat{\mathbb{I}} \bigg\rbrack \Psi^{k} \\
\end{multline}
and, setting $\mathbf{Q} = \frac{1}{2}[\hat{\mathbb{I}} + i\hat{H}\frac{\Delta t}{2}] $, it can be written as
\begin{eqnarray}
\Psi^{k+1} = (\mathbf{Q}^{-1} - \hat{\mathbb{I}})\Psi^k \notag \\
=  \mathbf{Q}^{-1}\Psi^k - \Psi^k .
\end{eqnarray}
Finally, introducing an intermediary array $\chi$, then identifying the linear system on the form $Ax = b$, we have 
\begin{eqnarray}
\mathbf{Q} \chi = \Psi^k, \notag \\
\Psi^{k+1}=\chi - \Psi^k
\end{eqnarray}
where we can perform Gaussian elimination (avoiding a costly matrix inversion) to determine $\chi$, on a packed tridiagonal matrix, consisting of the main diagonal and the first sub and superdiagonals which have nonzeros elements (sparse matrix), known as Thomas Algorithm, see Fig. \ref{Fig_TridiagMatrix}. At last, to seed the algorithm, the initial wave function is chosen as a Gaussian wavepacket, centered at $x=0$.  
The numerical solution of the TDSE is hampered by the imperative lengthening of the spatial and temporal grids, because since our simulation domain $\Omega$ is necessarily of finite dimensions, its boundaries generate spurious unphysical reflections, by the traveling (outgoing) waves impinging the wall of the domain at $x = \pm L$. However, simply broadening the size of the box not only turns out to be inefficient, but also adds non negligible computation cost. One manner of freeing from this constraint is to impose artificial boundary conditions such as, amongst several other methods, a gobbler function, or else, absorbing boundary conditions (ABC) \cite{Fevens_Jiang_1999}, to dampen these numerical phenomena at the borders, rather than suffering reflections back in the interior of the domain. Then, they are incorporated into the CN scheme Eq. (\ref{eq:CN_Propagator}). We applied here a three order ($p=3$, where $p$ is the order in the sum of the reflection coefficient at the boundary, between incident and reflected waves with different group velocities) absorbing boundary condition. Finally, we activated the Parallel Pool of workers in Matlab to accelerate computing. 
\begin{figure}[ht!]
\centering\includegraphics[width=15.8cm]{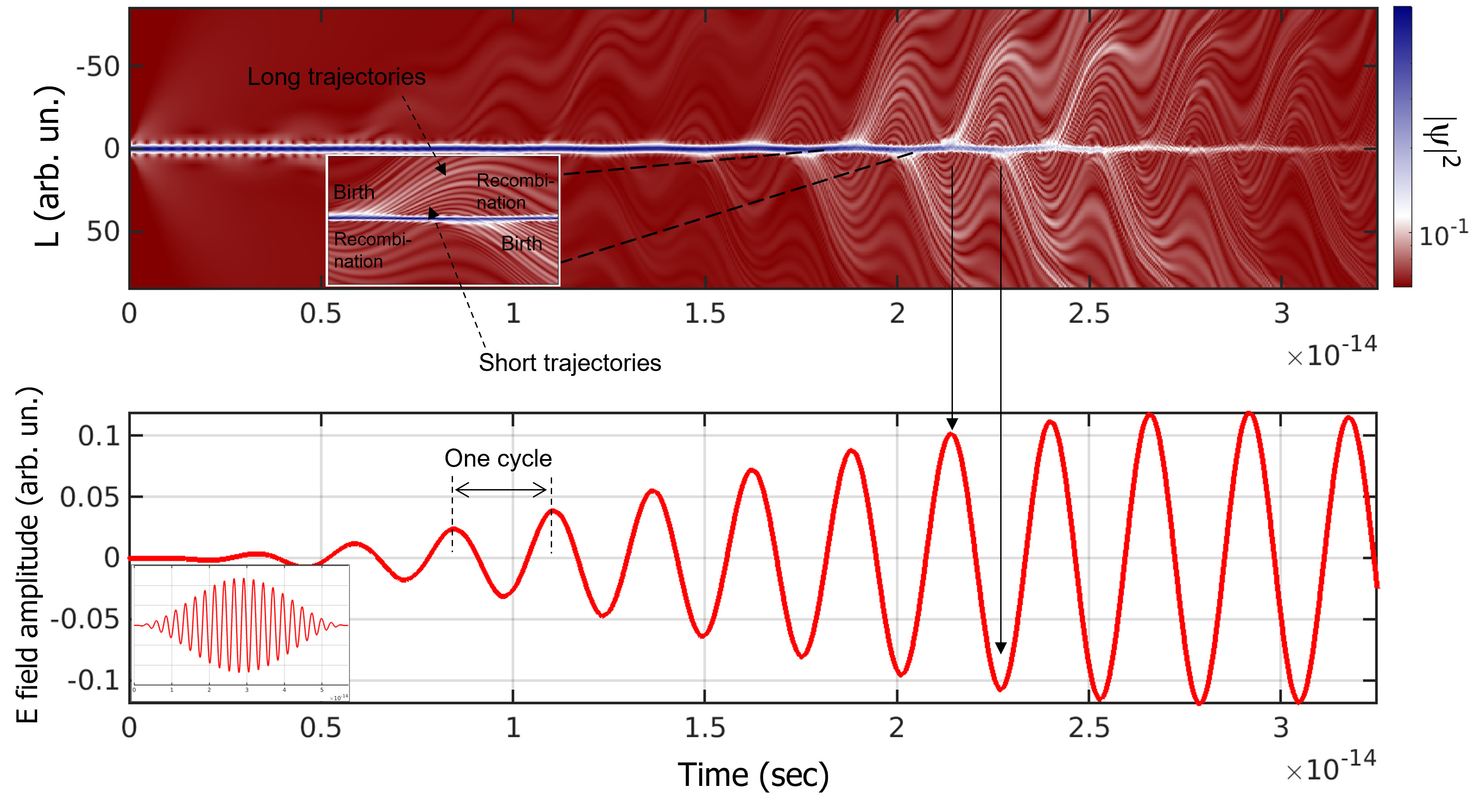}
\caption{Top figure: temporal evolution of the probability density from a soft-Coulomb potential from numerical simulation of 1D-TDSE solved with Crank-Nicolson scheme, and applying ABC conditions at the boundaries of the domain $\Omega$. $L$ is the distance in the continuum. It is observable that long paths can last one laser cycle (one period) before recombining with the nucleus (ion). Insert shows different times of birth from and recombination with the nucleus, as well as what it is defined as short and long trajectories. Bottom figure: shape of the first-half of the external Laser electric field, and insert is the full scale laser electric field. One cycle corresponds to one period of the sine wave. }
\label{Fig_TDSE_1D_withABC}
\end{figure}
Looking at Fig. \ref{Fig_TDSE_1D_withABC}, one can see that a small portion around a peak of the sinus of the laser field allows the recombination of the ionized electron. Also, long ($\ell$) and short ($\mathcal{s}$) trajectories are obvious (1- electron tunneling, called "birth", 2- electron in the continuum where it propagates in the strong laser electric field), for which their designation depends on the time of birth. Indeed, long trajectories are those created where ionization occurs closer to a field peak. Compared with long trajectories, short trajectories have later ionization and earlier recombination times. Long trajectories are also those that last one or more than one laser cycle. They have a weak probability of recombine, but they exhibit higher energies because of the acquired kinetic momentum, as they consume more time in the continuum, embedded in the strong electromagnetic field. In addition, the quantum nature of HHG, inherent in tunnel ionization, introduces a displacement (white bumpy in Fig. \ref{Fig_TDSE_1D_withABC}) of the electron birth position in the continuum from the atom position. These effects are visible during ionization (see top insert in Fig. \ref{Fig_TDSE_1D_withABC}) where electrons are liberated at different positions (a few arb. un. of the $L$ scale) from the parent atom depending on the laser field strength. 
The unbounded "hair in the wind" trajectories are those which do not recombine with the parent ion. 
\\
To sum up, the Fig. \ref{Fig_TDSE_1D_withABC} illustrates the fact that the trajectories that return back to zero are those which release the (quiver) kinetic energy gained in the continuum by the ionized (or freed) electron, accelerated by the laser electric field (in the sense of the Lorentz Force), under the form of a XUV radiation. 
%
%
\section{Ultrafast laser description}
\label{sec:Laser_SuppMat}
The standalone laser is based on a commercial system from Coherent consisting of a mode-locked oscillator (Vitara), which is a Ti:Sapphire crystal in cavity pumped by a continuous wave (CW) 5W optically pumped semiconductor laser (Verdi Green OPSL), then followed by a two-stage Ti:Sapphire amplifier (thermo-electrically cooled regenerative and single-pass cavities) pumped with two diode-pumped water-cooled Q-switched Nd:YLF laser heads with LBO crystals for frequency doubling, which, combined with a stretcher and a compressor diffraction gratings, are considered as the chief constituents of the so-called chirped pulse amplification (CPA) scheme. The whole sensitive optical parts is chilled with our homemade distilled water. 
\section{Basic optics into the capillary waveguide}
\subsection{Preamble}
\begin{figure}[ht!]
\centering\includegraphics[width=12.9cm]{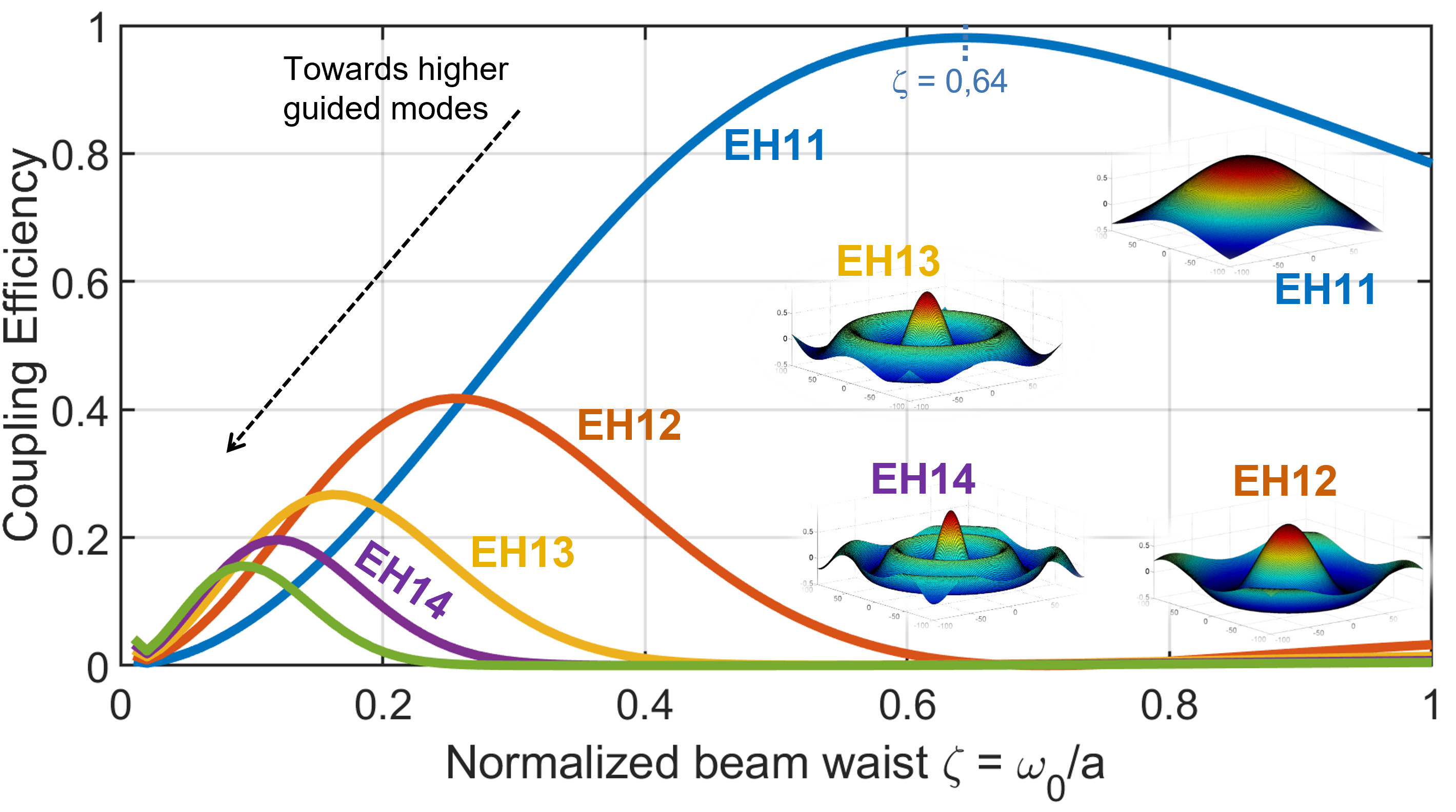}
\caption{Coupling efficiency of a linearly polarized laser with Gaussian transverse
profile (TEM$_{00}$) as a function of the beam waist $\omega_0$ for the first five guided modes of a HCW with a core radius $a$, empty of any gas. Simulated transverse intensity profiles of the guided mode (electric hybrid) EH$_{1j}$ are also shown, where $j$ is the order of the guided mode, indicating the $j^{th}$ root of the zero-order Bessel function, that enters in the definition of the intensity profiles \cite{Marcatili_1964}. }
\label{Fig_CouplingWaveguide}
\end{figure}

A campaign of experiments has been performed with HCWs continuously filled with Argon or Helium, under phase matching conditions described in section \ref{sec:Theoretical considerations} of the manuscript to account for a proof-of-principle of the beamline. Several HCW geometries (inner diameters $2a= 150, 200, 250, 300, 400$ µm and lengths L$_\textrm{med}=39, 31, 25, 23, 16, 13, 11, 10$ mm) have been successfully tested. But, for the sake of simplicity, the scope of the main paper has been focused on a HCW with $2a=300$ µm, although the coupling parameter $\eta_{\mathrm{CouplEfficiency}}$ which is defined as
\begin{equation}
\eta_{\mathrm{CouplEfficiency}} = \frac{[\int_0^a \exp(-r^2/w^2)J_0(u_{0j}\frac{r}{a})r\mathrm{d}r]^2}{\int_0^\infty \exp(-2r^2/w^2)r\mathrm{d}r\int_0^aJ_0^2(u_{0j}\frac{r}{a})r\mathrm{d}r}.
\label{eqn:Coupling_Efficiency}
\end{equation}
is not optimized, because in that case $\zeta = w_0/a \approx 0.5$ where $w_0$ is the waist of the incident laser pulse. Indeed, it should approach $\zeta=0.64$ in order to couple the fundamental guided mode $EH_{1,1}$ in the waveguide with an efficiency of $98\%$, see Fig. \ref{Fig_CouplingWaveguide}. But it has been shown in \cite{vConta} that for a broad range of $0.3 > \zeta/a > 0.7$, the first three of the $EH_{1,m}$ modes cover almost $100\%$ of the guided power (in absence of gas), see Fig. \ref{Fig_CouplingWaveguide}. 
Whereas, in the presence of gas, and in long enough HCWs, the pulse undergoes a spatial redistribution into higher-order modes \cite{Anderson2014} or the pulse can even splits \cite{Wagner2004}, or suffered beam reshaping (due to a longitudinal gradient of gas ionization), such that it can enhance fluctuations in the HH output \cite{Goh2015} and affect the laser intensity in the direction of propagation. But, following simulations for a varying gas pressure in the HCW and laser intensity below saturation level, from \cite{Nurhuda2003}, nonlinear optical effects such as Kerr-induced self-phase modulation (SPM), mode beating or self-steepening do not significantly affect the pulse propagation in the first $5$ to $10$ cm, which is even longer than the HCWs considered throughout this paper. In other words, at longer pulse duration or reasonable intensity level, one can ignore the effects of ionization (responsible for the nonlinear optical processes previously cited) on the propagation of the driving laser field and on the phase matching of the generated XUV radiation \cite{Gaarde2008}. Indeed, the dependence on the intensity is a rather thorny problem due to the highly nonlinear dependence of $\eta_{\mathrm{CouplEfficiency}}$ on the laser intensity.
This is the case here for Helium, which is rather in a weakly ionized regime. Nevertheless, for Argon, the contribution of free electrons dominates that of neutral atoms in Eq. (\ref{eqn:eqn1}). The phase mismatch $\Delta k$ is therefore positive and, again, strongly limits the HH efficiency. 
\subsection{Ray tracing}
\begin{figure}[ht!]
\centering\includegraphics[width=15.1cm]{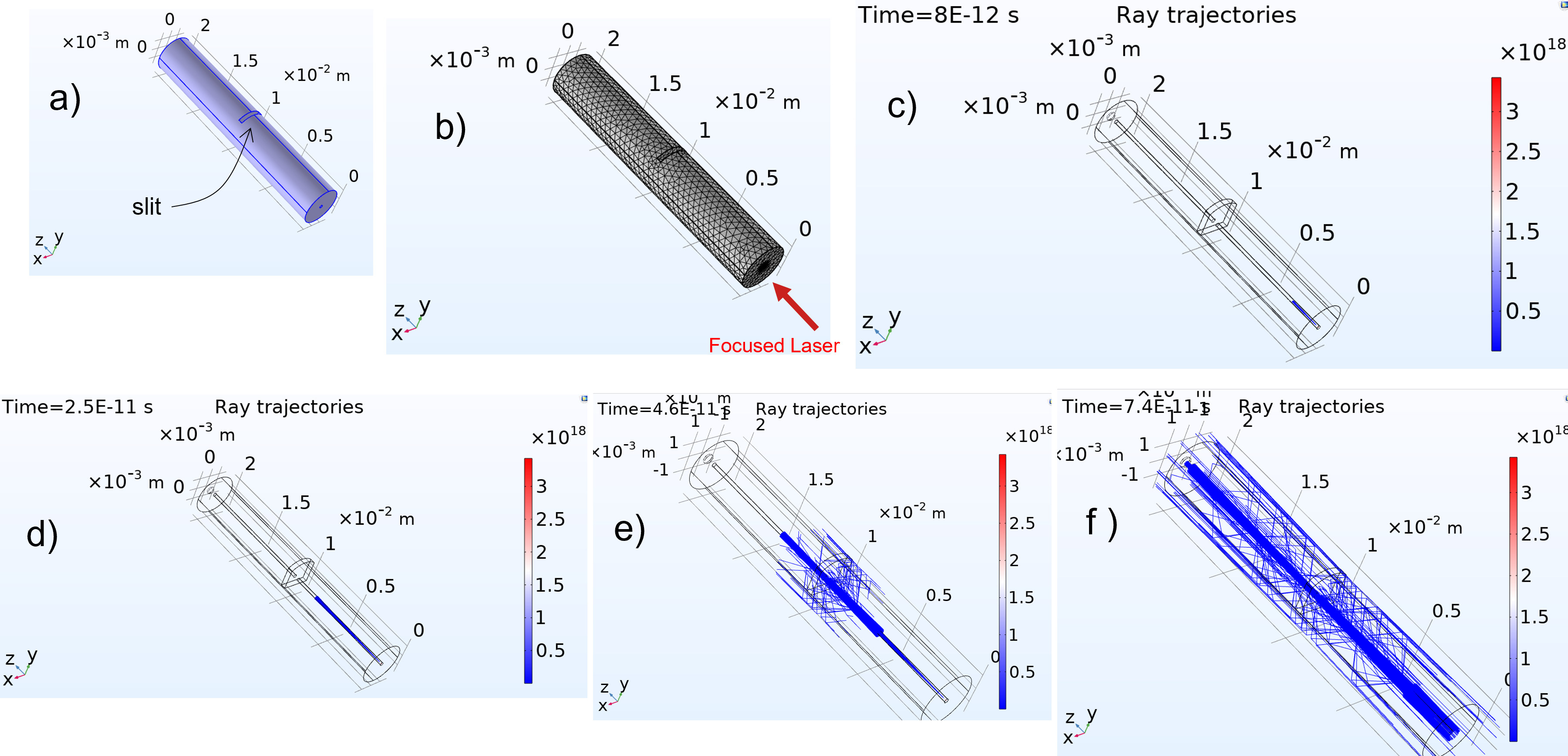}
\caption{A Comsol simulation showing the propagation of the laser (red arrow) to the aim of observing the reflections on the slit, and at the air-glass interfaces all along the capillary dimensions. The chosen numerical aperture of the beam at the capillary entrance is equivalent to our experiment, with focal length $400$ mm, and worth about $0.02$ radians. A few hundreds of rays are launched. In reality, slit has unavoidable hard roughness due to the abrasive disc and difficulties to deburr the edges of the cut slit, that is difficult to compute.}
\label{Ray_Tracing_HCW}
\end{figure}

We have seen in Section \ref{sec_HCW_Descritpion} that a portion of the guided laser light in the hollowed capillary is scattered by the slit, or is reflected out of the air (gas)-glass interface, and this can be observed on the simulation of geometrical optics (ray-tracing approach) of Fig. \ref{Ray_Tracing_HCW}.
There is no gas inside, but it is clearly seen that the capillary is the seat of multiple reflections at the air-glass interfaces. That is a source of unavoidable optical power losses. In practice, the presence of gas, and thus plasma, helps to confine and guide the laser pulse. 
%
%
\section{Coexisting coupled multiphysics phenomena in the capillary environment}
\label{sec:Thermal_MechanicStress_Optics}
\begin{figure}[ht!]
\centering\includegraphics[width=14.7cm]{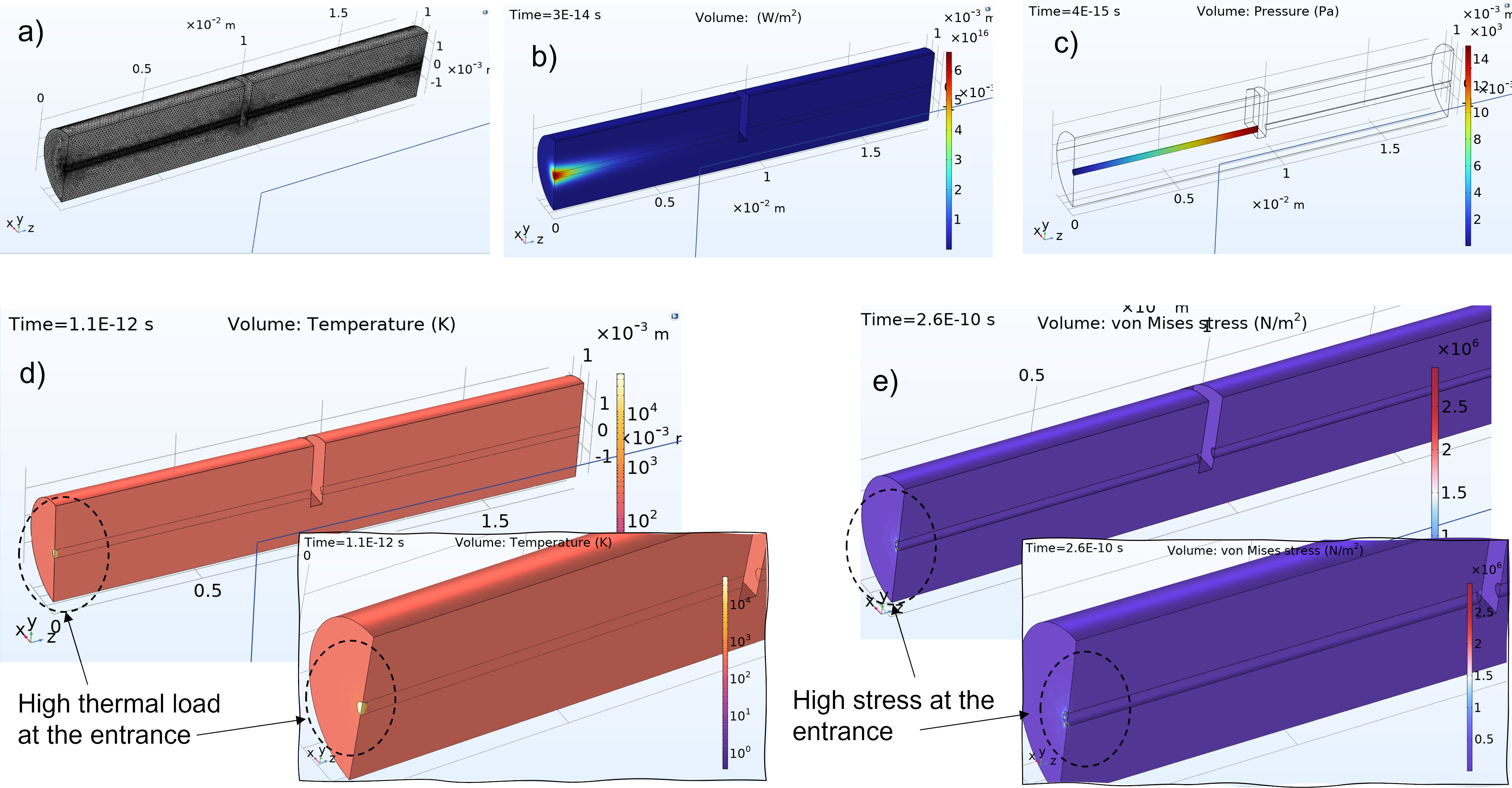}
\caption{Thermal and stress endured by the HCW during the laser passage. Comsol Non-isothermal Flow between laminar Argon flow and glass solid body, and Thermal Expansion Multiphysics Interfaces. Gaussian Pulse of $45$ fs, with spatial characteristics as given in \cite{Vora_2013}. Reference Pressure level is set to $10$E$-3$ mbar. Meshing is chosen "Physics Controlled" for selecting the type "Laminar boundary layer". In the interaction zone at the capillary entrance, let $1/3$ mm (where about $\pm 2 \sigma$ of the Gaussian pulse intensity is deposited), thus an interaction volume of about $10$E$-11$ m$^3$. a) meshing of the capillary. b) gaussian laser pulse focused at the entrance (left-side) of the capillary. c) View of the pressure distribution in one capillary side (the one in contact with the laser focus) for an inlet of $150$ mbar and an outlet boundary condition of $0.1$ mbar. d) Temperature elevation at $t\approx 1$ ps following the laser irradiation at the capillary entrance. e) Mechanical stress due to temperature increase at the entrance at $t\approx 260$ ps .  }
\label{Thermal_ConstraintStress_HCW}
\end{figure}

As discussed in subsection \ref{sec_HCW_Descritpion} and in the introduction of section \ref{sec:Results}, one can also move the investigation to the thermal load and stress phenomena that occur when the ultra-short laser pulse hits the entrance of the capillary. As shown in Fig. \ref{Thermal_ConstraintStress_HCW}, a spatially Gaussian pulse is deposited at the entrance, with an optical absorption coefficient in low-pressure of argon set at $7$ cm$^{-1}$ as the input of the Beer-Lambert law, but considered as constant, at $300$ K. Temperature and mechanical stress values are in relative agreement with those of \cite{Vora_2013}.
It is seen from these results that a very high thermal load is sustained at the capillary entrance and that impacts the capillary structure in terms of mechanical stresses. It is easily verified by looking at the capillary after hours of operation, where it can be locally molten or abraded. As the capillary must be changed regularly, our modular design allows for this requirement. 
%
%
\section{Basic of hydrodynamics: gas distribution and vacuum levels in the capillary and in its environment}
\subsection{Inside the capillary}
A waveguide with L$_{\textrm{med}}=31$ mm has been studied, with respect to the criterion L$_\textrm{med}$ $\geq$ $3$L$_\textrm{abs}$ discussed in Section \ref{sec:Theoretical considerations}, for Argon. Also, as re-absorption limits the useful medium length to $5-10$ absorption lengths \cite{PhysRevLett.82.1668}, L$_\textrm{med} = 31$ mm is within the range. HH spectra have been recorded on a scaled scale with the input pressure in the HCW. 
In the experiments, after evacuation, we measured the backing pressure of the gas upstream from the HCW. 
The gas pressure between the gauge and the injection into the slit has been estimated with a gas model \cite{VST21} and we found a negligible pressure drop between the gauge and the entrance slit of the HCW. Then, the pressure gradient in the HCW, from the slit up to the tips, can be evaluated with a simple model of finite volume element \cite{vConta}, under assumptions where the so-called Hagen-Poiseuille formulation is valid, that is to say, the laminar flow (see Table \ref{table:2} with respect to the relevant parameters for Argon and Helium) and with the use of the Courant-Friedrichs-Lewy (CFL) convergence condition $C_{CFL}= V.\Delta t/\Delta x$, with $V$ the uppermost gas speed, $\Delta t$ and $\Delta x$ the time and spatial ($i$) steps of the grid, respectively. The gas flow rate $\Phi_G$ is then given by
\begin{equation}
\Phi_G= \frac{\Delta P}{\Delta x}\frac{\pi R^4}{8\eta_f} \simeq \frac{\Delta V_{\textrm{HCW}}.}{\Delta t}
\label{eqn:VolumFlow}
\end{equation}
along with
\begin{equation*}
\Delta P = \Delta N \frac{R T}{V_{\textrm{HCW}}}
\end{equation*}
\begin{equation*}
\Delta P_i = \Delta P_{i-1 \rightarrow i} + \Delta P_{i+1 \rightarrow i}
\end{equation*}
where $V_{\textrm{HCW}}$ is the volume of the object considered, $P$ the gas pressure, $T$ the temperature of the fluid, $N$ the gas density number and $R$ the ideal gas constant.
\begin{table}[ht]
\centering
\begin{tabular}{||c c c||} 
 \hline
 Parameters & Argon@ P$_{\textrm{in}}=150$ mbar & Helium@ P$_{\textrm{in}}=400$ mbar\\ [0.5ex] 
 \hline\hline
  K$_n$ \cite{Kunze22} & $0.0035$ & $0.0029$  \\
 R$_e$ \cite{TISON19931171} & $325$ & $269$ \\  
   Regime & Hydrodynamical & Hydrodynamical  \\
 Flow & Laminar & Laminar \\ [1ex] 
 \hline
\end{tabular}
\caption{Mean Knudsen K$_n = \lambda_g/L_c$ and Reynolds R$_e = u_gL_c/\nu_g$ numbers for our experimental cases, where $\lambda_g$ is the mean free path of the gas molecule at the indicated pressure and at T$=293°$K, $L_c$ is the characteristic length of the dimension that constrains the flow, $u_g$ is the flow speed and $\nu_g$ is the kinematic viscosity of the fluid (in m$^2$/s). L$_\textrm{med}=31$ mm for Argon and L$_\textrm{med}=13$ mm for Helium, $2a=300$ µm. The Mach number is estimated to be $M_\textrm{a}<<0.3$, entering the Low-Mach Number Flow (subsonic regime) of the physical interface in Comsol computation .}
\label{table:2}
\end{table}
\begin{figure}[ht!]
\centering\includegraphics[width=11.4cm]{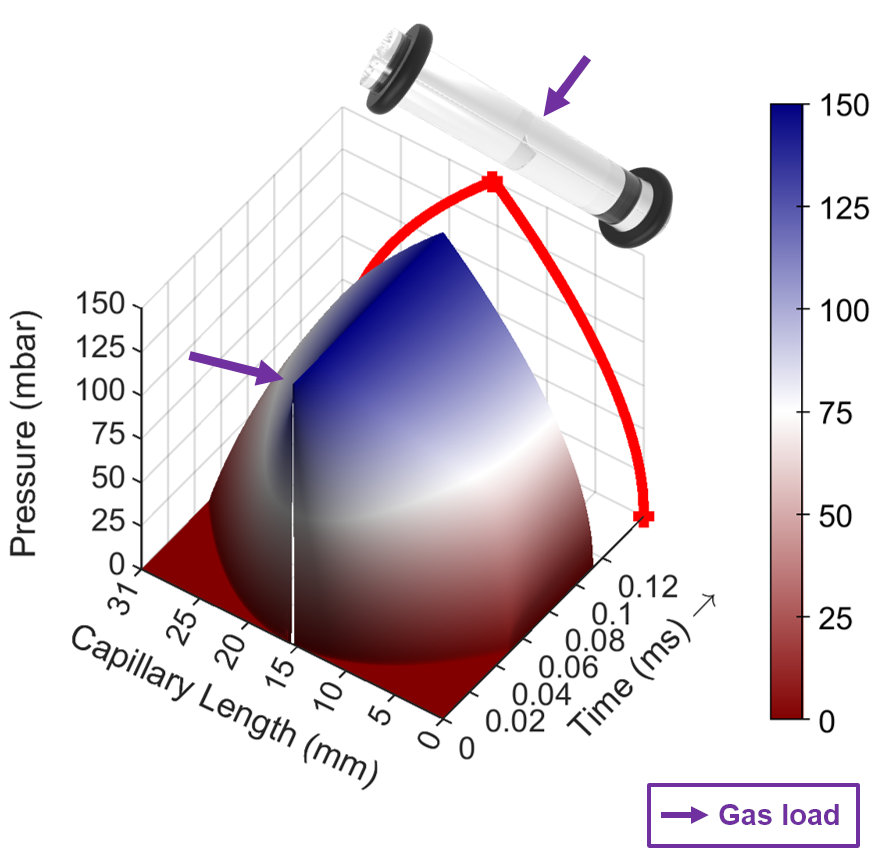}
\caption{Left: map of time-space pressure distribution in the HCW filled with Argon, in the case of laminar flow, $2a=300$ µm, length L$_\textrm{med}=31$ mm, one slit at the center, viscosity $\nu=22.9$x$10^{-6}$ Pa/s and V$_{\textrm{rms}} = 427 $ m/s. Wall-projected curve is the longitudinal (along the waveguide length) pressure profile at the steady-state. The initial and boundary conditions are $P_{\textrm{in}} = 150$ mbar (gas load) and $t=0$ s, and $P=1$ mbar (at the exits), respectively. Purple arrow indicates the locus of gas inlet into the capillary (through the notch), which is represented both on the top capillary picture, with its two o'ring seals, and on the graphic. White line denotes pressure at half-maximum. }
\label{fig:Distri_PressureArgon}
\end{figure}
Thus, for an injected (continuously) backing pressure $P_{\textrm{in}} =150$ mbar in Argon, at $1$ mm from the fiber entrance (where the laser is supposed to be focused), and at the steady-state, the effective pressure is less than $\sim 50$ mbar, see Fig. \ref{fig:Distri_PressureArgon}.
For Helium, the HCW has L$_\textrm{med} = 13 $ mm with respect to the criterion L$_\textrm{med}$ $\geq$ $3$L$_\textrm{abs}$ discussed in Section \ref{sec:Theoretical considerations}. The effective pressure is $\sim 150$ mbar for $P_{\textrm{in}} =350$ mbar continuously injected, see Fig. \ref{fig:Distri_PressureHelium}. 
and agrees reasonably well with the expectations from our numerical analysis from Section \ref{sec:Theoretical considerations}, Fig. \ref{fig:fig1} and Fig. \ref{fig:fig2}, although experimental ionization levels $\eta$ for Ar and He are different and not optimal. It is also obvious that steady-state is quickly reached (after a few µsec), thus we always worked in equilibrium state, since laser pulses come one after another (the repetition rate) every $1$ ms, and can confidently claim that, given the time constant, the gas flux is continuous during our measurements. 
%
\subsection{The capillary environment}
\label{Gas_Fluidic_Comsol}
\begin{figure}[ht!]
\centering\includegraphics[width=11.4cm]{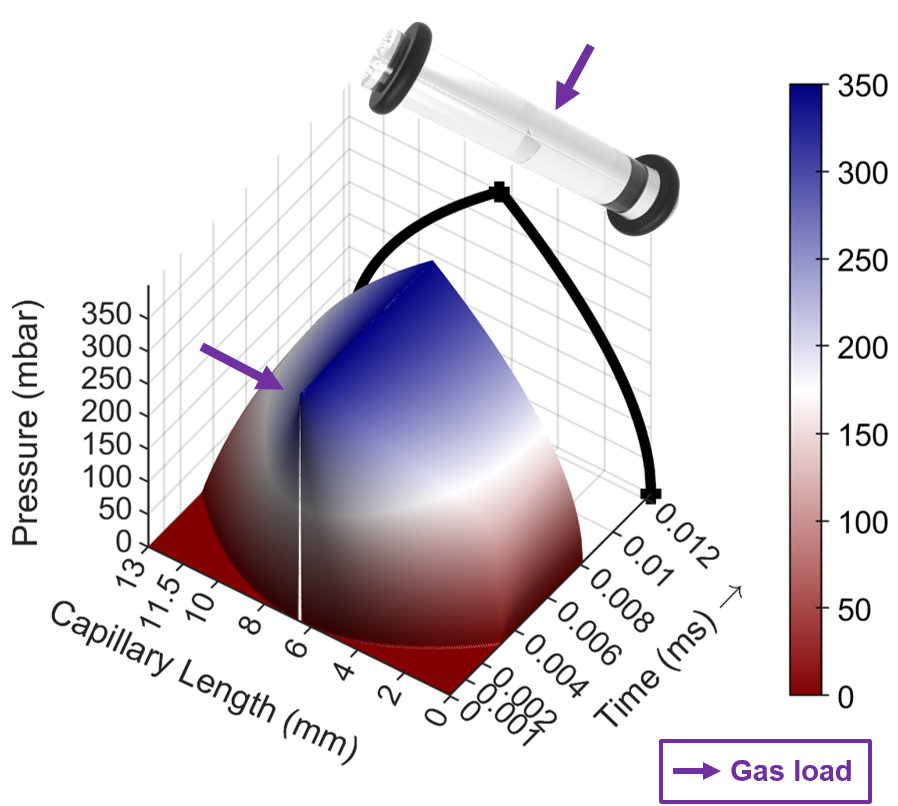}
\caption{Left: Same as Fig. \ref{fig:Distri_PressureArgon}, for Helium, length L$_\textrm{med}=13$ mm, viscosity $\nu=19.6$x$10^{-6}$ Pa/s and V$_{\textrm{rms}} = 1304 $ m/s. Wall-projected curve is the longitudinal pressure profile at the steady-state. The initial and boundary conditions are $P_{\textrm{in}} = 350$ mbar (gas load) and $t=0$ s, and $P=1$ mbar (at the exits), respectively. Purple arrow indicates the locus of gas inlet into the capillary (through the notch), which is represented both on the top capillary picture, with its two o'ring seals, and on the graphic. White line denotes pressure at half-maximum.}
\label{fig:Distri_PressureHelium}
\end{figure}

\begin{figure}[ht!]
\centering\includegraphics[width=15.4cm]{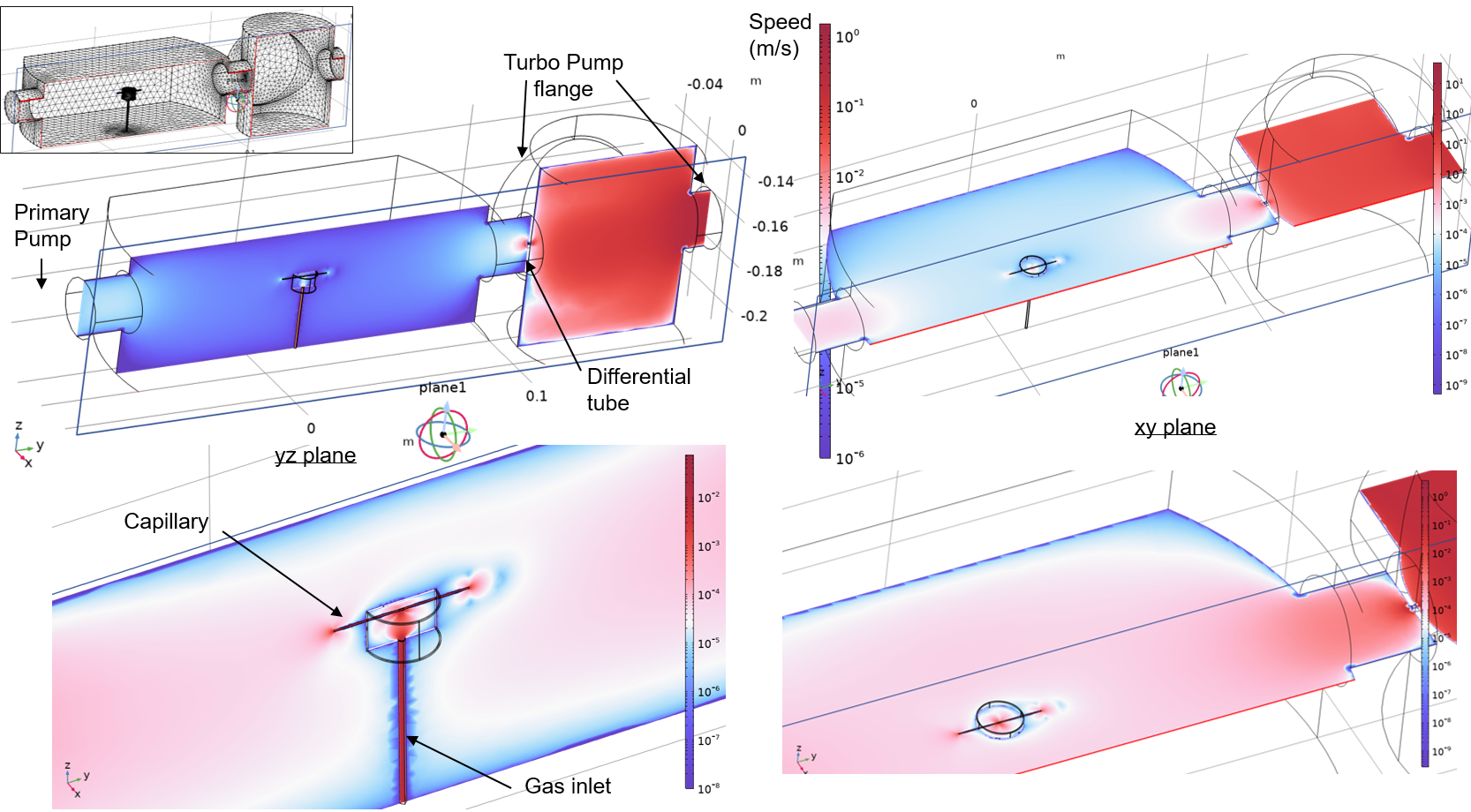}
\caption{Representation in (Incompressible flow) Laminar Flow of the gas speed field distribution (m/s) in the capillary vessel under Argon inlet and in the first vessel after the differential pumping tube, without laser. For visual convenience, and by symmetry, the vessels are cut at the middle in the \textit{yz} plane. Wireframe rendering shows the edges of the vessels. Input parameters are the followings: volumetric speed of primary pump$= 25$m$^3$/h, volumetric speed of TurboMolecular pump$ = 80$L/sec, speed of gas inlet $= 1$cm/s (estimated order of magnitude of about $10$ ml/min of volumetric flow), chosen boundary conditions for primary vacuum = $7$E$-3$ mbar, and for secondary vacuum (at differential tube output) $= 2$E$-4$ mbar). Vacuum for CF63 flange (turbomolecular pump) is set at $= 2$E$-3$ mbar, capillary diameter $= 300$ µm with slit (about $500$ µm wide) in the middle, gas inlet tube diameter $4$ mm (standard from Festo/Legris). The reference pressure $P_{\textrm{ref}}$ is taken equal to $10$E$-3$ mbar. Comsol Solver characteristics are: strong coupling, stationary study, AMG algorithm, and, by default, iterative GMRES, Multigrid preconditionner, tolerance error for convergence criterion for the solver $ = 5.95$E$-3$, coarse meshing (physics controlled) is shown in the top-insert. Sliced results of the left figures are a cut in the \textit{yz} plane, while those of the right figures correspond to a cut in the \textit{xy} plane. Logarithmic color scale has been adapted for the two bottom zoomed windows, for visual convenience. To situate the reader, laser comes from the left.}
\label{Laminar_Flow_andPressure}
\end{figure}

In order to strengthen the latter experimental results shown in Table \ref{table:3} in Section \ref{Multi_Atmo}, we performed simulations using Comsol MultiPhysics$\textsuperscript{\textregistered}$ to model the gas journey in the vessels, to complete simulations hereinabove presented into the capillary, as seen in Fig. \ref{fig:Distri_PressureArgon} and Fig. \ref{fig:Distri_PressureHelium}. We thus obtain an emulation of the gas behavior, under Laminar flow assumption, in the capillary vessel, surrounded the capillary, and under vacuum. For instance, in Fig. \ref{Laminar_Flow_andPressure}, we present simulation of the gas speed in the HHG heart source, \textit{i.e.}, the capillary box and the first vessel that follows (see Fig. \ref{fig:figCapillary}). The darkest red loci depict highest gas speeds, and as expected, are close to openings connected to vacuum pumps or where reduced gas conductance is encountered. In addition, a denser zone of higher gas speed at the capillary tips, which explains why it is tricky to precisely evaluate the HHG flow and to master the fluctuations in the output spectra. Moreover, this zone corresponds to the begotten plasma plume we distinguished during the experiments. In the bottom-left figure, one can clearly see that the capillary end that faces the Turbo pump flange describes a different gas speed behavior from the one facing the laser. 
\begin{figure}[ht!]
\centering\includegraphics[width=15.4cm]{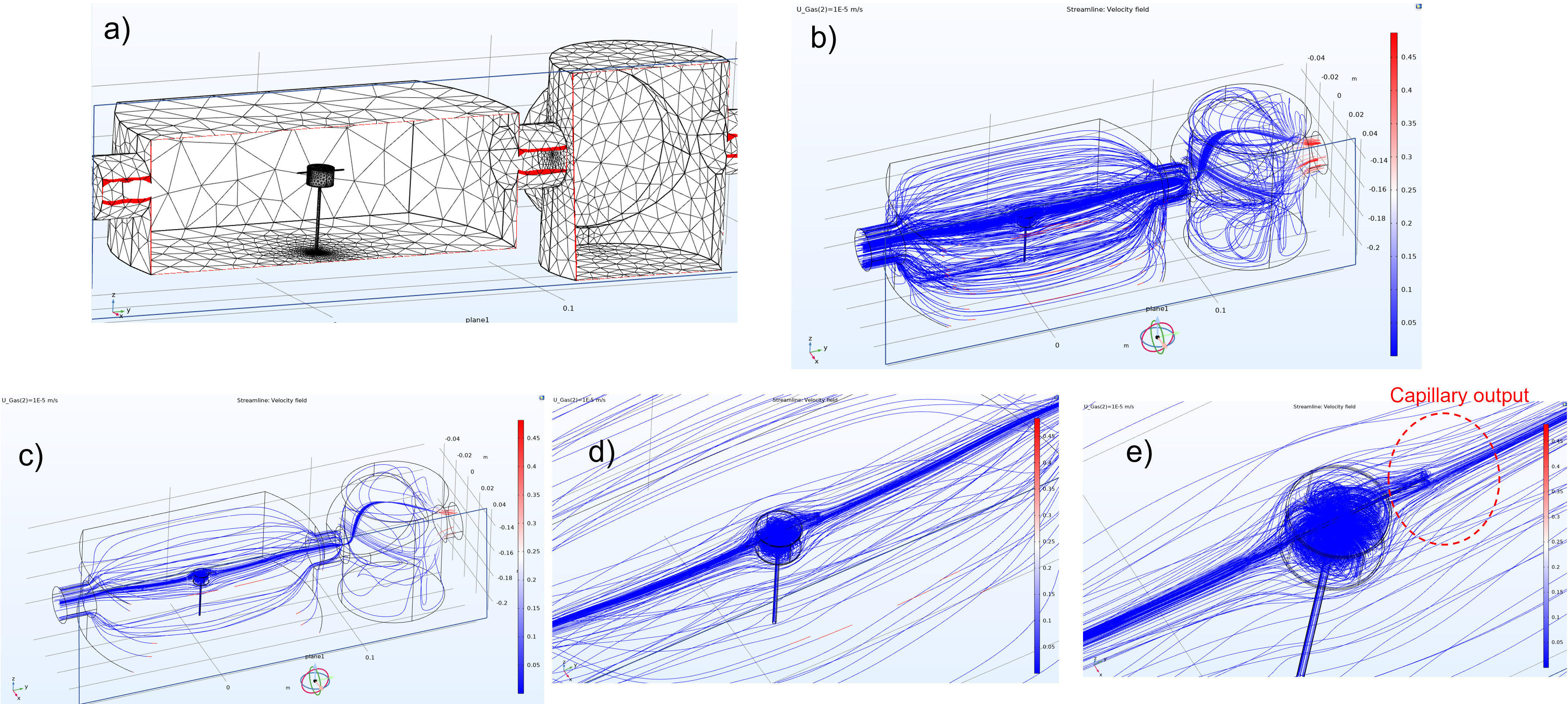}
\caption{Simulation from Comsol with Turbulent (Incompressible) Flow Interface, selecting the L-VEL model, one can trace the turbulent flow (several streamlines) at the passage in the differential pumping tube/at the ends of the HCW. It is then shown that this vortices behavior could be a drawback regarding the optical path. We chose to work on pressure levels instead of the mass flow volumetric since we know from the experiments the vacuum levels, and given the inlet pressure, we want to observe the turbulence flows. With a characteristics length of $200$ mm (the vessel chamber, seen by the capillary, which is, as seen above, in laminar flow, inside), and a gas speed injection of at least $1$ cm/s at the capillary output (see Fig. \ref{Laminar_Flow_andPressure}), then it is found Re $>3000$, so Turbulence flow can be considered. A Wall Distance Initialization step is first calculated. The reference pressure $P_{\textrm{ref}}$ is taken equal to $10$E$-3$ mbar. a) Meshing of the whole device. b) global view of the device with a few hundred of gas streamlines drawn. c) Only a few gas streamlines. d) A $1^{\textrm{st}}$ zoom. e) A $2^{\textrm{nd}}$ closer-view of the capillary. }
\label{Turbulent_Flow_andPressure_Streamlines}
\end{figure}
In Fig. \ref{Turbulent_Flow_andPressure_Streamlines}, the streamlines of the gas circulation into the both vessels are drawn. These results illustrate the earlier discussion about Fig. \ref{Fig_Jitter_Helium}. Indeed, the behavior of the gas in a turbulent structure, at the output of the capillary, could be one of the explanations of the jitter observed in the measured XUV spectra. Here, let us insist that this is only a qualitative observation and that the results might be seen as a first approximation. In the sense of the Knudsen number, the flow is Continuum, and Navier-Stokes equations are applied. Yet, Intermediate Regime\footnote{Is part of the Rarefied Gases Dynamics field. In substance, gas molecules should here be seen as ballistic and with non collective trajectories. We thus rather have a statistical behavior, with dominant wall interactions, than continuum hydrodynamical flow (with collisions between molecules of gas and fluid viscosity, and well directional flows, instead of diffuse jets in Transitional Regime). We will consider that in future simulations, with sufficient computational resources, or by using OpenFOAM, the open source CFD software, with its Direct Simulation probabilistic Monte Carlo (DSMC) Module. } would intuitively seem to be the right one to consider, but we are confident in the latter results that turbulences and vortices probably seat at some places into the vessels and into the optical path, and participate to perturb the HHG signal. 
\begin{figure}[ht!]
\centering\includegraphics[width=14.9cm]{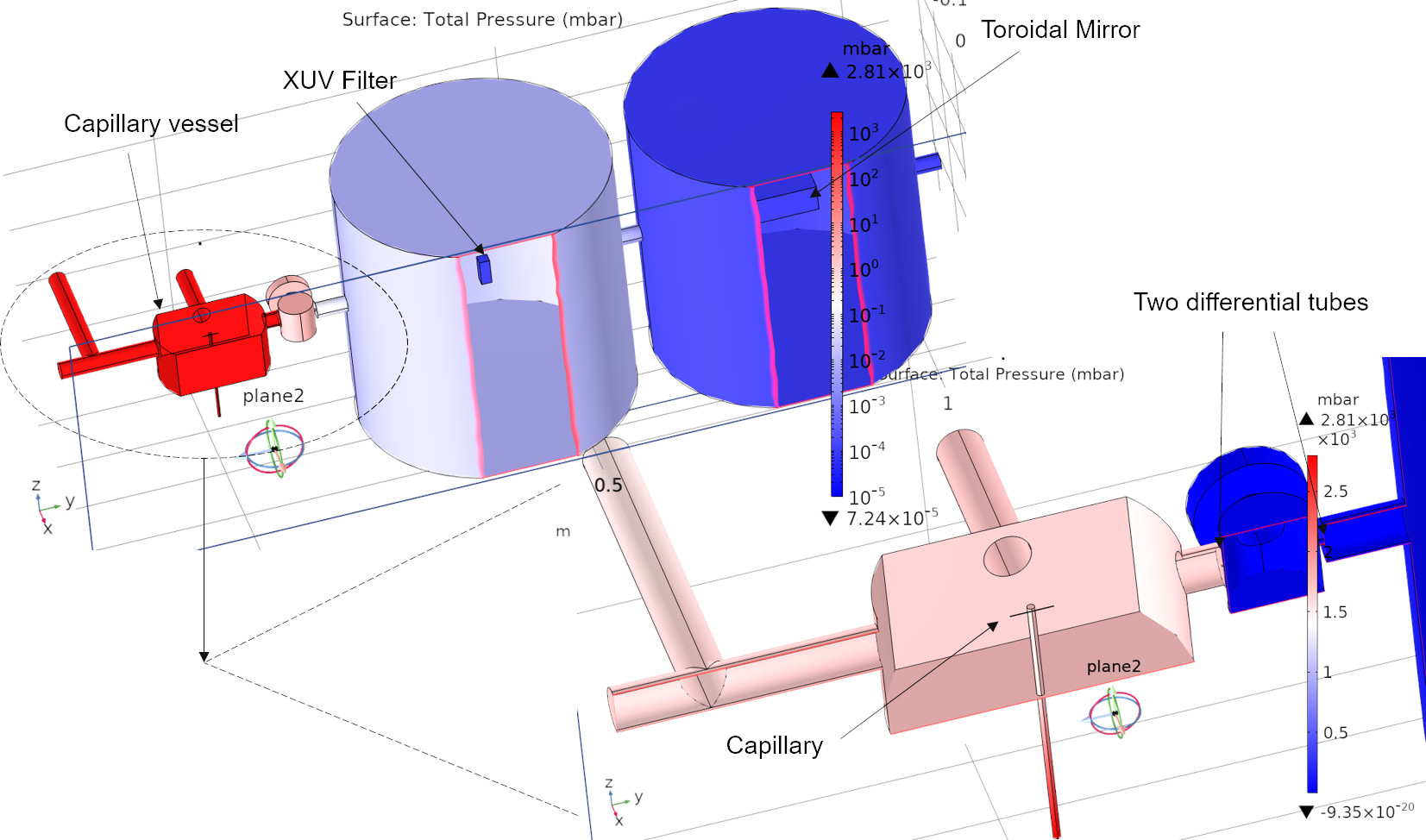}
\caption{Top figure in logarithmic scale shows the vacuum pressure in the capillary vessel and the first two chambers (Filtering and toroidal mirror), with their corresponding pumping devices entering into the input parameters of the simulation (Free Molecular regime). Bottom figure is a close-view of the capillary chamber (corresponding to Fig. \ref{Laminar_Flow_andPressure} which deals with gas speed distribution), with a half-cut in $yz$ plane, and with colormap in linear scale, for convenient clarity, so as to highlight pressure gradients. Two differential pumping (tubes) are set. Inlet (backing) pressure (from the bottom of the vertical tube feeding the capillary) of He is $2.8$ bars. This inlet pressure is approaching the limit in our configuration, because the turbomolecular pump connected to the chamber right downstream the capillary box is then beginning to work at the lowest limit of its nominal operating curve. To situate the reader, laser comes from the left. To account for ratio between drawn volumes, dimensions are at scale $1:1$, hence, for instance, rigid hosepipes (those connecting chambers) are $25$KF standard diameter.}
\label{Molecular_Flow_andPressure}
\end{figure}
In Fig. \ref{Molecular_Flow_andPressure} is displayed the steady-state vacuum pressure in each chamber that constitutes the beginning of the beamline, for a backing inlet of Helium at $2.8$ bars set to feed the capillary. It is clearly seen that from the filtering chamber, a vacuum pressure of $\sim 10^{-2}$ mbar is already reached. The similar pressure obtained in our experiment allows us to rely on these presented Comsol simulations in Free Molecular Flow.
\\
Finally, let us notice that here we considered Laminar flow into the HCW, nevertheless, we have tested in Turbulent flow, and cost in time is more important. Also, we have tried in Transitional (or Intermediate) flow (valid for Knudsen number about $0.1< $K$_n<10$), but the computational complexity in that case is too high (a computational resource of more than $120$GB of RAM is necessary in that case, because of a million Degrees of Freedom -DoF - induced by the nodes in the meshing), and exceeds our available computational resources. At least, at the HCW exits, where the gas expands and forms diffuse free jets into the surrounding vacuum, it likely falls into the Intermediate regime \cite{Mercier2000}. Finally, we shall consider Time-Dependent calculations where the steady-state solution could be obtained using a time-dependent compressible Laminar Flow (with velocity slip) and/or Free Molecular Flow formulations. Boundary conditions (inlet pressure, for instance) could also be gradually ramped in time to ensure numerical stability, and time-averaged quantities will be used to represent the steady-state regime.
\section{Conclusion}
To conclude this Supplementary Material, we have conducted numerical simulations that contribute to strengthening our understanding of the technical aspects of the beamline. 
%